\documentclass[preprint,aps,prd,superscriptaddress,showpacs,nofootinbib,eqsecnum,floats]{revtex4}
\usepackage{graphicx,color}
\usepackage{float}
\usepackage{amsmath}
\usepackage{graphicx}
\usepackage{bm}
\usepackage{color,colordvi}
\usepackage{epsfig}
\usepackage{natbib}
\usepackage{mciteplus}
\usepackage{mathcomp}
\begin{document}
\begin{flushright}
\end{flushright}
\newcommand  {\ba} {\begin{eqnarray}}
\newcommand  {\ea} {\end{eqnarray}}
\def\cM{{\cal M}}
\def\cO{{\cal O}}
\def\cK{{\cal K}}
\def\cS{{\cal S}}
\newcommand{\mh}{m_{h^0}}
\newcommand{\mw}{m_W}
\newcommand{\mz}{m_Z}
\newcommand{\mt}{m_t}
\newcommand{\mb}{m_b}
\newcommand{\be}{\beta}\newcommand{\al}{\alpha}
\newcommand{\lam}{\lambda}
\newcommand{\no}{\nonumber}
\def\ga{\mathrel{\raise.3ex\hbox{$>$\kern-.75em\lower1ex\hbox{$\sim$}}}}
\def\la{\mathrel{\raise.3ex\hbox{$<$\kern-.75em\lower1ex\hbox{$\sim$}}}}

\title{Type II Seesaw Higgsology and LEP/LHC constraints}


\author{Abdesslam Arhrib}
\affiliation{D\'epartement de Math\'ematiques, Facult\'e 
des Sciences et Techniques, Tanger, Morocco}
\author{Rachid Benbrik}
\affiliation{Facult\'e Polydisciplinaire, Universit\'e 
Cadi Ayyad, Sidi Bouzid, Safi-Morocco}
\author{Gilbert Moultaka\footnote{corresponding author}}
\affiliation{Laboratoire Charles Coulomb (L2C) UMR 5221 CNRS-Univ. Montpellier 2, Montpellier, F-France}
\author{Larbi Rahili}
\affiliation{Laboratoire de Physique des Hautes Energies 
et Astrophysique, Universit\'e Cadi-Ayyad, FSSM, Marrakech, Morocco} 
\date{\today}

\begin{abstract}
In the {\sl type II seesaw} model, if 
spontaneous violation of the lepton number conservation 
prevails over that of 
explicit violation,   
a rich Higgs sector phenomenology is expected to arise 
with light scalar states having mixed 
charged-fermiophobic/neutrinophilic properties. 
We study the constraints  on these light CP-even ($h^0$) 
and CP-odd ($A^0$) states from LEP exclusion limits, combined
with the so far established limits and properties of 
the  $125-126$~GeV  ${\cal H}$ boson discovered at the LHC. 
We show that, apart 
from a fine-tuned region of the parameter space, masses 
in the $\sim 44$ to $80$ GeV range escape from the LEP 
limits if the 
vacuum expectation value of the Higgs triplet is 
$\lesssim {\cal O}(10^{-3})$GeV, that is comfortably in
the region for 'natural' generation of Majorana neutrino masses 
within this model. In the lower part of the scalar mass spectrum
the decay channels ${\cal H} \to h^0 h^0, A^0 A^0$ lead 
predominantly to 
heavy flavor plus missing energy or to totally invisible Higgs 
decays, 
mimicking dark matter signatures without a dark matter candidate. 
Exclusion limits at the percent level of these (semi-)invisible 
decay channels would be 
needed, together with stringent bounds on the (doubly-)charged
states, to constrain significantly this scenario.
We also revisit complementary constraints from ${\cal H} \to 
\gamma \gamma$ and ${\cal H}  \to Z \gamma$ channels on the 
(doubly)charged scalar sector of the model, pinpointing non-sensitivity regions, 
and carry out a likeliness study for the theoretically allowed
couplings in the scalar potential.
 
\end{abstract}

\maketitle
\section{Introduction}
\noindent
Since the major discovery of a new bosonic particle at the LHC, 
\cite{Aad:2012tfa}, 
\cite{Chatrchyan:2012ufa}, denoted hereafter by ${\cal H}$, 
evidence has been accumulating in favor of it being Standard 
Model (SM) Higgs-like that culminated by the analyses of the 
full data sets of the LHC Run 1, both of the fit of its couplings 
to the SM gauge bosons and fermions \cite{ATLAS:2013sla}, 
\cite{CMS:yva} and of the determination of its intrinsic 
spin and parity properties \cite{ATLAS:2013mla}, 
\cite{CMS:xwa,CMS:bxa}, and is being continuously confirmed 
through the most recent results
\cite{Aad:2014aba,Khachatryan:2014ira}. 
On the other hand, the so far negative direct searches for physics 
beyond the SM (BSM) tend to become somewhat intriguing
 as the emblematic 
TeV scale limits are crossed. 
Nonetheless, various degrees of model assumptions go into these 
exclusion limits, and the LHC Run 2 is feverishly awaited, to 
improve on them or, better, to discover new physics
that could be revealed through more subtle kinematic  
configurations. 

In either case, one of the most important task in 
the coming years, with accumulated data and increased 
precision at the LHC, will be to improve on our experimental 
understanding of the properties of the ${\cal H}$ boson, 
particularly in the not yet well tested heavy fermion sector;
should they come ever closer to the SM expectation, then
BSM models that predict naturally this behavior
would become particularly attractive.      


The Higgs sector of the {\sl type II seesaw} model of Majorana
neutrino masses
\cite{Konetschny:1977bn, Cheng:1980qt, Lazarides:1980nt, 
Schechter:1980gr, Mohapatra:1980yp} provides
such a behavior, making of it an
interesting phenomenological setting for an extended Higgs 
sector. Indeed, the interplay between the $SU(2)_L$ doublet and
triplet scalar states present in this model, together with
 the large hierarchy between the associated vacuum 
 expectation values 
 accounting for the hierarchy between the neutrino masses and the 
 electroweak scale, imply {\sl naturally} that one of
 the scalar states is almost a SM Higgs, i.e. with tree-level 
 couplings to matter and gauge bosons deviating only by 
 $\displaystyle {\cal O}(m_{\nu}/M_{top})$ from the SM ones. 
 We stress that this comes about without the need of a 
 'decoupling regime' entailing heavy BSM degrees of 
 freedom, [in contradistinction with some of the other
 fashionable BSM models relying on supersymmetric, extra 
 space dimensional, or compositeness scenarios.] 
 The other predicted scalar states of the model could thus 
 still be accessible at 
 the LHC energies, in particular a distinctive doubly 
 charged state,  as well as a singly charged, a CP-even 
 and a CP-odd neutral states, {\sl even if ${\cal H}$ would become more and more 
 compatible with the SM}. However these new states, whether 
 charged or neutral, are typically not easy to produce and correspondingly 
 to exclude. Their single production cross-sections are suppressed by 
 $\displaystyle {\cal O}(m_{\nu}^2/M_{top}^2)$ for the same reason as 
 above, and one has to resort to the unsuppressed 
 pair production with less available phase space.

The discovered ${\cal H}$ boson can be identified either
with the lighter CP-even scalar state  of the model, $h^0$,
or with the heavier $H^0$.\footnote{Note that there is 
also the possibility
that these two states be essentially degenerate, each carrying
an equal fraction of the SM-like couplings, however this occurs 
in an
extremely fine-tuned region of the parameter space.}
In the following we will refer to these two possibilities respectively as the
{\sl $h^0$-, $H^0$-scenario}. 
The distinction between the {\sl $h^0$-scenario} and
the {\sl $H^0$-scenario} is parametrically controlled by the relative magnitudes of the 
explicit and spontaneous lepton number violating (LNV) parameters present in 
the model \cite{Arhrib:2011uy}. It can thus have a 
bearing on the ultraviolet (UV) completion that underlies the
dynamical origins of these two sources of LNV. 
Choosing phenomenologically the magnitude of 
explicit violation to be of the same order or (much) smaller than spontaneous 
violation, resulting respectively in the {\sl $h^0$-}, {\sl $H^0$-scenarios} with
electroweak scale scalar masses, can be 
theoretically justified in models where the two sources of LNV are dynamically related
(with, for instance, a loop induced effective LNV). In contrast, the
{\sl $h^0$-scenario} was initially motivated by the 
conventional assumption that the mass parameters in the triplet sector are
of order the grand unification (GUT) scale, leading to a triplet sector (including
explicit LNV) that is too heavy to be relevant at the 
electroweak scale. Even in such settings where the origin of the explicit lepton number
violation is dynamically unrelated to that of spontaneous violation,  one can obviously
still motivate light triplet states by assuming an UV completion much lighter than 
the GUT scale. In this case the magnitude of the explicit violation
can again be comparable to or smaller than that of spontaneous violation, so that
both {\sl $h^0$-, $H^0$-scenarios} can occur.

Most of the recent phenomenological studies have assumed the 
{\sl $h^0$-scenario}  
\cite{Akeroyd:2012ms},\cite{Wang:2012ts}, \cite{Akeroyd:1900zz}, 
\cite{Aoki:2011pz},
\cite{Dev:2013ff}, \cite{Akeroyd:2010je}.
In the present paper we consider both scenarios but focus more on the 
{\sl $H^0$-scenario}.
In the latter, the lighter CP-even and CP-odd
states, $h^0, A^0$, become essentially degenerate with a mass
below $125$GeV, while the
charged and doubly-charged states $H^\pm, H^{\pm \pm}$ can lie 
anywhere above their present exclusion limits. It will thus be
important to take into account all present direct
and indirect experimental constraints in the Higgs sector in order
to narrow down the viable parameter space regions of the {\sl $H^0$-scenario}. 
In particular, stringent exclusions come from the LEP limits on direct 
searches for light scalar and pseudo scalar states as well as from the
Z-boson width. Improving the measurement of the 
${\cal H}$ production
cross-sections and decays into $W,Z$ gauge bosons and fermions, 
as expected at the LHC Run 2,  
will not constrain significantly the model. 
The $\gamma \gamma$ (and $Z \gamma$)
decay channels will be in principle more sensitive to the extended scalar sector 
through the charged and doubly charged states loop effects.  We will show, however, 
that these loop effects can still be blind to nearby (doubly) charged  states due to 
a somewhat generic regime of destructive interference present 
in both  {\sl $h^0$-}, 
{\sl $H^0$-scenarios}. Direct searches for (doubly) charged states become then
particularly compelling. In the {\sl $H^0$-scenario}, though,
further constraints will come from LHC limits on invisible/undetected ${\cal H}$ 
decays into pairs of $h^0$ or $A^0$. 
One of the main results of this paper will be that the {\sl $H^0$-scenario} 
is particularly difficult to exclude. 
Parts of its parameter space that are favored by the smallness of neutrino masses 
totally evade the LEP exclusion limits on light states. 
To narrow down the allowed parameter space
would require  achieving very strict limits on invisible/
undetected ${\cal H}$ decays  combined with increased lower bounds 
on the (doubly)charged Higgs masses states from direct searches at 
the LHC, while the other decay channels of ${\cal H}$ will have a 
marginal impact as they remain essentially SM-like.
   

The rest of the paper is organized as follows: in Section
\ref{sec:model} we recall the main ingredients of the scalar potential, the
physical scalar states and their mass spectrum and
couplings, distinguishing explicitly the model parameter domains corresponding to
the {\sl $h^0$-} and {\sl $H^0$-scenarios}. We also describe briefly possible 
UV-completions of the less conventional {\sl $H^0$-scenario}.  
Section \ref{sec:HZgaga} is devoted to a re-analysis of the $\gamma \gamma$ and 
$Z \gamma$ Higgs decay channels within our model. We provide a 
general parameterization that encompasses 
the two channels and the two {\sl ${\cal H}$-scenarios} and 
discuss  the charged and doubly charged scalar states loop 
effects. 
We revisit the correlation between the two decay channels and its 
phenomenological incidence, given the foreseen future 
experimental low precision on the $Z \gamma$ channel.
We also demonstrate the existence of a generic screening of the 
charged scalar loop effects. Section \ref{sec:lam1lam4}  is 
devoted to a study of the likeliness of the various parameter 
space regions of the relevant couplings as dictated by 
the unitarity and vacuum stability (U-BFB) 
constraints in the scalar potential.   
A detailed phenomenological study of the {\sl $H^0$-scenario} with 
the CP-even and CP-odd scalar states lighter
than ${\cal H}$,  is carried out
in section \ref{sec:lightS}, taking into account LEP and LHC
constraints. Section \ref{sec:conclusion} contains the
conclusions, and some technical material is given in the
appendices. 


\section{The model \label{sec:model}} 

The {\sl type II seesaw} model\footnote{It 
has become customary in the literature to dub this model HTM or 
DTHM when focusing exclusively on the scalar sector. 
We will however stick here to the original name as the coupling
to fermions will be of an issue in our study.} consists of the
Standard Model with an additional colorless  $SU(2)_L$ triplet 
scalar field $\Delta$ with hypercharge $Y_\Delta=2$. 
Denoting by $H$ the standard  scalar field $SU(2)_L$ doublet
and taking the $2 \times 2$ traceless matrix representation for 
the triplet $\Delta$ we write the two multiplets in terms of
complex valued scalar components as,  
\begin{eqnarray}
H=\left(\begin{array}{c}
                      \phi^+ \\
                      \phi^0 \\
                    \end{array}
                  \right),~~~~~~~~\Delta &=\left(
\begin{array}{cc}
\delta^+/\sqrt{2} & \delta^{++} \\
\delta^0 & -\delta^+/\sqrt{2}\\
\end{array}
\right) \label{eq:HDelta}
\end{eqnarray}
with the conventional electric charge assignment for the doublet
and following
 $Q= I_3 + 
\frac{Y}{2}$ with $I_3= -1,0, 1$ for the triplet. 
The Lagrangian of the model reads
\begin{eqnarray}
\mathcal{L} &=&
(D_\mu{H})^\dagger(D^\mu{H})+Tr\{(D_\mu{\Delta})^\dagger(
D^\mu{\Delta})\} -V(H, \Delta) + \mathcal{L}_{\rm Yukawa} + 
\mathcal{L}_{\rm Gauge}
\label{eq:DTHM}
\end{eqnarray}
where
\begin{eqnarray}
D_\mu{H} &\equiv& \partial_\mu{H}+igT^a{W}^a_\mu{H}+i\frac{g'}{2}B_\mu{H} \label{eq:covd1}\\
D_\mu{\Delta} &\equiv&
  \partial_\mu{\Delta}+ig[T^a{W}^a_\mu,\Delta]+ig' \frac{Y_\Delta}{2} B_\mu{\Delta} \label{eq:covd2}
\end{eqnarray}
(${W}^a_\mu$, $g$), and ($B_\mu$, $g'$) denoting respectively the 
$SU(2)_L$ and $U(1)_Y$ gauge fields and couplings
and $T^a$ the $SU(2)$ generators in the fundamental representation.
The most general renormalizable potential consistent with all
the symmetries of the model is given by 
\begin{eqnarray}
V(H, \Delta) &=& -m_H^2{H^\dagger{H}}+\frac{\lambda}{4}(H^\dagger{H})^2+M_\Delta^2Tr(\Delta^{\dagger}{\Delta})
+[\mu(H^{\rm T} {i}\sigma_2\Delta^{\dagger}H)+{\rm h.c.}] \nonumber\\
&&+\lambda_1(H^\dagger{H})Tr(\Delta^{\dagger}{\Delta})+\lambda_2(Tr\Delta^{\dagger}{\Delta})^2
+\lambda_3Tr(\Delta^{\dagger}{\Delta})^2 +\lambda_4{H^\dagger\Delta\Delta^{\dagger}H}
\label{eq:Vpot}
\end{eqnarray}

\noindent
where $Tr$ is the trace over $2\times2$ matrices. 
$\mathcal{L}_{\rm Yukawa}$ contains on top of the Yukawa sector
of the SM the extra term 
\begin{equation}
- L^{\rm T} Y_{\nu}  \otimes C \otimes i \sigma_2 \Delta L  + {\rm 
h.c.} \subset \mathcal{L}_{\rm Yukawa} \label{eq:yukawa}
\end{equation}

\noindent
where $L$ denotes the $SU(2)_L$ doublets of the 
left-handed leptons, $Y_
{\nu}$ denotes a $3\times3$ matrix of Yukawa couplings in the 
lepton flavor space, suppressing flavor indices for simplicity, 
$C$ the charge conjugation operator, and $\sigma_2$ the second 
Pauli matrix.  The tensor product stresses the fact that these
operators act on different spaces. In this paper we assume $Y_{\nu}
$ to be diagonal, ignoring possible lepton flavor violation that 
could originate from the above term.

The spontaneous electroweak symmetry breaking is triggered by
the structure of the minima of $V(H, \Delta)$ in the ten
dimensional space of real valued scalar fields Eq.(\ref{eq:HDelta}
). The physically interesting minimum will correspond to non-zero
vacuum expectation values $v_d$ and $v_t$ respectively for
$H$ and $\Delta$ along electrically neutral directions with
$v_d \gg v_t$, such that the electroweak scale, 
$\sqrt{v_d^2 + 2 v_t^2} \equiv 246$GeV, is given
essentially by $v_d$ while $v_t$ induces neutrino Majorana
masses through the Yukawa coupling Eq.(\ref{eq:yukawa}). At the
tree-level, the {\sl
necessary} electroweak symmetry breaking conditions read

\begin{eqnarray}
  M_\Delta^2 &=& \frac{2\mu{v_d^2}-\sqrt{2}(\lambda_1+
\lambda_4)v_d^2v_t-2\sqrt{2}(\lambda_2+\lambda_3)v_t^3}{2\sqrt{2}v_t} \label{eq:ewsb1}\\
  m_H^2 &=& \frac{\lambda{v_d^2}}{4}-\sqrt{2}\mu{v_t}+\frac{(\lambda_1+\lambda_4)}{2}v_t^2  \label{eq:ewsb2}
\end{eqnarray}
We are not interested here in fine-tuning issues related to these
equations when requiring $v_t \ll v_d$ in order 
to cope with the neutrino masses. 
We only note here that the regimes $M_\Delta \sim \mu \gg
v_d$ or $\mu \ll v_t$ would not require fine-tuning. 
The first leads to a seesaw effect but to a BSM sector totally
out of reach at the LHC, while the second does not feature a
seesaw effect but is naturally compatible
with neutrino masses and implies electroweak scale BSM physics.

After electroweak symmetry breaking the 10 scalar states decompose
into 7 massive physical Higgses, $h^0, H^0, A^0, H^{\pm}, H^{\pm 
\pm}$ and 3 Goldstone bosons, with
3 angles $\alpha, \beta, \beta'$ mixing the neutral 
and singly charged doublet and triplet states, 
\begin{eqnarray}
h^0 = \cos\alpha \, h + \sin\alpha \, \xi^0, 
~~~~~&&~~~H^0 = - \sin\alpha \, h + \cos\alpha \, \xi^0 
\label{eq:mixh0H0}\\
A^0 = - \sin \beta \, Z_1 + \cos \beta \, Z_2,
\,&&~~~G^0 =  \cos \beta \,  Z_1 + \sin \beta \, Z_2 
\label{eq:mixA0G0}\\
G^\pm = \cos\beta^{'} \phi^{\pm}+ \sin\beta^{'} \delta^{\pm},
~\;&&~~~H^\pm = -\sin\beta^{'} \phi^{\pm}+ \cos\beta^{'}
\delta^{\pm} \label{eq:mixG+H+} \\
H^{\pm\pm} = \delta^{\pm\pm} ~~~~~~~~~~~~~~~~~~~~~~~&&
\label{eq:mixH++}
\end{eqnarray}
with the definitions 
$\phi^0=\frac{1}{\sqrt{2}} (v_d+h+i Z_1), \delta^0=\frac{1}{\sqrt{2}} (v_t+\xi^0+i Z_2)$. 
Hereafter we make some general comments and then focus on the 
features directly related to the scenario
under consideration. [For more details about the Higgs spectrum 
and couplings the reader may refer to \cite{Perez:2008ha}, 
\cite{Arhrib:2011uy}.] The large hierarchy between $v_d$ and $v_t$
implies that $\sin \beta$ and $\sin \beta'$ 
in Eqs.(\ref{eq:mixA0G0},~\ref{eq:mixG+H+}) are always suppressed,
that is $A^0$ and $H^\pm$ carry essentially triplet components
while the neutral and charged Goldstone bosons are essentially
parts of the doublet. In contrast, $\sin \alpha$ scans all 
possible values but is  
either close to $0$ or close to $\pm 1$, apart from a fine-tuned 
region with maximal mixing $\sin \alpha \simeq 1/\sqrt{2}$, again 
due to the smallness of $v_t/v_d$.
 As a consequence, one has
generically two possibilities which we will dub  `${\cal H}${\sl -scenarios'}: 
\begin{itemize}
\item[] \underline{{\sl $h^0$-scenario}}: the lighest CP-even state $h^0$ is
SM-like ($\sin \alpha \simeq 0$); this occurs typically when
\begin{eqnarray}
\mu & \gtrsim& \hat{\mu}'_{(\mp)} \equiv
(\lambda  - \frac{2 (\lambda - \lambda_1 - \lambda_4)}{2 \mp 
\sqrt{12 + 2 k}}) \frac
{v_t}{\sqrt{2}} + {\cal O}(\frac{v_t^3}{v_d^2})
\label{eq:mucritical1}
\end{eqnarray}
where $k>0$ is defined implicitly through 
\begin{equation}
\cos \alpha|_{\mu =\hat{\mu}'_{(\mp)}}  = 1 - k \frac{v_t^2}
{ v_d^2} + {\cal O}(\frac{v_t^3}{v_d^3})
\label{eq:cosk}
\end{equation}
\item[]  \underline{{\sl $H^0$-scenario}}: the heaviest CP-state $H^0$ is SM-like
 ($|\sin \alpha| \simeq 1$); this occurs 
typically when
\begin{eqnarray}
0< \mu & \lesssim& \hat{\mu}_{(\pm)} \equiv  (\frac{\lambda}{\sqrt{2}} - \frac{\lambda - 
\lambda_1 - \lambda_4}{\sqrt{2} \pm \sqrt{k}}) v_t +
{\cal O}(\frac{v_t^3}{v_d^2}) \label{eq:mucritical2}
\end{eqnarray}
where $k>0$ is defined implicitly through
\begin{equation}
|\sin \alpha|_{\mu =\hat{\mu}_{(\pm)}} = 1 - k \frac{v_t^2}{v_d^2} + {\cal O}(\frac{v_t^3}{v_d^3})
\label{eq:sink}
\end{equation}
\end{itemize}

The upper bound in  Eq.(\ref{eq:mucritical2}) has been derived
in \cite{Arhrib:2011uy} to which we refer the reader for more
details, while the bound in 
Eq.(\ref{eq:mucritical1}) is new.\footnote{In both cases, the 
symbols
$\pm$ and $\mp$ correspond to the sign of 
$\lambda - \lambda_1 - \lambda_4$ and select the relevant 
(sufficient and necessary)    
bound to be used depending on this sign. 
In particular, this implies that the relevant bounds
always satisfy 
$\hat{\mu} \lesssim \frac{\lambda}{\sqrt{2}} v_t \lesssim \hat{\mu}'$.} 
As can be seen from Eqs.(\ref{eq:cosk}, \ref{eq:sink}), $k$ 
parameterizes the purity of the 
SM-like Higgs state in both scenarios; for $k=0$ all
the couplings  of the ${\cal H}$ state are strictly those of the 
SM. 
Phenomenologically, $\delta \equiv k v_t^2/v_d^2$ can 
be identified with a given precision at which the SM couplings 
are measured. It is readily related to the general effective 
parameterization
\cite{LHCHiggsCrossSectionWorkingGroup:2012nn}. Thus $\hat{\mu}$
and $\hat{\mu}'$ determine the domains of 
$v_t, \mu, \lambda, \lambda_1 + \lambda_4$ that will not 
be excluded by merely narrowing down the experimental 
determination of the couplings, 
should they remain consistent with the SM within a given  projected
precision, \cite{CMS:2013xfa},\cite{ATLAS-projections}.
In fact, given the suppression factor $v_t^2/v_d^2$, it is only
for very high precision, consistent with $k \sim {\cal O}(1)$,
that $\hat{\mu}$ and $\hat{\mu}'$ become explicitly sensitive
to $v_t$ and $\lambda_1 + \lambda_4$, c.f. 
Eqs~.(\ref{eq:mucritical1},
\ref{eq:mucritical2}). This sensitivity is quickly lost 
for a precision $\delta$ of order a few percent. 
An implicit dependence on $v_t$ and 
$\lambda_1 + \lambda_4$ will remain, however, in
$\lambda$ itself, when $m_{\cal H}$ is fixed at its observed
value. For instance, in the 
{\sl $H^0$-scenario} it will be given by Eq.(\ref{eq:lambda}).

We discuss now a little further some qualitative features
of the two scenarios:
\begin{itemize}
\item in the {\sl $h^0$-scenario} the new scalar states are of the 
same order or
heavier than the SM-like Higgs state. The SM tree-level 
predictions {\sl in the SM sector} are then generically only 
slightly modified. 
Signatures of the model can then come only from direct evidence 
for the new scalar states including the doubly charged one, 
together with evidence for Majorana neutrino masses and lepton 
number violating processes. However, the 
rationale for a seesaw mechanism operates strictly speaking when 
$M_\Delta$ and $\mu$ are assumed to have high scale (perhaps
GUT scale) origin, leading to
$\mu \sim M_\Delta \gg {\cal O}(1)$TeV. 
The case of such very large $\mu (\sim M_\Delta \sim M_{\rm GUT})$ 
is however of no relevance  if one
is interested at all in testable Higgs 
phenomenology at the colliders, since all the non-
standard Higgs states decouple from the low energy (TeV) sector.
In fact, taking $\mu \gtrsim {\cal O}(20) \, v_t$ for typical 
values of the $\lambda_i$'s, already leads to states heavier than 
${\cal O}(1)$ TeV and virtually out of reach at the LHC!\footnote{It is to be noted 
that even in this case low energy precision observables 
are not trivially consistent with the SM predictions.
For instance deviation of the $\rho$-parameter from its SM tree-
level value, $\rho \approx 1 - 2 v_t/v_d < 1$, requires 
non-standard contributions to the radiative corrections in order to
restore $\rho \gtrsim 1$ to reach consistency with the 
experimental value \cite{Nakamura:2010zzi}, \cite{Kanemura:2012rs}
.}

\item in the {\sl $H^0$-scenario} the $H^\pm$
and $H^{\pm \pm}$ masses are bounded from above disfavoring 
configurations with $\lambda_1 + \lambda_4 -\lambda <0$ in order
to cope with the present experimental lower bounds on these
masses. Furthermore, the $h^0$ and $A^0$ states are lighter
than the SM-like Higgs state and their masses will decrease
with decreasing $\mu$; the ensuing phenomenological issues
will be addressed in section \ref{sec:lightS}. Here we comment
on the plausibility of this $\mu \sim v_t$ or $\ll v_t$ scenario
from the model-building point of view. 

As stated in the 
introduction, the relative magnitude of $\mu$ and $v_t$ can be
related to the status of the UV origin of lepton number 
conservation
whose violation at the electroweak scale is triggered 
by these two parameters independently. 
Examples of scenarios 
where small $\mu$ is generated through one-loop suppressed 
effective operator, have been given in
non-supersymmetric \cite{Kanemura:2012rj} or supersymmetric \cite
{Franceschini:2013aha} extensions of the {\sl type II seesaw} 
Lagrangian 
Eqs.~(\ref{eq:DTHM} -- \ref{eq:Vpot}). Although different, 
these scenarios have in common the assumption that $\mu$ 
triggers non-vanishing $v_t$, leading typically to
$v_t \sim {\cal O}(\mu)$. However, the latter assumption is
not necessary and in fact does not fully account for the general
structure of the scalar potential Eq.~(\ref{eq:Vpot}); 
as one can see from Eqs.(\ref{eq:ewsb1}, \ref{eq:ewsb2}), there 
are also $\mu$ independent contributions from the 
dimensionless couplings between $H$ and $\Delta$, mainly through 
the combination $\lambda_1 + \lambda_4$,  so that the relative 
size $\mu/v_t$ can be much smaller than ${\cal O}(1)$ and 
still remain consistent with electroweak symmetry breaking.
Furthermore, the loop suppressed mechanisms of  
\cite{Kanemura:2012rj} and
\cite{Franceschini:2013aha} can operate even in this context.
 
Before ending this discussion we sketch 
yet another possibility, namely  that
lepton number violation be seeded by gravitational effects.
We note first that the $\mu$ parameter is natural in the sense 
that putting it to
zero increases the symmetry of the Lagrangian Eq.~(\ref{eq:DTHM})
 [that is to a  {\sl global} $U(1)$ symmetry associated 
with the lepton number with charge assignments $l_{\Delta} = -2$, 
$l_H =0$, $l_{l} = -l_{\bar{l}}=1, l_{q}=l_{\bar{q}} =0$].
A corollary is that a small $\mu$ remains small against radiative
corrections before any spontaneous symmetry breaking, 
since as can be seen from Eq.~(\ref{eq:Vpot}) loop corrections to 
the operator``$H^{\rm T} \Delta^\dagger H$" 
will be proportional to $\mu$ 
itself. These properties are preserved in extensions of the model
to larger gauge groups where $H$ and $\Delta$ would be parts
of some multiplets $\Phi$ and $\Sigma$ respectively in the 
fundamental and adjoint representations of this gauge group. 
One can thus require consistently the conservation of the lepton
number at the Lagrangian level, and this symmetry will be exactly
preserved in the full theory, as far as energy scales above the 
spontaneous breaking of some of the gauge symmetries are 
concerned. However, this global symmetry is expected to be broken
when (quantum) gravitational effects are switched on
(see \cite{Banks:2010zn} for a recent reappraisal). This breaking
would manifest itself at scales lower than the Planck scale $M_{Pl}
$,  
through the presence of higher dimensional operators suppressed
by powers of $M_{Pl}$. The leading effect originates from the
dimension-5 gauge invariant and lepton number violating operator, 
$\Phi^\dagger \Phi (\Phi^{\rm T} \Sigma^{\dagger} \Phi)$ (where 
we suppressed the group indices for simplicity).
An effective $\mu$ parameter of order 
$\langle \Phi \rangle^2/M_{Pl}$ will thus be generated 
at an intermediate scale where spontaneous breaking of (some of
) the gauge symmetries takes place, triggered by the vacuum 
expectation value of the field $\Phi$. This mechanism has two
nice features --for one thing, it relates the magnitude of $\mu$ 
to the scale at which the underlying gauge symmetry is broken;
obviously $\langle \Phi \rangle \sim {\sl O}(M_{GUT})$ leads back
to {\sl $h^0$-scenario}, while assuming a desert between the electroweak
and the Planck scales leads to a $\mu$ of order 
$10^{-14} - 10^{-15}$GeV at the edge of the phenomenologically 
acceptable values in {\sl $H^0$-scenario} --for the other, by
choosing $246$GeV $\ll \langle \Phi \rangle \ll M_{GUT}$ one can 
arrange to have all the extended sector too heavy to be accessible 
to the colliders but still keeping $\mu$ very small independently
of the value of $v_t$. We do not
dwell further here on an explicit building of the model which 
is out of the scope of the present paper. We only take the above
general arguments as a motivation to study  the phenomenology      
of configurations satisfying $\mu/v_t \ll 1$.
\end{itemize}
 
When dealing with the {\sl $H^0$-scenario} it will be instructive 
to expand the various scalar field masses
in terms of the small ratio $\mu/v_t$. Starting from the
exact tree-level expressions \cite{Arhrib:2011uy}, one
finds,

\begin{eqnarray}
&& m_{h^0}^2 = 2\,(\lambda_{2 3}^+ \!-\! 
\frac{(\lambda_{1 4}^+)^2}{\lambda})\,v_t^2 +
\frac{1}{\sqrt{2}}\frac{\mu}{v_t}\,(v_d^2 + 4\,(2\,\lambda 
\!- \!\lambda_{1 4}^+)\,\lambda_{1 4}^+ \,\frac{v_t^2}{\lambda^2}) + {\cal O}(\frac{\mu^2}{v^2_t}) \label{eq:mh0}\\
&& m_{H^0}^2 = \lambda\,\frac{v_d^2}{2} + 
2\,\frac{(\lambda_{1 4}^+)^2}{\lambda}
\,v_t^2 - 2\,\sqrt{2}\,\frac{\mu}{v_t} 
\, (2\,\lambda \!-\! \lambda_{1 4}^+)\,\lambda_{1 4}^+\, \frac{v_t^2}
{\lambda^2} 
+{\cal O}(\frac{\mu^2}{v^2_t}) \label{eq:mH0}\\
&& m_{A^0}^2 = \frac{1}{\sqrt{2}} \frac{\mu}{v_t}\,(v_d^2 + 4\,v_t^2) \label{eq:mA0}\\
&& {m^2_{H^+}} = (-\frac{\lambda_4}{4} + 
      \frac{1}{\sqrt{2}} \frac{\mu}{v_t} 
      ) \, (v_d^2 + 2 \, v_t^2) +{\cal O}(\frac{\mu^2}{v^2_t}) 
      \label{eq:mH+}\\
&& {m^2_{H^{++}}} = -\lambda_4\,\frac{v_d^2}{2} 
- \lambda_3\,v_t^2
+ \frac{\mu}{v_t}\,
\frac{v_d^2}{\sqrt{2}}  +{\cal O}(\frac{\mu^2}{v^2_t}) \label{eq:mH++}
\end{eqnarray}
where we used the shorthand notation
$\lambda_{1 4}^+ \equiv \lambda_1 + \lambda_4$,
$\lambda_{2 3}^+ \equiv \lambda_2 + \lambda_3$, and kept 
negligible $v_t^2$ contributions in order to assess small
splitting as well as no-tachyon conditions.

The splitting $\Delta m^2_0 \equiv m_{h^0}^2 - m_{A^0}^2$
verifies,

\begin{equation}
\Delta m^2_0 \leq 2 \,(\lambda_2 + \lambda_3 
- \sqrt{2} \, \frac{\mu}{v_t}) \, v_t^2 + 
{\cal O}(\frac{\mu^2}{v^2_t})  \label{eq:splitting}
\end{equation}
and thus remains very small as compared to the allowed values
of $m_{h^0}$ and $m_{A^0}$ (see section \ref{sec:lightS}), even 
for the largest allowed values of $v_t (\sim 1 GeV)$ and 
$(\lambda_2 + \lambda_3)|_{\rm max} = \kappa \frac{\pi}{5} \approx 
5$ (see the discussion in appendix \ref{app:lambda1} and 
\cite{Arhrib:2011uy}). In the sequel we will always assume 
$m_{h^0} \simeq m_{A^0}$. Note also that the exact form for
$m_{A^0}$ comes naturally proportional to $\mu/v_t$. This is
reminiscent of the would-be-Goldstone character of $A^0$
in the limit $\mu \to 0$, $v_t \neq 0$ of spontaneous breaking
of the continuous lepton number symmetry. It stresses as well the
fact that the magnitude of $m_{A^0}$ is not controlled solely
by $\mu$, but also by $v_t$ as two independent parameters. 

As can be seen from Eq.~(\ref{eq:mh0}), one retrieves consistently
an approximate lower bound on $\mu$ that ensures a non-tachyonic 
$h^0$ (see \cite{Arhrib:2011uy} for the exact bounds).
Also this bound evaporates if
$\lambda_{2 3}^+ \!-\! 
\frac{(\lambda_{1 4}^+)^2}{\lambda} > 0$, the latter being
consistent with a bounded from below potential, see appendix 
\ref{app:lambda1}. In any case, the no-tachyon issue is
 independent of the sign of
of $\lambda_{1 4}^+$, thus illustrating the fact
that $M_\Delta^2 <0$, c.f. Eq.~(\ref{eq:ewsb1}), is 
perfectly compatible
with consistent electroweak symmetry breaking, contrary to what 
is sometimes stated in the literature. Even more so, for
$\mu \ll v_t$, requiring $M_\Delta^2 > 0$ boils down to
requiring $\lambda_{1 4}^+ < 0$ which in turn excludes the
{\sl $H^0$-scenario} altogether for any value of $\mu (>0)$ in the 
strict SM-like limit with $k=0$,
c.f. Eq.~(\ref{eq:mucritical2}). Put differently, insisting on
$M_\Delta^2 > 0$ excludes {\sl a priori} a viable scenario rather 
than non-physical configurations!  Obviously another
no-tachyon constraint will be $\lambda_4 <0$ as can be seen
from Eqs.~(\ref{eq:mH+}, \ref{eq:mH++}). The discussion of
the more stringent experimental exclusion constraints is deferred
to the subsequent sections.

Finally, inverting Eq.(\ref{eq:mH0}) one obtains

\begin{equation}
\lambda =
   \frac{2 m_{H^0}^2}{v_d^2} - \frac{\lambda_{1 4}^+ 
   v_t^2}{m_{H^0}^2} 
   \; (2 \lambda_{1 4}^+  + 
         \sqrt{2} \, (\lambda_{1 4}^+ \frac{v_d^2}{m_{H^0}^2}- 4) 
         \; \frac{\mu}{v_t})  + {\cal O}(\frac{\mu^2}{v_t^2})
         \label{eq:lambda}
\end{equation}
which gives the precise correlation of $\lambda$ with the other
parameters of the model for fixed $m_{H^0}$. In practice this relation will be useful
for consistent scans keeping the SM-Higgs-like scalar mass at it's
experimentally measured value within ${\cal O}(1)$GeV precision. 

We end this section by recalling the mixing angles and 
various couplings of the model 
that will be relevant for the phenomenological study 
in the rest of the paper.

\noindent
{\sl \underline{mixing angles}:}
using the shorthand notations $s_x, c_x$ for $\cos x, \sin x$,
the angles $\beta$ and $\beta'$ are given by 

\begin{equation}
s_\beta =  \frac{2 v_t}{\sqrt{v_d^2 + 4 v_t^2}}~,~~~~
c_\beta = \frac{v_d}{\sqrt{v_d^2 + 4 v_t^2}}
\label{eq:scb}
\end{equation}
\begin{eqnarray}
s_{\beta^{'}}&=&  \frac{\sqrt{2}v_t}{\sqrt{v_d^2+2v_t^2}} 
\qquad , \qquad
c_{\beta^{'}}=  \frac{v_d}{\sqrt{v_d^2+2v_t^2}} \label{eq:scbprime}
\end{eqnarray} 
up to an arbitrary global sign (but the same for $s_x$ and $c_x$)
independently of the considered {\sl ${\cal H}$-scenario}. The 
expressions for the mixing angle $\alpha$ are more involved (
see \cite{Arhrib:2011uy}). We give them here for simplicity 
in the limits $\mu \to \infty$ and $\mu \to 0$,  illustrating two 
special cases of respectively the $h^0$-, $H^0$-{\sl scenarios},

$\mu \to \infty$:
\vspace{-.5cm}
\begin{eqnarray}
s_{\alpha} &=& \frac{\bar{\epsilon}}{\sqrt{2}}\sqrt{1 - 
\frac{v_d}{\sqrt{v_d^2 + 16 v_t^2}}} \approx 2 \frac{v_t}{v_d} + 
{\cal O}(\frac{v_t^3}{v_d^3})
, ~~
c_{\alpha} = \sqrt{\frac{1}{2} + \frac{v_d}
{2 \sqrt{v_d^2 + 16 v_t^2}}} \approx 1  \label{eq:scalim1}
\end{eqnarray}

$\mu \to 0$:
\vspace{-.5cm}
\begin{eqnarray}
s_\alpha &=& \bar{\epsilon}(1 
- 2 \frac{{\lambda_{14}^+}^2}{\lambda^2} \frac{v_t^2}{v_d^2}) + {\cal O}(\frac{v_t^3}{v_d^3})
, ~~~~~~~~~~~~~~~~~~~~~~
c_\alpha = 2 \frac{|\lambda_{14}^+|}{\lambda} \frac{v_t}{v_d} + {\cal O}(\frac{v_t^3}{v_d^3})  \label{eq:scalim2}
\end{eqnarray}

where, adopting the convention $c_\alpha >0$, 
the sign $\bar{\epsilon}$ is given by 
\begin{eqnarray}
\bar{\epsilon} = 1, \mbox{ [{\sl $h^0$-scenario}]}  \qquad , \qquad
\bar{\epsilon} = {\rm sign}[ \sqrt{2}\mu - (\lambda_1+\lambda_4) v_t], \; \mbox{[{\sl $H^0$-scenario}]} \label{eq:epsilonbar}
\end{eqnarray} 

\noindent
In the following and throughout the paper we refer to the
couplings as they appear in the Lagrangian, [i.e. no extra $i$
factors or symmetry factors of Feynman rules]:

\noindent
{\sl \underline{gauge boson(-gauge boson)-scalar-(scalar) couplings}:}
it is easy to see from the structure of the kinetic terms 
Eqs.~(\ref{eq:DTHM} -- \ref{eq:covd2})
and the scalar field components that develop vacuum expectation 
values, that couplings involving one scalar boson and two gauge 
bosons are $v_t$ suppressed if the scalar is essentially triplet; 
couplings involving one or two gauge bosons and two scalars
are $v_t$ suppressed only if one of the scalars is triplet-like
and the other doublet-like; all other cases feature SM-like
couplings. For instance the magnitudes of the derivative 
couplings $Z^0h^0A^0, Z^0H^0A^0$ are ${\cal H}$-{\sl scenario}
 dependent and read (skipping the Lorentz structure for simplicity):

\begin{eqnarray}
&&g_{Z^0h^0A^0}= -{g\over2c_W}(c_\alpha s_\beta-2c_\beta s_\alpha)
\;\; \approx  {g\over c_W} {v_t\over v_d} (2 \bar{\epsilon} -1), 
\, [\mbox{{\sl $h^0$-scenario}}]; \;\;\approx 
\bar{\epsilon} {g\over c_W}, \, [\mbox{{\sl $H^0$-scenario}}] 
\label{eq:coup5}\nonumber \\
&& \\
&&g_{Z^0H^0A^0}={g\over2c_W}(s_\alpha s_\beta+2 c_\alpha c_\beta)
\;\; \approx  {g\over c_W}, \, [\mbox{{\sl $h^0$-scenario}}];
\;\; \approx 
{g\over c_W} {v_t\over v_d} ( \bar{\epsilon} + 2 \frac{|\lambda_{14}^+|}{\lambda}), \, \, [\mbox{{\sl $H^0$-scenario}}]
\label{eq:coup6} \nonumber \\
\end{eqnarray} 
while the magnitudes of the $Z$-boson 
to the (doubly-)charged Higgs bosons derivative couplings
 are ${\cal H}$-{\sl scenario} independent and given by 
\begin{eqnarray}
g_{ZH^{+}H^{-}}  & = &
+\frac{1}{2}[-(c_W^2\,s_{\beta^{'}}^2)+(2c_{\beta^{'}}^2+s_{\beta^{'}}^2)\,s_W^2]/(s_W\,c_W)\approx
s_W/c_W \label{eq:couplages_Z_Hp}\\
g_{ZH^{++}H^{--}} & = & -[c_W^2-s_W^2]/(s_W\,c_W) = - 2 \cot 2 \theta_W
\label{eq:couplages_Z_Hpp}
\end{eqnarray}
[Note that Eq.(\ref{eq:couplages_Z_Hp}) differs from
the one  given in our 
Eq.(C.20) of ref.\cite{Arhrib:2011uy}, due to a typo in the 
latter.]
The $\gamma H^{+}H^{-}$ and $\gamma H^{++}H^{--}$ couplings 
are obviously those of (scalar) QED and are given by the 
$H^+$ and $H^{++}$ electric charges.

\noindent
{\sl \underline{triple-scalar couplings}:}
again, one easily sees from the structure of the potential Eq.~(\ref{eq:Vpot}) and the vacuum expectation values, that 
triple scalar couplings are $v_t$ suppressed when only one of
the three scalars is triplet-like, and not suppressed when
only one of the three scalars is doublet-like.  
For instance the couplings $h^0h^0H^0$, $A^0A^0H^0$,
$h^0H^{++}H^{--}, h^0H^{+}H^{-}, H^0H^{++}H^{--}, H^0H^{+}H^{-}$ 
are given by 

\begin{eqnarray}
g_{h^0 h^0 H^0} \!\!&=&\!\!\!
    \sqrt{2}c_\alpha\mu(1 \!-\! 3s_\alpha^2) + 
    (\frac{3}{2}c_\alpha^2\lambda \!+\! (1 \!-\! 3c_\alpha^2)\lambda_{1 4}^+)s_\alpha v_d \!-\! 
    c_\alpha(6\lambda_{2 3}^+s_\alpha^2 \!+\! \lambda_{1 4}^+(1 \!-\! 3s_\alpha^2)) v_t  \label{eq:hhH} \\
g_{A^0 A^0 H^0} \!\!&=&\!\!\!\sqrt{2}\,\mu\,s_\beta\,(2\,c_\beta\,s_\alpha - c_\alpha\,s_\beta) + 
    s_\alpha\,(c_\beta^2\,\lambda_{1 4}^+ + \frac{\lambda}{2}\,s_\beta^2) v_d - 
    c_\alpha (2\,c_\beta^2\,\lambda_{2 3}^+ + \lambda_{1 4}^+\,s_\beta^2) v_t  \label{eq:AAH}
\end{eqnarray}

\begin{eqnarray}
g_{h^0H^{++}H^{--}} &=&-\{2\lambda_2v_ts_\alpha+\lambda_1v_dc_\alpha\} 
 \label{eq:ghHpp}\\
g_{h^0H^+H^-} &=&-\frac{1}{2}
\bigg\{\{4v_t(\lambda_2 + \lambda_3) c_{\beta'}^2+2v_t\lambda_1s_{\beta'}^2-
\sqrt{2}\lambda_4v_dc_{\beta'}s_{\beta'}\}s_\alpha \nonumber \\
&&+\{\lambda\,v_ds_{\beta'}^2+{(2\lambda_{1}+\lambda_{4}) }v_dc_{\beta'}^2+
(4\mu-\sqrt{2}\lambda_4v_t)c_{\beta'}s_{\beta'}\}c_\alpha\bigg\}
\label{eq:ghHp}
\end{eqnarray}
\noindent
and 
\begin{eqnarray}
g_{H^0H^{++}H^{--}}&=&  g_{h^0H^{++}H^{--}}  [c_\alpha \rightarrow -s_\alpha, s_\alpha \rightarrow c_\alpha]
\label{eq:gHHpp}\\
g_{H^0H^+H^-}&=& g_{h^0H^+H^-}  [c_\alpha \rightarrow -s_\alpha, s_\alpha \rightarrow c_\alpha]
\label{eq:gHHp} 
\end{eqnarray}
\noindent
These couplings are phenomenologically interesting in both
$h^0$-, $H^0${\sl -scenarios}. In the former they will trigger
decays of the heavy non-standard CP-even state. In the latter
they will trigger non-standard decays of the SM-Higgs-like
state. The couplings to (doubly-)charged states will be important
in both ${\cal H}${\sl -scenarios} when studying the $\gamma \gamma$ and $Z \gamma$
Higgs decay channels.
We give in table \ref{table_couplings0} the limiting behavior for all these couplings in the two 
 ${\cal H}${\sl -scenarios}  for later reference.


\begin{table}[!h]
\begin{center}
\renewcommand{\arraystretch}{1.5}
\begin{tabular}{|c|c|c|} \hline\hline
\ \ $\approx$ \ \  & {\sl $h^0$-scenario}& ~~~~{\sl $H^0$-scenario}~~~~ \\ \hline\hline
$g_{h^0 h^0 H^0}$\ & $  \sqrt{2} \mu - (5 \lambda_{14}^+ - 
3 \lambda)v_t $ &  $ \bar{\epsilon} \lambda_{14}^+ v_d$ \\ \hline
$g_{A^0 A^0 H^0}$\ & $ \displaystyle 2 ( \lambda_{14}^+- \lambda_{23}^+) v_t + 4 \sqrt{2} \mu 
\frac{v_t^2}{v_d^2}$ & $ \bar{\epsilon} \lambda_{14}^+ v_d$ \\ \hline
$g_{h^0H^{++}H^{--}}$ &  $-\lambda_1 v_d$ &  $ \displaystyle 
-2 (\bar{\epsilon} \lambda_2 + \lambda_1 \frac{|\lambda_{14}^+|}{\lambda}) v_t$  \\ \hline
$g_{h^0H^{+}H^{-}}$ & $\displaystyle -(\lambda_1 + \frac{\lambda_4}{2}) v_d - 2 \sqrt{2} \mu  \frac{v_t}{v_d}$ &   
$\displaystyle (-2 (\frac{{\lambda_{14}^+}|\lambda_{14}^+|}{\lambda} + \bar{\epsilon} \lambda_{23}^+) + (\bar{\epsilon} + \frac{|\lambda_{14}^+|}{\lambda}) \lambda_4) v_t$\\ \hline
$g_{H^0H^{++}H^{--}}$ & $2 (\lambda_1 - \lambda_2) v_t$ &  $\bar{\epsilon} \lambda_1 v_d$\\ \hline
$g_{H^0H^{+}H^{-}}$   & $\displaystyle  2 v_t ( \lambda_{14}^+ - \lambda_{23}^+ ) + 4 \sqrt{2} \mu \frac{v_t^2}{v_d^2}$& 
$\displaystyle \bar{\epsilon} (\lambda_1 + \frac{\lambda_4}{2}) v_d$  
    \\ \hline\hline
\end{tabular}
\end{center}
\caption{ Approximate expressions in the $h^0$-, $H^0${\sl -scenarios}, for a selected list of three-scalar couplings as they 
appear in the Lagrangian.}
\label{table_couplings0}
\end{table}

\noindent
{\sl \underline{Yukawa couplings}:}
it is straightforward to obtain these couplings from the Yukawa
sector of the SM and the extra Yukawa terms Eq.~(\ref{eq:yukawa}),
upon use of Eqs.~(\ref{eq:mixh0H0} --\ref{eq:mixH++}). For
instance the $A^0 \bar{f} f$ and $H^{\pm} \bar{f}' f$ couplings
are given by the corresponding SM neutral and charged 
Goldstone bosons, suppressed respectively by $-\sin \beta$,
and $-\sin \beta'$. Similarly,
 the $h^0 \bar{f} f$, $H^0 \bar{f} f$ couplings are given by 
 the SM Higgs coupling to fermions suppressed respectively
 by $\cos \alpha$ and $-\sin \alpha$. Also charge conservation
 forbids $H^{\pm \pm}$ from inheriting from any of the SM 
 Yukawa couplings. Taking into account Eqs.~(\ref{eq:scb} --
 \ref{eq:scalim2}), one retrieves that all the new scalar states
 become increasingly fermiophobic with decreasing $v_t$, except
 for one CP-even state that becomes increasingly SM-like.
 In contrast, the new Yukawa terms of Eq.(\ref{eq:yukawa})
 induce couplings of the scalar states, through their triplet
 components, to same lepton number lepton and neutrino pairs.
 All these coupings are proportional to 
 $(Y_{\nu})_l \equiv m_{\nu_l}/\sqrt{2}v_t$, (we take for
 simplicity here diagonal flavor-conserving $Y_\nu$ matrix),
 and the mixing to the triplet components are not suppressed
 except for the SM-Higgs-like state, c.f. 
 Eqs.~(\ref{eq:mixh0H0} -- \ref{eq:mixH++}, \ref{eq:scb} -- \ref{eq:scalim2}). 
 Thus all the new scalar states
 become increasingly `same-lepton-number-philic' with decreasing 
 $v_t$, excepted the SM-Higgs-like state whose coupling to
 pairs of (anti-)neutrinos is ${\cal O}(m_{\nu_l}/v_d)$ suppressed
 as can be seen from $s_\alpha$, Eq.(\ref{eq:scalim1}),
 and $c_\alpha$, Eq.(\ref{eq:scalim2}). We give the magnitudes of
 these couplings in table \ref{table_couplings1}.
 [An extended list of couplings can be found elsewhere in the 
literature; see for instance the appendix of \cite{Arhrib:2011uy}
and \cite{Perez:2008ha}, albeit with different notations
in the latter. We note however a disagreement in the relative sign 
in the $H^+ \bar{f}' f$ coupling given in table 
\ref{table_couplings1}  
as compared to the one given in Table VII of ref.\cite{Perez:2008ha}.]  
 
 \begin{table}[!t]
\begin{center}
\renewcommand{\arraystretch}{1.5}
\begin{tabular}{|c|c|} \hline\hline
\ \ $\approx$ \ \  & ~~~~{\sl $h^0$-scenario}/{\sl $H^0$-scenario}~~~~ \\ \hline\hline
$g_{h^0 \bar{f} f }$\ &  $\displaystyle -\frac{m_f}{v_d}$ / 
$\displaystyle -2 \frac{|\lambda_{14}^+|}{\lambda}\frac{m_f v_t}{v_d^2}$  \\ \hline
$g_{H^0 \bar{f} f }$\ &  $\displaystyle -2 \frac{m_f v_t}{v_d^2}$  /$\displaystyle -\bar{\epsilon} \frac{m_f}{v_d}$\\ \hline
$g_{A^0 \bar{f} f }$\ &  $\displaystyle 2 i\frac{m_f v_t}{v_d^2} \gamma_5$   \\ \hline
$g_{H^+ \bar{f}' f }$\ &  $ \displaystyle -2 \frac{v_t}{v_d^2}
(m_{f'} P_L - m_{f} P_R)$  \\ \hline\hline
$g_{h^0 \nu_l \nu_l  (\bar{\nu}_l \bar{\nu}_l) }$\ &  
$ \displaystyle  - \frac{m_{\nu_l}}{v_d} C P_L$ / 
$\displaystyle  -\bar{\epsilon} \frac{m_{\nu_l}}{2 v_t} C P_L$\\ \hline
$g_{H^0 \nu_l \nu_l (\bar{\nu}_l \bar{\nu}_l) }$\ & 
 $\displaystyle  -\frac{m_{\nu_l}}{2 v_t} C P_L$ / 
 $ \displaystyle  - \frac{|\lambda_{14}^+|}{\lambda} \frac{m_{\nu_l}}{v_d} C P_L$
   \\ \hline
$g_{A^0 \nu_l \nu_l (\bar{\nu}_l \bar{\nu}_l) }$\ &  
$\displaystyle  - i \frac{m_{\nu_l}}{2 v_t} C P_L$  \\ \hline
$g_{H^{+} l^- \nu_l }$\ &   $\displaystyle  \frac{m_{\nu_l}}{ v_t} C P_L$  \\ \hline
$g_{H^{++} l^- l^- }$\ &  $\displaystyle  \frac{m_{\nu_l}}{ \sqrt{2} v_t} C P_L$  \\ \hline\hline
\end{tabular}
\end{center}
\caption{ The Yukawa couplings as they appear in the Lagrangian
after electroweak symmetry breaking. $P_L, P_R$ denote the left,right chirality 
projectors, and we have substituted the values
of the various mixing angles, in the $h^0$-, $H^0$-{\sl scenarios}. The upper 
(resp. lower) block corresponds to the operators involving opposite (resp. same) 
lepton-number fermions.}
\label{table_couplings1}
\end{table}

\section{$\mathcal{H} \to \gamma \gamma , Z \gamma$ \label{sec:HZgaga}} 
In this section we study the effects of the {\sl type II seesaw} 
model on the diphoton and $Z \gamma$ decay channels of the
SM-Higgs-like scalar. These channels have been considered in the 
literature in various BSM scenarios as they can probe new heavy 
degrees of freedom through loop effects. 
In the {\sl type II seesaw} they probe the presence of
the $H^\pm$ and $H^{\pm\pm}$ states as well as non-zero coupling
between the doublet and triplet scalar sectors. 
Note that in the model under consideration, 
the gluon fusion production channel remains essentially standard
since the new states are not colored and, furthermore, the top and 
bottom quark Yukawa couplings are very close to SM-like in both ${\cal H}${\sl -
scenarios}. 

The relative tension between
ATLAS \cite{ATLAS:2013oma} and CMS  \cite{CMS:ril}
regarding the diphoton channel has essentially evaporated \cite{Khachatryan:2014ira,Aad:2014aba}. 
Anticipating that the future analyses with accumulated
luminosity and increased C.M. energy at the LHC will 
confirm further the SM predictions for this channel, 
the $Z \gamma$ channel could still provide independent 
and complementary information on BSM physics. Hereafter, we first 
recall the theoretical structure of these two channels and then 
discuss their phenomenological features and correlations. 
\subsection{$\mathcal{H} \to \gamma \gamma$ \label{sec:Hgaga}} 

The structure of the diphoton decay channel width in the 
{\sl type II seesaw} model can be summarized as follows:


\begin{eqnarray}
\Gamma({\mathcal{H}} \rightarrow\gamma\gamma)
& = & \frac{G_F\alpha^2 M_{{\mathcal{H}}}^3}
{128\sqrt{2}\pi^3} \bigg| \sum_f N_c Q_f^2 \tilde{g}_{{\mathcal{H}} ff} 
A_{1/2}^{{\mathcal{H}}}
(\tau_f) + \tilde{g}_{{\mathcal{H}} WW} A_1^{{\mathcal{H}}} (\tau_W) \nonumber \\
&& + Q_+^2 \tilde{g}_{\mathcal{H} H^+\,H^-}
A_0^{{\mathcal{H}}}(\tau_{H^{+}})+
 Q_{++}^2 \tilde{g}_{\mathcal{H} H^{++}H^{--}}
A_0^{{\mathcal{H}}}(\tau_{H^{++}}) \bigg|^2
\label{eq:Htogammagamma}
\end{eqnarray}
where we have introduced the units of electric charge of $H^+$
and $H^{++}$, namely $Q_+=1$ and $Q_{++}=2$, 
 for later use in the next section.
The scalar functions $A_0^{{\mathcal{H}}}, A_{1/2}^{{\mathcal{H}}}$ 
and $A_1^{{\mathcal{H}}}$ corresponding to spin-$0, 1/2,1$ contributions 
in the loops are defined as,

\begin{eqnarray}
A_{1/2}^{{\mathcal{H}}}(\tau) &=& 2 [\tau + (\tau -1) f(\tau)] \, \tau^
{-2} \label{eq:A1/2gg} \\
A_1^{{\mathcal{H}}}(\tau) &=& - [ 2 \tau^2 + 3 \tau + 3 (2\tau -1) f
(\tau)]\, \tau^{-2}  \label{eq:A1gg} \\
A_{0}^{{\mathcal{H}}}(\tau) &=& -[\tau -f(\tau)]\, \tau^{-2} 
\label{eq:A0gg}
\end{eqnarray}
with $\tau_{i}=m^2_{\mathcal{H}}/4m^2_{i}$ $(i=f,W,H^{+},H^{++})$ 
and the function $f(\tau)$ is given by
\begin{eqnarray}
f(\tau)=\left\{
\begin{array}{ll}  \displaystyle
\arcsin^2\sqrt{\tau} & \tau\leq 1 \\
\displaystyle -\frac{1}{4}\left[ \log\frac{1+\sqrt{1-\tau^{-1}}}
{1-\sqrt{1-\tau^{-1}}}-i\pi \right]^2 \hspace{0.5cm} & \tau>1
\end{array} \right. 
\label{eq:ftau} 
\end{eqnarray}
while the {\sl reduced} trilinear couplings of $\mathcal{H}$ 
to $H^+$ and $H^{+ +}$ are given by

\begin{eqnarray}
\tilde{g}_{\mathcal{H} H^{++}H^{--}}  & = & - \frac{s_W}{e} \frac{m_W}{m_{H^{+ +}}^2} g_{\mathcal{H} H^{++}H^{--}}
 \label{eq:redgcalHHpp}\\
\tilde{g}_{\mathcal{H} H^+H^-} & = & - \frac{s_W}{e} \frac{m_W}{m_{H^{+}}^2} g_{\mathcal{H} H^+H^-} \label{eq:redgcalHHp}
\end{eqnarray}
where $g_{\mathcal{H} H^{++}H^{--}}, g_{\mathcal{H} H^+H^-}$
can be found in table \ref{table_couplings0} for the corresponding
${\cal H}$-scenario. When $\mu v_t \ll v_d^2$, they can be summarized as
\begin{eqnarray}
g_{\mathcal{H} H^{++}H^{--}}  &\approx &  - s \lambda_1v_d \label{eq:gcalHHpp}\\
g_{\mathcal{H} H^+H^-} & \approx & - s (\lambda_{1} + \frac{\lambda_{4}}{2}) v_d \label{eq:gcalHHp}
\end{eqnarray}
 by defining $s=1$ in the $h^0$-{\sl scenario} (${\cal H} \equiv h^0$), and $s = - \bar{\epsilon}$
in the  $H^0$-{\sl scenario} (${\cal H} \equiv H^0$)
with $\bar{\epsilon}$ as given by 
Eq.(\ref{eq:epsilonbar}).\footnote{Note an overall sign mismatch 
between Eq.(\ref{eq:gcalHHp}) above and Eq.(3.16) 
of \cite{Arhrib:2011vc}. This is just due to a notational 
confusion between $s$ and $\bar{\epsilon}$ in the latter paper, 
but which did not enter nor affect the physics analysis!}

\begin{table}[!h]
\begin{center}
\renewcommand{\arraystretch}{1.5}
\begin{tabular}{|c|c|c|c|c|} \hline\hline
\ \ $\mathcal{H}$ \ \  &$\tilde{g}_{\mathcal{H} \bar{u}u}$&
                  $\tilde{g}_{\mathcal{H} \bar{d}d}$&
$\tilde{g}_{\mathcal{H} W^+W^-} $ \\ \hline\hline
$h^0$  & \ $\; c_\alpha/c_{\beta'} \; $ \
     & \ $ \; c_\alpha/c_{\beta'} \; $ \
     & \ $ \; +e(c_\alpha\,v_d+2s_\alpha\,v_t)/(2s_W\,m_W) \; $ \ \\
$H^0$  & \ $\; - s_\alpha/c_{\beta'} \; $ \
     & \ $ \; - s_\alpha/ c_{\beta'} \; $ \
     & \ $ \; -e(s_\alpha\,v_d-2c_\alpha\,v_t)/(2s_W\,m_W) \; $ \ \\ \hline\hline
\end{tabular}
\end{center}
\caption{ The CP-even neutral Higgs {\sl reduced} couplings to 
fermions and gauge bosons in the {\sl type II seesaw} model
  relative to the SM Higgs couplings,
$\alpha$ and $\beta'$ denote the mixing angles respectively in the CP-even and
charged Higgs sectors, $e$ is the electron charge, $m_W$ the $W$ gauge boson
mass and $s_W$ the weak mixing angle.}
\label{table_couplings}
\end{table}

\subsection{$\mathcal{H} \to Z \gamma$ \label{sec:HgaZ}}

Since the original work \cite{Cahn:1978nz}, 
\cite{Bergstrom:1985hp, *Bergstrom:1986err}, various notations and 
normalizations have been adopted by different reviewers (see e.g.  
\cite{Gunion:1989we}, \cite{Spira:1997dg}, \cite{Djouadi:2005gj})
as well as by authors of very recent studies specific to the 
{\sl type II seesaw} model \cite{Dev:2013ff}, 
\cite{Chen:2013dh, Chen:2013vi} 
(not to mention different notational
conventions for the scalar couplings in the {\sl type II seesaw}
model potential). This 
unfortunately makes comparisons among different forms
for $\Gamma(\mathcal{H} \to Z \gamma)$  unnecessarily tedious 
and a direct
use of the literature to  include new loop contributions far from 
straightforward, leading occasionally to erroneous factors. We 
have thus recomputed $\Gamma ({\cal H} \to Z\gamma)$ from scratch, 
using the FeynArts and
FormCalc \cite{Hahn:2000kx,Hahn:1998yk} packages for the one-loop
amplitudes, for which we provided a {\sl type II seesaw} model 
file. We then compared with \cite{Gunion:1989we}, 
\cite{Spira:1997dg} and checked the 
consistency of the different
normalizations. In the form we give below, we adopt conventions 
that are natural in the following sense: 
\begin{itemize}
\item[-]the couplings are 
identified as the ones read directly from the Lagrangian (up to an 
electric charge factor $e$) 
\item[-] the defined functions correspond 
directly to the loop form factors 
\item[-] the partial width 
$\Gamma ({\cal H} \to \gamma
\gamma)$ of Eq.(\ref{eq:Htogammagamma}) is obtained 
straightforwardly from $\Gamma ({\cal H} \to Z \gamma)$ in an 
appropriate formal limit with $M_Z \to 0$.
\end{itemize} 

With these conventions we find, 
\begin{eqnarray}
\Gamma ({\cal H} \to Z\gamma) &=& 
\frac{G^2_F M_W^2\,\alpha\,M_{\cal H}^{3}}{64\,\pi^{4}} 
\left(1-\frac{M_Z^2}{M_{\cal H}^2} \right)^3 s_W^2
\bigg| 
\sum_{f} N_c^f g_{\gamma f f} g^v_{Zff} \,\tilde{g}_{\mathcal{H}ff}\, 
{A}^{\cal H}_{1/2} (\tau_f,\lambda_f) \nonumber
\\ & & + g_{\gamma WW} g_{ZWW} \tilde{g}_{\mathcal{H}WW}\,{A}^{\cal 
H}_1 (\tau_W,\lambda_W) +
 Q_{+} g_{ZH^+H^-}\tilde{g}_{\mathcal{H} H^+\,H^-}\,{A}^{\cal H}_0
 (\tau_{H^{+}},\lambda_{H^{+}})\nonumber\\ && + 
Q_{+ +} g_{ZH^{++}H^{--}}\tilde{g}_{\mathcal{H} H^{++}\,H^{--}}\,{A}^{\cal H}_0 (\tau_{H^{++}},\lambda_{H^{++}}) \bigg|^2
\label{eq:HtogammaZ}
\end{eqnarray}
where $\tau_{i}$ is defined as in section \ref{sec:Hgaga} and 
$\lambda_i=M_Z^2/4m^2_{i}$ $(i=f,W,H^{+},H^{++})$,
$g^v_{Zff}= -\frac{(I_f^3- 2  s_W^2 Q_f)}{2 s_W c_W} $, with $Q_f$ denoting the fermions
electric charges and $I_f$ their weak isospin, $g_{\gamma f f} =
-Q_f$, $g_{ZWW}= -\cot \theta_W, g_{\gamma WW} = -1$
and

\begin{eqnarray}
{A}_{1/2}^{\cal H} (\tau,\lambda) & = & -4 \left[I_1(\tau,\lambda) - I_2(\tau,\lambda)
\right]  \label{eq:A1/2gZ} \\
{A}_1^{\cal H} (\tau,\lambda) & = & 2 \left\{ [2(1 + 2\tau )(1 - \lambda) + (1 - 2 \tau )]I_1(\tau, \lambda) 
- 8(1 - \lambda)I_2(\tau, \lambda) \right\} \label{eq:A1gZ}\\
{A}^{\cal H}_0 (\tau,\lambda) & = & 2 I_1(\tau,\lambda) 
\label{eq:A0gZ}
\end{eqnarray}
%

 
%
The functions $I_1$ and $I_2$ are given by 
\begin{eqnarray}
I_1(\tau,\lambda) & = & \frac{1}{2(\lambda-\tau)}
+ \frac{1}{2(\tau-\lambda)^2} \left[ f(\tau)-f(\lambda) 
\right] + \frac{\lambda}{(\tau-\lambda)^2} \left[ g(\tau) - 
g(\lambda) \right] \nonumber \\
I_2(\tau,\lambda) & = &  \frac{1}{2(\tau-\lambda)}\left[ f(\tau
)- f(\lambda) \right]
\end{eqnarray}

where $f(\tau)$ is given in Eq.(\ref{eq:ftau})and $g(\tau)$ is defined as
\begin{equation}
g(\tau) = \left\{ \begin{array}{ll}
\displaystyle \sqrt{\tau^{-1}-1} \arcsin \sqrt{\tau} & \tau \le 1 \\
\displaystyle \frac{\sqrt{1-\tau^{-1}}}{2} \left[ \log \frac{1+\sqrt{1-\tau
^{-1}}}{1-\sqrt{1-\tau^{-1}}} - i\pi \right] & \tau  > 1
\end{array} \right.
\label{eq:gtau}
\end{equation}

The {\sl reduced} couplings $ \tilde{g}_{\mathcal{H} f f}, \tilde{g}_{\mathcal{H} W W}, \tilde{g}_{\mathcal{H} H^+H^-}$ and 
$\tilde{g}_{\mathcal{H} H^{++}H^{--}}$ are as given in table (\ref{table_couplings}) and 
Eqs.(\ref{eq:redgcalHHpp} - \ref{eq:gcalHHp}). The $g_{ZH^{+}H^{-}},\,g_{ZH^{++}H^{--}}$ couplings are given in Eqs.~(\ref{eq:couplages_Z_Hp}, \ref{eq:couplages_Z_Hpp}).
We adopted for the definition of
 ${A}_1^{\cal H}$ that of  eq.(\ref{eq:A1gZ}), see also \cite{Chen:2013dh},
rather than the more often used one,   
\begin{equation}
{A}_1^{\cal H} (\tau_W,\lambda_W) = -\left\{ 4\left(3-\frac{s_W^2}{c_W^2} \right)
I_2(\tau_W,\lambda_W) + \left[ \left(1+ 2 \tau_W \right) \frac{s_W^2}{c_W^2}
- \left(5+ 2 \tau_W \right) \right] I_1(\tau_W,\lambda_W) \right\}
\label{eq:A1primegZ}
\end{equation}
Both coincide only when the $W$-boson is circulating in the loop 
and upon use of the tree-level relation $M_W^2 = c_W^2 M_Z^2$. Eq.
(\ref{eq:A1gZ})
is obviously more transparent if one wishes to include effects of heavier new gauge bosons, or for that matter to retrieve the diphoton channel by simply taking $\lambda \to 0$.\footnote{If 
Eq.(\ref{eq:A1primegZ})  were to be used instead, then one would
have to make the formal and unintuitive replacement 
$\displaystyle \frac{s^2_W}{c^2_W} \to -1 $.}
Note also that with our conventions the amplitude in Eq.(\ref{eq:HtogammaZ}) has a global minus sign as compared to 
\cite{Spira:1997dg}.

\subsection{Correlations 
\label{sec:HgagagaZ}}

Hereafter we examine some model-dependent properties of the 
$\mathcal{H}
\rightarrow \gamma\gamma, Z \gamma$ branching ratios as well as
the correlation between the two channels.

The behavior of the branching ratio ${\rm Br}(\mathcal{H}
\rightarrow \gamma\gamma)$ has already been 
studied in \cite{Arhrib:2011vc}, (see 
also \cite{Kanemura:2012rs}, \cite{Akeroyd:2012ms} and the 
discussion in section \ref{sec:lam1lam4}). 
Here we discuss this behavior
in somewhat more details taking into account
the realistic $m_{\cal H}= 125-126$GeV mass. 
The main message is that, for
not too heavy $H^{++}$ and $H^{+}$, the virtual effects
of these states 
bring in a high sensitivity to $\lambda_1$ and $\lambda_{14}^+$, 
(on top of an implicit sensitivity to $\lambda_4$ through the
 $H^{++}$, $H^{+}$ masses themselves). The quadratic
 dependence on $\lambda_1$ implies generically the 
 existence of  two-fold $(\lambda_1, \lambda_4$) values 
 that are compatible 
 with the SM prediction for 
 ${\rm Br}(\mathcal{H} \rightarrow \gamma\gamma)$; that is,
 for any given value of $\lambda_4$,  
 the branching ratio as a function of $\lambda_1$ crosses 
 twice the SM value, once for
 $\lambda_1$ very close to zero and once for $\lambda_1$ of
 order a few units. 
 This is of course a direct consequence of
 interference effects involving the (doubly)-charged scalars and 
 the quasi-SM $W$ and top-, bottom-quark loops. In fact, since 
 $\lambda_1, \lambda_4$ are real-valued 
 (see Eq.(\ref{eq:Vpot})), and taking into account that only 
 the bottom-quark loop develops an absorptive imaginary contribution, a close look at the
 structure of  Eq.(\ref{eq:Htogammagamma}) allows to trace the
 origin of small (resp. large) values of 
 $\lambda_1$ compatible with the SM prediction of  
 ${\rm Br}(\mathcal{H} \rightarrow \gamma\gamma)$, to a destructive
 interference in the $H^+$, $H^{++}$ sector alone 
 (resp. to a substantial
 interference between $H^+$, $H^{++}$ and the $W$ and top loops).
 Note that since present experimental constraints imply
 $m_{\cal H} < 2 m_{H^+}, 2 m_{H^{++}}$, the $H^+$, $H^{++}$
 loops do not have imaginary contributions (we are neglecting
 the widths of particles propagating in the loops), which would
 have otherwise destroyed the generic cancellations that we are 
 discussing. 
 We stress that none of these 
 two $(\lambda_1, \lambda_4)$ regions that are compatible with the
 SM prediction for 
 ${\rm Br}(\mathcal{H} \rightarrow \gamma\gamma)$  
 correspond to any sort of decoupling regime. Indeed,  
 they occur for moderate values of $\lambda_4$, hence for 
 relatively light $H^+,H^{++}$.  
 This means that the confirmation of a SM-
 like value for ${\rm Br}(\mathcal{H} \rightarrow \gamma\gamma)$ 
 will not suffice by itself to exclude  the existence of nearby 
 new charged scalar states, nor even the possibility of 
 relatively large $\lambda_1, \lambda_4$ values. 
 One should however keep in mind that a consistent 
 interpretation in terms of the $h^0$- or $H^0${\sl -scenarios}
 will bring a further constraint due to Eqs.(\ref{eq:mucritical1},
 \ref{eq:mucritical2}). 
 Moreover, although we require the range of variation of 
 $(\lambda_1, \lambda_4)$ to respect the perturbative 
 unitarity and boundedness from below
 (U-BFB) constraints (with $\kappa=8$, see next section for a full
 discussion), still some values of $\lambda_1$ compatible 
 with   ${\rm Br}(\mathcal{H} \rightarrow \gamma\gamma)^{\rm (SM)}$
 can be relatively large, possibly questioning the validity 
 of the perturbative evaluation. A more sophisticated treatment
 would be called for in this case, resumming for instance some
 of the higher order effects.
 We illustrate the above features in Figs.\ref{fig:BRgaga_BRgaZ}
 (a), (b) respectively in the $h^0${\sl -scenario} and
 $H^0${\sl -scenario} where we depicted also the SM model
 line.  We allow conservatively a $10 \%$ 
 uncertainty on the future determination of the SM Higgs couplings
 to fermions and gauge bosons, \cite{CMS:2013xfa},
 \cite{ATLAS-projections}. This amounts to requiring $0.9 \leq c_\alpha \leq 1
 $, respectively $0.9 \leq |s_\alpha| \leq 1$, in the 
 $h^0${\sl -scenario}, respectively $H^0${\sl -scenario}.
 Note that in terms of the $\kappa$-coupling scale factors 
 \cite{LHCHiggsCrossSectionWorkingGroup:2012nn}, one has, depending
 on the $h^0$- or $H^0${\sl -scenario} under consideration,
 respectively
 $\kappa_F = c_\alpha$, or $s_\alpha$, and 
 $\kappa_Z= c_\alpha c_\beta + 2 s_\alpha s_\beta$ or
 $s_\alpha c_\beta - 2 c_\alpha s_\beta$, and
 $\kappa_W= c_\alpha c_\beta + s_\alpha s_\beta$ or
 $c_\alpha s_\beta - s_\alpha c_\beta$. The constant 
 $\lambda_4$-lines in the figures are cut at some high values of
 $\lambda_1$, correspondingly to the assumed $10 \%$ precision 
 on the $\kappa$'s. This prevents the lines with large 
 $\lambda_4$ values from reaching 
 the SM line. Such an effect is generic and implies 
 that an increased future 
 precision will tend to eliminate the large 
 $(\lambda_1, \lambda_4)$ configurations that are compatible with
 ${\rm Br}(\mathcal{H} \rightarrow \gamma\gamma)^{\rm (SM)}$. 
  It should be noted, though, that the `$H^0${\sl -scenario}' 
 features smaller values of $\lambda_1$ than does the 
  `$h^0${\sl -scenario}' and would be thus somewhat simpler to 
  interpret theoretically. 
  
  Of more interest are the small values of  $\lambda_1$
  that are compatible with ${\rm Br}(\mathcal{H} \rightarrow \gamma
 \gamma)^{\rm (SM)}$. As stated previously these values correspond
 to zeroing the interference within the $H^+$, $H^{++}$ sector 
 itself, that is when
 
  \begin{equation}
 \lambda_1 \simeq \lambda_1^0 \equiv -\frac{\lambda_4 m_{H^{++}}^2 
 Q_{+}^2 
  A_0^{\cal H
  }(m_{\cal H}^2/4 m_{H^+}^2)}{
    2 (m_{H^+}^2 Q_{++}^2 A_0^{\cal H}(m_{\cal H}^2/4 m_{H^{++}}
    ^2) 
    + m_{H^{++}}^2 Q_{+}^2 A_0^{\cal H}(m_{\cal H}^2/4 m_{H^+}^2))}
  \label{eq:SM-compatible}
  \end{equation}
  as can be easily seen from Eqs.(\ref{eq:Htogammagamma}, 
  \ref{eq:redgcalHHpp} - \ref{eq:gcalHHp}).
 Note that $\lambda_1^0$ has a non-trivial dependence on 
 $\lambda_4, \mu, v_t$ through $m_{H^+}, m_{H^{++}}$.
 A somewhat striking behavior is found when $\lambda_1$ lies
 in the vicinity of $\lambda_1^0$;   
 as illustrated numerically in Figs.\ref{fig:BRgaga_BRgaZ}
 (a), (b), an essentially unique value of $\lambda_1^0$ 
 reproduces the SM diphoton branching ratio 
 irrespective of $\lambda_4$. Clearly the $\lambda_4=0$
 curve should cross the SM value when $\lambda_1=0$ since the couplings 
 to $H^+, H^{++}$ vanish in this case, 
 Eqs.(\ref{eq:redgcalHHpp} - \ref{eq:gcalHHp}).
 In contrast, when $\lambda_4 \neq 0$, the low sensitivity to 
 variations of $\lambda_4$ shown in the figures is far from 
 obvious. 
 A technical discussion of this point is relegated to 
 appendix \ref{app:efp}. We insist here on the phenomenological 
 consequences: in the previously discussed set of $(\lambda_1, 
 \lambda_4)$ with large values of $\lambda_1$ compatible with the
 SM prediction, a slight variation of $\lambda_4$ and thus of the
 $H^+$ and $H^{++}$ masses would require a very different value of
 $\lambda_1$ to fine-tune to the SM value. When $\lambda_1$ is 
 small we have the opposite situation, the SM compatible 
 configuration becoming much more stable against the variation of 
 $H^+$ and $H^{++}$ masses through variations of $\lambda_4$. This
corresponds to a domain with relatively light $H^+$ and $H^{++}$ 
but still very difficult to exclude solely by the $\gamma \gamma$ 
 (and $Z \gamma$) decay channels. Moreover this domain 
 corresponding to small $\lambda_1^0$ does not require 
 fine-tuning to retrieve the SM value.
  For instance one finds from
  Eq.(\ref{eq:SM-compatible}) that 
  $\lambda_1^0 \simeq -\lambda_4/6$ in the regime $\mu \ll v_t$
  and $\lambda_1^0 \simeq -\lambda_4/10$, in the
  regime $\mu \gg v_t$ (see appendix \ref{app:efp} for further discussion).
    
  The above features translate into two-fold domains 
  in the $(\lambda_1, m_{H^{\pm\pm}})$ parameter space as
 illustrated in the upper plots of 
 Fig.\ref{fig:Rgaga_RgaZ_scatter} in terms of ratios of branching ratios,
 \begin{equation} 
R_{\gamma\gamma, Z\gamma}(\mathcal{H}) \equiv
\frac{{\rm Br}_{\mathcal{H}\rightarrow \gamma\gamma, Z\gamma}^{
{}^{\rm type \, II \, seesaw}}}
{ {\rm Br}_{\mathcal{H} \rightarrow \gamma\gamma, Z\gamma}^{{}^{\rm SM}}}
\approx
\frac{(\Gamma_{\mathcal{H}\rightarrow gg}
\times {\rm Br}_{\mathcal{H}\rightarrow \gamma\gamma, Z\gamma})^{{}^{
\rm type \, II \, seesaw}}}
{(\Gamma_{\mathcal{H}\rightarrow gg}\times 
{\rm Br}_{\mathcal{H} \rightarrow \gamma\gamma, Z\gamma})^{\rm SM}}
\label{eq:Rgg}.
\end{equation}
  The domain of large $\lambda_1$
 is relevant for the lighter part of the $H^{\pm}, H^{\pm \pm}$
 spectrum, while the small $\lambda_1$ domain is more substantial
 and extends to heavier charged scalar masses. The scans in 
 Fig.\ref{fig:Rgaga_RgaZ_scatter} are consistent with the 
 U-BFB constraints as well as a loose lower bound of $110$GeV
 on the $H^\pm, H^{\pm \pm}$ masses. Note that the latter bound
 does not conflict with \cite{ATLAS:2012hi}, 
 \cite{Chatrchyan:2012ya}, as far as $v_t$ 
 is large enough so that the same-sign dilepton decay channels 
 of $H^{\pm\pm}$
 are not dominant (see also the discussion at the end of
  sec.\ref{sec:lightS}).
 
 The $Z \gamma$ channel enjoys qualitatively the same properties 
 as the ones 
 discussed above; we note only some quantitative differences such 
 as the absence
 of the large $\lambda_1$ SM-like solution which lies well above 
 the perturbative unitarity constraint as shown in 
 Figs.\ref{fig:BRgaga_BRgaZ} (c),(d), which explains the absence 
 of two distinct domains in Figs.\ref{fig:Rgaga_RgaZ_scatter} (c),
 (d).
 
 One also sees on Figs.\ref{fig:BRgaga_BRgaZ}, 
 \ref{fig:Rgaga_RgaZ_scatter} that  
changing the sign of $\lambda_1$ from positive to negative changes the 
interference effects from destructive to constructive. We discuss however more 
in detail the likeliness of $\lambda_1 <0$ in the next section. 

Finally it is interesting to understand the structure of
 the correlation between ${\rm Br}(\mathcal{H}
\rightarrow \gamma\gamma)$ and ${\rm Br}(\mathcal{H}
\rightarrow Z \gamma)$.  We illustrate this correlation in 
Fig.\ref{fig:gg_gZ_correlation} for fixed values
of $\mu$ and $v_t$ and a scan over $\lambda_1, \lambda_4$ 
(and $\lambda_2, \lambda_3$ as well, the 
latter being however less relevant). The overall conical shape 
of the allowed domain traces the variation of $\lambda_1$, 
while the band results from the scan over $\lambda_4$.
This behavior is generic: the
two physical observables being of the form $|a + b \lambda_1|^2$,
their parametric correlation through $\lambda_1$ will always
be parabolic (rather than elliptic or hyperbolic). 
Thus, for fixed $\lambda_4$, the model predicts for 
each experimentally determined value of $R_{\gamma  \gamma}$  
two possible values of $R_{Z \gamma}$.\footnote{Note 
that the opposite is true too; for each $R_{Z \gamma}$ the model 
predicts two possible values of $R_{\gamma \gamma}$, since in
fact the parabola is always tilted. This does not show on the plot
because the tilt is extremely small, 
typically of order $1\%$, due to the  smallness 
of ${A}^{\cal H}_0 (\tau, M_Z^2/4 m_i^2)$ as compared
to ${A}^{\cal H}_0 (\tau)$ entering respectively in 
$R_{Z \gamma}$ and $R_{\gamma \gamma}$. It follows that the 
second possible value for $R_{\gamma \gamma}$ is very large and
totally irrelevant phenomenologically.}  
Present limits on $\mathcal{H} \to Z \gamma$ from ATLAS 
\cite{ATLAS:2013rma,Aad:2014fia} and CMS 
\cite{Chatrchyan:2013vaa} are still very weak.  
Given the projections for future precisions on the measurement
of the signal strength for this decay channel, putting them
in the ballpark of 20\% - 60\% 
\cite{CMS:2013xfa},\cite{ATLAS-projections}, it is worth noting
that typically an $R_{\gamma \gamma} \gtrsim 1$
will be consistent with the model either for 
$R_{Z \gamma} \gtrsim 1$ or for $R_{Z \gamma} \lesssim 0.2$.
Thus the  projected low precision on the 
$Z \gamma$ decay channel will nevertheless be sufficient to lift 
this degeneracy. However, as stesssed previously, should 
future data favor both $R_{\gamma \gamma}$ and $R_{Z \gamma}$ to 
be very close to the SM predictions, this by itself would 
neither constrain $\lambda_4$ to be close to zero nor the (doubly)-
charged Higgs masses to be very heavy and lying in the decoupling 
limit.

\section{Theoretical constraints on the Higgs self-couplings
\label{sec:lam1lam4}}
%

\begin{figure}[h]
\hspace*{-0.5cm}
\resizebox{84mm}{!}{\includegraphics{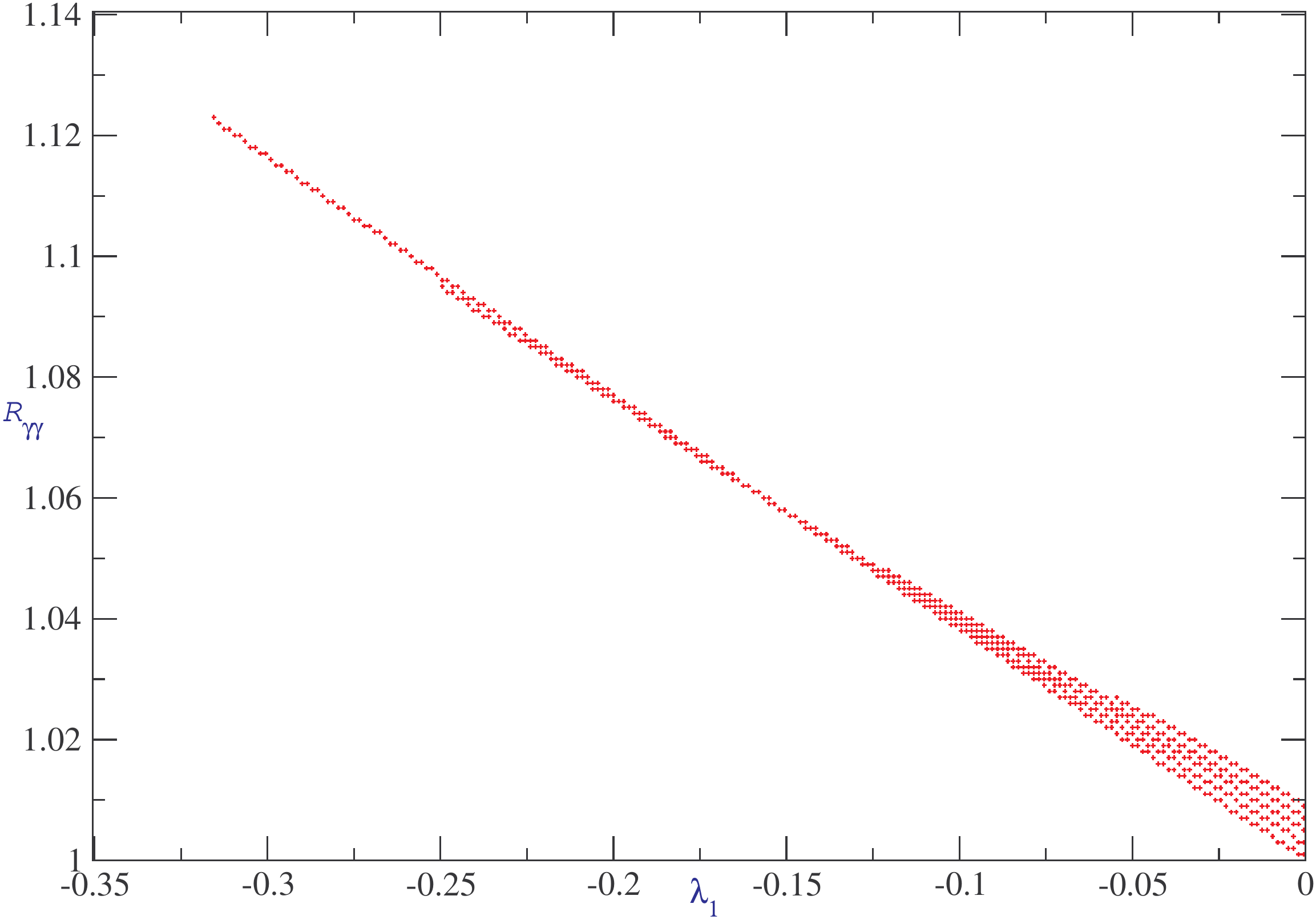}}
\vspace{-0.41cm}
\caption{\label{fig:lamd1_Rgaga}
Scatter plot for $R_{\gamma\gamma}$ versus $\lambda_1$ in 
the $\lambda_1 <0$
plane with
$\lambda=0.52, \lambda_3=2, \lambda_2=0.2,v_t\,=\,1$ GeV, 
$-10 \le \lambda_4 \le 2$ and $\mu =1$GeV. 
One clearly sees that the U-BFB constraints
reduce drastically the allowed $\lambda_1$ values which
become increasingly scarce with increasingly negative $\lambda_1$.
}
\end{figure}

In this section we would like to clarify the issue of the allowed 
regions in the $[\lambda_1, \lambda_4]$ parameter space when taking into account the full set of tree-level U-BFB
constraints established in\cite{Arhrib:2011uy}. 
The virtual contributions of (doubly-)charged Higgses  enhancing or suppressing the $\mathcal{H} 
\to \gamma \gamma$ decay channel branching ratio, first noted in   \cite{Arhrib:2011vc}
for  $\lambda_1 > 0$, were reassessed in \cite{Akeroyd:2012ms}, \cite{Melfo:2011nx} 
and \cite{Chen:2013dh} in the case $\lambda_1 < 0$  leading to stronger constraints on the 
model.  Although we agree that $\lambda_1 < 0$ configurations are 
not strictly forbidden by the U-BFB constraints, these constraints have been
only partially taken into account in the latter studies relying essentially only
on two of the BFB constraints Eq.(\ref{eq:A10}).
In fact  the full set of U-BFB constraints strongly disfavors the $\lambda_1 < 0$ 
configurations. We provide hereafter and in appendix \ref{app:lambda1} a general
proof of this property, but let us first give a numerical illustration:
Fig.\ref{fig:lamd1_Rgaga} shows the ratio $R_{\gamma\gamma}$
as defined in Eq.(\ref{eq:Rgg})
versus $\lambda_1$, for a chosen set of the remaining $\lambda_i$ 
parameters.
While $R_{\gamma\gamma}$ is indeed increased for $\lambda_1 <0 $, 
one clearly sees that 
the U-BFB constraints reduce drastically the allowed points which
become increasingly scarce with increasingly negative $\lambda_1$.
As was recognized
in \cite{Akeroyd:2012ms}, to reach more negative $\lambda_1$ values 
one has to increase
$\lambda_2$ and $\lambda_3$, see Eq.(\ref{eq:A10}). However, the 
point is that the scarcity of the allowed points
will remain.
More generally, the fully analytical form of the U-BFB constraints 
as given
in \cite{Arhrib:2011uy} (see also Eqs.(\ref{eq:A1} - \ref{eq:A13}) 
) allows an exact evaluation
of the allowed hyper-volume in the four dimensional $\lambda_i$
space.  
We state here the result, deferring the details of this somewhat 
tedious evaluation to appendix
\ref{app:lambda1}:  on the basis of a flat prior in the full 
$\lambda_i$ space  one finds that 
$\lambda_1 < 0 $ accounts for  $\sim 10\%$ of 
the allowed parameter space volume. Requiring $\lambda_1 < -0.5$  
or $\lambda_1 < -1$
as considered in \cite{Akeroyd:2012ms}, \cite{Melfo:2011nx}, 
reduces the contribution to  $3 \%$ for the former and  
 down to  $\sim 3 \tcperthousand$ for the latter.
 On the other edge, $\lambda_1 > 10$ accounts for $\sim 9\%$
 while $\sim 80 \%$ of the hyper-volume corresponds
 to $0 < \lambda_1 < 10$.\footnote{These figures are obtained for
 $\kappa =8$ and can be somewhat sensitive to $\kappa$ without
 changing though the overall conclusion as far as
 $\lambda_1 < 0 $ is concerned. For instance taking $\kappa =16$
 reduces the probability of the latter to 7\% while increasing
 tremendously that of $\lambda_1 > 10$ up to 40\%. See 
 table \ref{table:volumes} of appendix 
 \ref{app:lambda1}.} 

Thus, in the absence of any 
underlying theoretical assumptions, based 
possibly on some UV completion of the model and 
favoring $\lambda_1 <0$ or $ \lambda_1 \gtrsim 10  $, the 
above results should be taken at fair value. 
In particular the very strong constraint on the model 
inferred from $\mathcal{H} \to \gamma \gamma$ data in the
regions $\lambda_1 < -1$ or $\lambda_1< -0.5$, should be 
convoluted by the percent to 
per thousand probability of their occurrence!  
This comes for instance in contrast with the issue of taking 
values of the $\mu$ parameter much larger
or much smaller than the electroweak scale, where in either
case one can provide theoretical motivations, as discussed
in section \ref{sec:model}. 

%

\section{Invisible/undetected Higgs decays \label{sec:lightS}}


In this section we examine the possible existence of non standard 
scalar states lighter than the observed $\sim 125$GeV
SM-Higgs-like state. Such a configuration has attracted much 
attention in the recent literature on BSM physics, 
but has seldom been addressed in the context 
of the {\sl type II seesaw} model.  
It corresponds to the {\sl $H^0$-scenario} 
described in section \ref{sec:model}, where the heavier CP-even 
state $H^0$ becomes SM-like due to small values of the $\mu$ 
parameter, of order $v_t$ or smaller. Such $\mu$ configurations
should not be considered as marginal,
even though they  correspond to small parts of the parameter space.
Possible model settings motivating these configurations have been
briefly discussed in section \ref{sec:model}.

The would-be Majoron due to spontaneous
violation of the lepton number induced by $v_t$  [if $\mu$ 
were vanishing], receives then a small ${\cal O}(\mu)$ mass, 
Eq.~(\ref{eq:mA0}), 
and is identified with the CP-odd physical state $A^0$. 
For such small $\mu$ the lighter CP-even state $h^0$ will have
mainly a triplet component and is typically very light too, 
Eq.(\ref{eq:mh0}). 
The heavier CP-even state $H^0$ becomes essentially SM-like. 
Its mass can be made to match the observed value
with arbitrary precision by fixing the parameter $\lambda$
according to Eq.~(\ref{eq:lambda}).
In contrast, 
the charged and doubly charged states can be made (very) heavy
by choosing sufficiently large (and negative) values of $\lambda_4$, Eqs.~(\ref{eq:mH+}, \ref{eq:mH++}). 
This freedom allows to match the present experimental 
lower bounds, in particular on the doubly-charged state mass,
that is of order $410$ GeV, \cite{ATLAS:2012hi, 
Chatrchyan:2012ya}. 
We will come back to this point at the end of the present section.

The dependence of the Higgs spectrum on the parameters of the 
model in the regime $\mu \lesssim v_t$ is given in 
Eqs.(\ref{eq:mh0} - \ref{eq:mH++}). 
Sufficiently small $\mu$ offers a rich 
phenomenology as the decay channels $H^0 \to h^0 {h^0}^{(*)}, A^0 
{A^0}^{(*)}$ become kinematically favored with significant 
branching ratios. Indeed in the considered limit where $H^0$ 
carries essentially an $SU(2)_L$ doublet component,
$|s_\alpha| \approx 1, c_\alpha \approx 0$, the $h^0h^0H^0$ and 
$A^0A^0H^0$ couplings become
\begin{equation}
g_{h^0 h^0 H^0} = g_{A^0 A^0 H^0} \simeq (\lambda_{1 4}^+) v_d + {\cal O}(v_t)
\label{eq:hhH-couplings}
\end{equation}
see Eqs.(\ref{eq:hhH}, \ref{eq:AAH}), 
leading typically to electroweak scale enhanced Higgs into Higgs
decays.
The subsequent decays of $h^0$ and $A^0$ into fermions, gluons, or 
photons, will lead either to invisible or undetected 
$H^0$ decays that can be constrained by the global fit of the 
present ATLAS  and CMS data to Higgs 
couplings \cite{ATLAS:2013sla}, \cite{CMS:yva, Chatrchyan:2013lba}
, or to four photon final states that are also constrained when 
interpreted in terms of two photon final states 
(collimated photons) \cite{ATLAS:2012soa}. 
Searches for invisible decay of the Higgs boson have been 
carried out by CMS and
ATLAS from a variety of production processes.  
The two collaborations used the SM Higgs-strahlung $pp\to ZH$ 
cross section 
with SM Higgs boson at 125 GeV, and excluded an  
invisible branching ratio larger than 65\% with 95\% C.L. 
\cite{Aad:2014iia}, \cite{CMS:2013yda}. The CMS collaboration has 
also performed a search for the invisible decay
of the Higgs boson via the vector boson fusion process (VBF) and 
an upper limit
of 69\% with 95\% C.L was set \cite{CMS:HIG-13-013}. CMS did also 
a combination of Higgs-strahlung
and VBF process analysis which improved slightly the upper limit 
on the Branching ratio of the invisible decay to 
54\% at the 95\% C.L. \cite{Chatrchyan:2014tja}.
These limits are still rather weak and 
will improve with future LHC runs. 
 
On the other hand, global fit analyses 
performed on LHC data can in turn put limits on the invisible decay of the
 Higgs, 
\cite{Englert:2011aa, Cheung:2013kla, Belanger:2013kya,Belanger:2013xza, Espinosa:2012vu,
Englert:2012wf}. The outcome of these studies depends of course 
on the allowed deviation of the coupling of the SM Higgs to SM particles.
In a scenario where all couplings of the Higgs to SM particles are 
SM-like and the invisible decay of the Higgs allowed, the 
upper limit on the invisible branching ratio is 19\% .

We show in 
Fig.\ref{fig:invisible/undetected} contour plots of invisible/undetected 
decay branching ratios for a SM-like Higgs decaying into a pair of on-shell $h^0$ and 
$A^0$, in the $m_{h^0}\simeq m_{A^0}$ versus $|\lambda_{1 4}^+|$ 
plane, using the couplings given in Eq.~(\ref{eq:hhH-couplings}). 
In the sequel we shall assume the nominal bound
\begin{equation}
|\lambda_{1 4}^+| \lesssim 0.05
\label{eq:invisible/undetected}
\end{equation}   
to cope with the present LHC upper limits on invisible/undetected SM Higgs decays 
in our scenario. Similar limits
obtain if off-shell contributions are included, but would eventually be weakened
for much heavier $h^0, A^0$ that go beyond our scenario. 
We show in Fig.~\ref{fig:BRH} the various $H^0$ branching ratios when varying
the ratio $\displaystyle \frac{\mu}{v_t}$ or equivalently the $m_{h^0}\simeq m_{A^0}$
mass.

Moreover, due to their dominant triplet component, $h^0$
and $A^0$ are fermiophobic (except possibly for neutrinos), 
with couplings to up and down quarks and charged
leptons suppressed by a factor $2 v_t/v_d (\lesssim 8 \times 10^{-3})$ 
with respect to the SM Higgs and neutral Goldstone couplings. 
It follows that many of the exclusion limits on light 
($\lesssim 10$ GeV)
CP-odd or CP-even Higgs states from radiative decays of J/$\psi$ 
\cite{Ablikim:2011es} or 
$\Upsilon$ \cite{Lees:2012iw, Lees:2013vuj} do not apply, and 
obviously neither do 
the LHC limits from searches in the $\mu^+ \mu^-$ decay channels
 \cite{CMS:2013aga, ATLAS:2013qma}.  
More importantly, some of the upper bounds set by LEP 
on the cross-sections for the processes
 $e^+ e^- \to h^0 A^0, h^0 Z^0$, interpreted in the 
type II Two Higgs Doublet (2HDM(II))model and 
model-independently \cite{Abbiendi:2000ug}, 
\cite{Abdallah:2004wy}, \cite{Abbiendi:2004gn},
or in the minimal supersymmetric SM
\cite{LEPHiggsWorking:2001ab}, turn out to be 
partly relevant to the triplet-like $h^0$ and $A^0$
states as well. We note first that 
the $Z^0Z^0h^0$ coupling has a $4 v_t/v_d$ suppression with 
respect to 
the SM  $Z^0Z^0{\cal H}$ coupling, thus leading to a reduction 
of order $10^{-4}$ or less of the 
$e^+ e^- \to h^0 Z^0$ cross-section, two orders of magnitude 
smaller than the model-independent exclusion sensitivity  at LEP 
for $e^+ e^- \to h^0 Z^0$ \cite{Abbiendi:2000ug}. In contrast,   
the $Z^0h^0A^0$ derivative coupling in the {\sl type II seesaw} 
is of the same magnitude as the SM $Z^0Z^0{\cal H}$ coupling.
In fact the  $Z^0h^0A^0$ coupling is, for most of the parameter
space, given by $\frac{g}{c_w}$ as shown in Eqs.~(\ref{eq:coup5}, 
\ref{eq:coup6}), to be compared with the corresponding 
coupling in the case of the 2HDM(II),  that is given by 
$c \times \frac{g}{2 c_w}$ where
$c$ is a further mixing angle cosine  suppression. Note the 
factor 2 difference between the two couplings.  
Following \cite{Abbiendi:2000ug}, the $c^2$ parameter in terms of which 
limits have been presented on the  associated production of scalar/pseudo-scalar 
states with subsequent visible decays, can be re-expressed  
as the ratio of the cross-section 
$\sigma_{e^+ e^- \to h^0 A^0}$ to the SM cross-section
$\sigma_{e^+ e^- \to {\cal H} Z^0}$, 
up to a kinematic suppression factor; negative searches in the 
$e^+ e^- \to h^0 A^0$ channel have lead to exclusion domains
in the $m_{h^0}, m_{A^0}$ plane depending on the value of
$c^2$ and the subsequent hadronic or leptonic decay rates of 
$h^0$ and $A^0$. The maximal value
$c^2 = 1$ in the 2HDM(II)  can in principle exclude
large mass regions between $20$ and $120$GeV, \cite{Abbiendi:2000ug}. To read off the exclusion domain for
the {\sl type II seesaw} model requires an extrapolation up to
$c^2 = 4$, due to the factor 2 in the coupling noted above.
Moreover, the LEP precision measurements 
of the $Z$-boson total width $\Gamma_Z$ set stringent
and complementary bounds on 
any extra contribution  $\Delta \Gamma_Z$  
from new decay channels, irrespective of the final states.
From the quoted LEP value $\Gamma_Z= 2.4952 \pm 0.0023$GeV 
and the SM prediction $\Gamma_Z^{SM} = 2.4961 \pm 0.0010$GeV  
\cite{Beringer:1900zz},
one can estimate the maximum allowed non-standard contribution to 
be $\Delta \Gamma_Z^{max} \simeq 4.2$MeV at the 95\% C.L.
As we will see, the combination of the above constraints
leads to particularly strong upper bounds either on $v_t$ or on 
$v_t \times (\lambda_{1 4}^+)/\lambda $ if
$m_{A^0} \simeq m_{h^0} \lesssim 80$GeV.

In Fig.\ref{fig:ZAhwidth} we show
the mass region in the $(m_{h^0}, m_{A^0})$ parameter space
compatible with the constraint
$ \Gamma (Z \to h^0 A^0) \leq \Delta \Gamma_Z^{max}$
where we used the $Z^0h^0A^0$ tree-level coupling. 
Specifying to the {\sl type II seesaw}, $h^0$ and $A^0$
are essentially degenerate in mass, c.f. Eq.(\ref{eq:splitting}). 
The above bound translates then into the irreducible lower bound   
\begin{equation}
m_{h^0} \simeq m_{A^0} \gtrsim 44.3 {\rm GeV}
\label{eq:Zwidthbound}
\end{equation}
or equivalenty, in terms of the model parameters into
$\mu \gtrsim 4.6 \times 10^{-2} \; v_t$.
 On the other hand,
if $h^0$ and $A^0$ decay 100\% into a $b \bar{b}$ pair, 
then the most stringent limit from
$e^+ e^- \to h^0 A^0$
at LEP2 C.M. energies $\sqrt{s} = 183, 187$GeV  given
by \cite{Abbiendi:2000ug} for $c^2 \simeq 1$ will exclude
the mass range
$33$GeV $\lesssim m_{h^0} (\simeq m_{A^0}) \lesssim 78$GeV.
In our case $h^0$ and $A^0$ decay predominantly either into
$b \bar{b}$ for sufficiently large values of $v_t$, or invisibly 
into $\nu \nu + \bar{\nu} \bar{\nu}$ for much smaller values of 
$v_t$, see Fig.~\ref{fig:BrA0}. However, in the region dominated
by the $b \bar{b}$ channel, the corresponding branching ratio
quickly reaches, but does not exceed, $\sim 80 - 85\%$.
Furthermore, as can be seen from Figs.~\ref
{fig:BRA0}, \ref{fig:BRh0}, the next-to-dominant decay channel
is $\tau^+ \tau^-$ $\lesssim 9\%$, except for some parts of the 
parameter space where it can be dominated by $\gamma \gamma$ decays of 
$h^0$ , Fig.~\ref{fig:BRh0}(b). [In evaluating the branching ratios
we have taken into account the leading
perturbative QCD corrections to the CP-even and CP-odd Higgs
decays into hadronic two-body final states; see later discussion 
and appendix C for more details.] Given these typical branching 
ratios, one can not read off
directly the limits from the published results (see, e.g.  
\cite{Abbiendi:2000ug}, \cite{Ackerstaff:1998ms} and references 
therein) where a $100\%$ branching ratio into
$b \bar{b}$ or into $\tau^+ \tau^-$ was assumed for the decaying
(pseudo)scalars, and in some cases SM-like Higgs branching ratios. 
In our case, a complete study would require a statistical 
combination 
of the various decay channels, re-using LEP data. Since we are
merely interested here in how to evade these constraints in a
conservative way, we adopt the simplifying assumption of associating
the total decay width of either $h^0$, $A^0$ into visible final states, 
$b\bar{b}, \tau^+\tau^-, gg, q\bar{q}, \gamma \gamma,...$, 
exclusively with $b\bar{b}$ final state. This assumption leads to 
conservative limits from the LEP analyses since bounds from 
$100\%$ branching ratio into $b\bar{b}$ are stronger than combined 
bounds when a small fraction of decay into other final states is 
allowed. Thus, hereafter we will denote by 
$Br(A^0, h^0 \to b \bar{b})$  the 
{\sl total visible} decay branching ratios of the two light states.
Furthermore, $A^0$ and $h^0$ having a very small doublet component 
can feature substantial branching ratios into 
$\nu \nu + \bar{\nu} \bar{\nu}$ final state depending
on the magnitude of the neutrino Yukawa couplings ${Y_\nu}_i$
for the three neutrino flavors, Eq.~(\ref{eq:yukawa}). 
The corresponding decay width will scale
like $\displaystyle \sum_{i=1}^3 m_{\nu_i}^2/v_t^2$, to be contrasted with that of
visible decays which scale like $v_t^2$ and $v_t^2 \times (\lambda_{1 4}^+)^2/\lambda^2$ respectively for $A^0$ and $h^0$, see table \ref{table_couplings1}.\footnote{We will consider for the numerical
illustrations
the two extreme values $ (\sum_{i=1}^3 m_{\nu_i}^2) |_{min} 
=2.48 \times 10^{-21}$GeV${}^2$ and  $(\sum_{i=1}^3 m_{\nu_i}^2) |_{max}=1.78 \times 
10^{-20}$GeV${}^2$, compatible with neutrino oscillation limits
$|m_{\nu_2}^2 - m_{\nu_1}^2| \simeq 7.6 \times 10^{-23}$GeV${}^2$, 
$|m_{\nu_3}^2 - m_{\nu_1}^2|  \simeq 2.4 \times 10^{-21}$GeV${}^2$
and the cosmological
mass bound $\sum_{i=1}^3 m_{\nu_i} < 0.23 \times 10^{-9}$GeV. Note that the minimum value requires normal mass hierarchy and the maximum value inverted mass hierarchy.
For later discussions, we refer to these two extreme cases respectively as normal minimal (NMIN) and inverted maximal (IMAX).
\label{footnote:8}}
Since we associate all visible final states with 
$b (\bar{b})$, the relevant quantities  are
\begin{eqnarray}
 b^2 &\equiv& 
Br(A^0 \to b \bar{b}) \times Br(h^0  \to b \bar{b}) \nonumber \\ 
(b\nu)^2 &\equiv& Br(A^0 \to b \bar{b}) \times Br(h^0  \to 
\nu \nu + \bar{\nu} \bar{\nu}) + 
Br(A^0 \to \nu \nu + \bar{\nu} \bar{\nu}) \times Br(h^0  \to 
b \bar{b}) \label{eq:branchings} \\
(\nu\nu)^2 &\equiv& Br(A^0 \to \nu \nu + \bar{\nu} \bar{\nu}) 
\times Br(h^0  \to \nu \nu + \bar{\nu} \bar{\nu}) \nonumber
\end{eqnarray}
with $b^2 + (b\nu)^2 + (\nu\nu)^2=1$.
It follows that  we can re-interpret the quantity $b^2$
as a modification of the scaling factor $c^2$ used in the LEP 
analyses, since $b^2 < 1$ signals a reduction of the 
expected total number of {\sl detectable} signal events. This is 
so because either the two higgses decayed invisibly into 
neutrinos, when $(\nu\nu)^2$ is substantial,  or, when $(b \nu)^2$ 
is substantial, the SM Higgs LEP searches through the 
$b \bar{b}+ E_{\rm miss}$ final state cannot be re-interpreted as 
constraints on our model, apart from possibly a
region around $m_{A^0} \simeq m_{h^0} \simeq 90$GeV since the 
latter searches triggered on missing energy close to $m_Z$, see 
e.g. \cite{Ackerstaff:1997cza}.
More specifically, taking into account the factor 2 enhancement  
in the $Z^0 h^0 A^0$ coupling noted earlier, the proper 
identification is $c^2 \equiv 4 b^2$ and one can now use directly 
the exclusion domains given in fig.13~(b) of ref.
\cite{Abbiendi:2000ug} up to $c^2=1$. 
We reproduce in table \ref{table:OPAL-c2} an excerpt of these 
domains in the mass configuration $m_{A^0} \simeq m_{h^0}$ 
relevant for our model. Since in our case $c^2$ can take values up 
to $4$, we need to extrapolate these exclusion domains.
While theoretically the cross-section 
$\sigma_{e^+ e^- \to h^0 A^0}$ scales linearly with $c^2$, from 
which bounds on $m (\equiv m_{h^0} \simeq m_{A^0})$ can be easily 
extracted using  Eq.~(\ref{eq:eetohA}), one should keep in mind 
that the experimental bounds on $c^2$ depend on detection 
efficiencies of $h^0 A^0$ events and thus on $m$ itself. We have 
indeed checked that the set of upper (lower) mass bounds in table 
\ref{table:OPAL-c2} do not fit to a straight line in the $[c^2, 
\sigma_{e^+ e^- \to h^0 A^0}]$ plane. However, we find that the 
subset of the upper part of the sample points corresponding to 
$c^2=0.3, 0.5, 1$ lies on
a straight line within $1\%$. We take this as signaling a high 
detection efficiency in this part of the parameter space and rely 
on this linear fit to extrapolate to higher values of $c^2$ as 
given in table \ref{table:extrapolation}. 

\begin{table}[!h]
\begin{center}
\renewcommand{\arraystretch}{1.5}
\begin{tabular}{|c||c|c|c|c|c|c|} \hline\hline
 
$c^2$   &  $<0.1$  &  $<0.12$  &  $<0.15$  &  $<0.3$  &  $<0.5$ 
&  $<1$  \newline   \\  \hline\hline
\ LEP-excluded $m_{h^0} \simeq m_{A^0}$(GeV)  &  $[49.9, 56.8]$  &  $[40.3, 
63.7]$   
&  $[38.3, 67.4]$    &   $[35.9, 70.9]$    &  $[34.8, 75]$    &  $[33.0, 78.1]$  \\
\hline\hline
\end{tabular}
\end{center}
\caption{\label{table:OPAL-c2} approximate intervals of mass exclusion, extracted from
Fig.13 (b) of the OPAL analysis \cite{Abbiendi:2000ug}.}
\end{table}

\begin{table}[!h]
\begin{center}
\renewcommand{\arraystretch}{1.5}
\begin{tabular}{|c||c|c|c|} \hline\hline
 $c^2$   &  $<2$  &  $<3$  &  $<4$ \newline  \\  \hline\hline
 $m_{h^0} \simeq m_{A^0}$(GeV)   &  $[32.3, 79.8]$  &  $[32.1, 80.4]$  &  $[31.9, 80.7]$  \\
\hline\hline
\end{tabular}
\end{center}
\caption{\label{table:extrapolation} Mass exclusion intervals extrapolated from
Fig.13 (b) of \cite{Abbiendi:2000ug}; see text for more details.}
\end{table}
Note that apart from $c^2 \lesssim 0.1$, the lower edges of the excluded intervals
given in the tables are lower than the irreducible lower bound of 
Eq.~(\ref{eq:Zwidthbound}). The allowed regions are thus determined
solely by the upper edges of these intervals that correspond to lower mass bounds.
As noted earlier, these bounds are controlled by the relative magnitudes of 
the branching ratios $b^2 (=c^2/4),
(b\nu)^2, (\nu\nu)^2$ defined in Eq.~(\ref{eq:branchings}) that 
depend on $v_t$ and $\lambda_{1 4}^+$. However, the visible decay widths
of $A^0$ and $h^0$ will also be of an issue, as the bounds would be invalidated
if at least one of the two particles decays outside the detector. 
The decay length $c\tau$ of an $A^0$ decaying mainly into $b \bar{b}$, reads,
in the instantaneous decay approximation and at tree-level,
\begin{equation}
c\tau_{A^0} \simeq 3.44 \times 10^{-8} \times 
\frac{(s -4 \,m_{A^0}^2)^{\frac12}}{m_{A^0}^2 \, v_t^2}  \; [{\rm meters}]
\label{eq:ctauA0}
\end{equation}
in a reference frame where the $A^0$ energy is $E_{A^0} = \sqrt{s}/2$ [GeV],
(see also appendix \ref{app:ZhA}). Similarly, one finds for $h^0$
\begin{equation}
c\tau_{h^0} \simeq 0.94 \times 10^{-8} \times 
\frac{(s -4 \,m_{h^0}^2)^{\frac12}}{m_{h^0}^2 \, v_t^2 (\lambda_1 +\lambda_4)^2}  \; [{\rm meters}]
\label{eq:ctauh0}
\end{equation}
where we took into account the enhancing factor $(\lambda_{1 4}^+)^2/\lambda^2$
in the width, with $\lambda \simeq 0.52$ as dictated by the SM-like Higgs mass. 
We show in Fig.\ref{fig:distance} the $c\tau_{A^0}$ contours in the 
$(m_{A^0}, v_t)$ plane assuming  $A^0$ produced through $e^+ e^- \to h^0 A^0$
at the LEP2 C.M. energy $\sqrt{s}=183$GeV and a  
visible decay mainly into $b \bar{b}$ pairs. 
It is instructive to assess the effect of the QCD
corrections to $\Gamma(A^0 \to b \bar{b})$ which we included
in Fig.\ref{fig:distance} (right), as compared
to the naive tree-level width Fig.\ref{fig:distance} (left) given by Eq.~(\ref{eq:ctauA0}), see also appendix \ref{app:QCD}. In the mass parameter space
under consideration a fiducial $3$-meter decay length is reached  for $v_t$ in the 
range $\sim (1.8 - 4) \times 10^{-5}$GeV; but for such small values of $v_t$ the 
branching ratios $b^2$ and $(b \nu)^2$ are already largely overwhelmed
by the totally invisible decay branching ratio $(\nu \nu)^2$ irrespective of
the allowed values of $\lambda_{1 4}^+$. It follows that the 
$A^0$ decay length does not play a role here. In contrast, the extra dependence on 
$\lambda_{1 4}^+$ in $c\tau_{h^0}$ will bring $v_t$ back in ranges where
$b^2$ and $(b\nu)^2$ are dominant, as we shall discuss below.
Furthermore, we find that $(b\nu)^2$ is dominated by 
$Br(A^0 \to b \bar{b}) \times Br(h^0  \to \nu \nu + \bar{\nu} \bar{\nu})$ for 
$\lambda_{1 4}^+ \lesssim 0.55$, which is always satisfied due to
Eq.~(\ref{eq:invisible/undetected}). Thus a large $c\tau_{h^0}$ 
will not add new 
constraints when $(b\nu)^2$ dominates over $b^2$, since an $h^0$ not decaying in the 
detector or decaying into neutrinos lead to the same experimental (missing energy) signature. For a better understanding of the interplay between 
the various constraints it is worth noting that due to the huge hierarchy between
the neutrino mass scale and the electroweak scale, the relative magnitudes of the 
various branching ratios will involve large/small numbers in the $(v_t, \lambda_{1 4}^+
)$ plane. For instance, taking $m_{h^0} \simeq m_{A^0} = 80$GeV and
$\sum m_{\nu}^2|_{min}$ (see footnote \ref{footnote:8}) one finds the following
necessary and sufficient conditions: 

\begin{itemize}
\item[(I)] $b^2$ dominates $(b\nu)^2$ when $v_t \gtrsim 4 \times 10^{-4}$GeV 
and 
$(\lambda_{1 4}^+)^2 \gtrsim 
(-3.13 + 1.21 \times 10^{14} \times (v_t[{\rm GeV}])^4)^{-1}$ 
\item[(II)] $b^2$ dominates $(\nu\nu)^2$ when $|\lambda_{1 4}^+| \gtrsim 1.46 \times 10^{-14} \times (v_t[{\rm GeV}])^{-4}$
\item[(III)]  $(\nu\nu)^2$ dominates $(b\nu)^2$ when $v_t \lesssim 4 \times 10^{-4}$GeV and $ (\lambda_{1 4}^+)^2\lesssim 
-0.32 + 8.23 \times 10^{-15} \times (v_t[{\rm GeV}])^{-4}$
\end{itemize}
Although the figures will depend on the neutrino mass assumptions
as well as the $h^0, A^0$ mass, we find that this dependence remains moderate
allowing to draw generic conclusions by examining the relative magnitudes
of the bounds appearing in (I), (II) and (III):  

\noindent
-if (I) is satisfied then (II) is satisfied as well but (III) violated, leading to
the hierarchy $b^2 > (b\nu)^2 > (\nu\nu)^2$. 
However, taking into account the LHC inferred bound Eq.~(\ref{eq:invisible/undetected})
one finds from (I) that a window where $b^2$ starts dominating opens only when 
$v_t \gtrsim 1.3 \times 10^{-3}$GeV; in the domain 
$4\times 10^{-4}$GeV$\lesssim v_t \lesssim 1.3 \times 10^{-3}$GeV the $(b\nu)^2$
branching ratio will dominate, starting from 
$(b\nu)^2 \approx 50\% \approx (\nu\nu)^2  \gg b^2$  near the lower edge of the domain
for any $\lambda_{1 4}^+ \lesssim 0.05$, a reversed hierarchy 
$(b\nu)^2 \gtrsim b^2  \gg (\nu\nu)^2$ obtains near the upper edge. 
For smaller $\lambda_{1 4}^+ (\lesssim 0.01)$, 
$(b\nu)^2$ is above 90-95\% for most of the upper part of the domain.

\noindent
-if $v_t \gtrsim 1.3 \times 10^{-3}$GeV, condition (I) applies fully and $b^2$
quickly reaches 99\% for increasing $v_t$ and $\lambda_{1 4}^+ \lesssim 0.05$.
For smaller values of $\lambda_{1 4}^+$ as would be implied by improved future
LHC limits on invisible/undetected SM-Higgs decays, $(b\nu)^2$ becomes substantial
again and also an increased $c\tau_{h^0}$ will eventually contribute to weaken
the LEP constraints on scalar and pseudo-scalar states as discussed previously.
Since this is the region where the LEP2 constraints can be the most stringent, 
we illustrate in Fig.\ref{fig:branchings-LEP2} the rather busy configuration
of the interplay among $b^2$, $(b\nu)^2$ and $c\tau_{h^0}$.
Fig.\ref{fig:branchings-LEP2}(a) shows the $c\tau_{h^0}=3$meters
 lines for various $h^0$ masses, below which $h^0$ is long-lived and the LEP2 limits
 from jets and/or lepton decays do not apply.  One can read 
 from figures \ref{fig:branchings-LEP2}(b), (c), corresponding respectively to the
 NMIN and IMAX neutrino mass configurations, the relative contributions
 of the $bb\bar{b}\bar{b}$ final state as compared to the 
 $b\bar{b}, \nu \nu + \bar{\nu} \bar{\nu}$ final state and the effect of the $h^0$ 
 decay length, in the $v_t, \lambda_{1 4}^+$ parameter space. Although smaller neutrino 
 masses lead to larger visible decay branching ratios and thus in principle to stronger
 exclusion limits in a given part of the parameter space, the $h^0$ decay length
 reduces this effect, as can be seen by comparing figures (b) and (c).  For instance, 
 in the NMIN configuration $m_{h^0} \simeq 55$GeV would not be excluded by LEP even for 
 $b^2 \approx 98.75\%$ (corresponding to $c^2 \simeq 3.95$, see figure (b) and table 
 \ref{table:extrapolation})  unless $\lambda_{1 4}^+ \gtrsim 3.4 \times 10^{-4}$.
 Exclusion for smaller $c^2$ would require larger $\lambda_{1 4}^+$;
 e.g. $c^2 = 3.5$ would exclude  $m_{h^0} \simeq 55$GeV from LEP negative searches
 only if $\lambda_{14}^+ \gtrsim 1.3 \times 10^{-3}$ and 
 $v_t \gtrsim 1.4 \times 10^{-2}$GeV, whereas $c^2=1$ would do so for 
 $\lambda_{14}^+ \gtrsim 5.8 \times 10^{-3}$ and $v_t \gtrsim 3 \times 10^{-3}$GeV.
 Increasing $m_{h^0}$ reduces the decay length and thus the $v_t, \lambda_{14}^+$ bounds
 above which the LEP exclusions hold. For example the exclusion of  
 $m_{h^0} \simeq 75$GeV for $c^2=1$, c.f. table \ref{table:OPAL-c2}, applies only if
 $\lambda_{14}^+ \gtrsim 2.4 \times 10^{-3}$,  $v_t \gtrsim 4.7 \times 10^{-3}$GeV. 
 Comparing the two latter examples illustrates the existence of windows in the
 $(v_t, \lambda_{14}^+)$ parameter space where heavier masses are excluded and lighter
 ones still allowed (!), in contrast with the model-independent LEP exclusion domains
 \cite{Abbiendi:2000ug}. One should however keep in mind that whenever $b^2$ is reduced
 in favor of $(b\nu)^2$ or becomes ineffective due to large $h^0$ decay length, the same
 experimental signature of two b-jets and missing energy ensues. The SM-Higgs search
 through the $e^+ e^- \to Z H^0 \to \nu \bar{\nu} b \bar{b}$ channel at LEP can then in 
 principle be reinterpreted to exclude $A^0 (h^0)$ masses
 of order $m_Z$. (Conservatively one could assume an exclusion of the domain
 $(76, 120)$GeV whenever $(b\nu)^2$ becomes substantial, even though a dedicated study
 would be necessary to take properly into account the corresponding backgrounds 
 and rates, see \cite{Ackerstaff:1997cza}.) 
 
 In the IMAX neutrino mass configuration, the LEP exclusions apply for smaller parts 
 of the parameter space as the branching ratio into visible decays is smaller. 
 The effect of the decay length is however less important since the
 $c^2$ contour lines are pushed upwards, cf. Fig.\ref{fig:branchings-LEP2}(c).
 For instance $\lambda_{1 4}^+$ has now to be smaller than $1.4 \times 10^{-4}$ for 
 the $c^2 \simeq 3.95$ line to become ineffective regarding the LEP exclusions. 
 All in all, we do not expect the re-interpretation of the LEP analyses to depend
 too much on the neutrino mass assumptions. It should be noted, though, that the
 very large hierarchy between the neutrino and electroweak scales implies a large
 sensitivity to $v_t, \lambda_{14}^+$ in the vicinity of $b^2 \simeq 100\%$: the
 corresponding $c^2 = 4$ line lies way out of the plots in 
 Fig.\ref{fig:branchings-LEP2}(b),(c) and does not intersect anymore the fiducial 
 $c\tau_{h^0}=3$ line.  This again illustrates the fact that a smaller branching ratio
 for visible decays does not only imply smaller exclusion mass bounds for $h^0, A^0$
 but also allows for unexcluded domains even below these bounds. 
 
\noindent
-when (III) is satisfied $b^2$ becomes negligible compared
to $(\nu\nu)^2$ and $(b\nu)^2$; the latter reaching at best $50\%$, could
be used for partial exclusion as discussed above. {\sl Finally, when 
$v_t\!~\lesssim~\!10^{-4}$GeV, the invisible decay branching ratio 
$(\nu\nu)^2$ reaches 98-99\% even for the loose bound $\lambda_{1 4}^+ \lesssim 1$,
thus evading all LEP constraints on scalar and pseudo-scalar states.} This conclusion
holds irrespective of the size of $\lambda_{1 4}^+$, that is  even if further
reduced by future LHC limits on invisible/undetected decays of the (SM-like) Higgs. 
It is noteworthy that the tininess of $v_t$, required to account for (Majorana) 
neutrino masses in a natural setting of the model with $Y_\nu$ of order one, 
automatically invalidates the LEP bounds on light scalars. 

\vspace{.5cm}
We close this section with some comments on the (doubly-)charged
states in the {\sl $H^0$-scenario}. As stated previously
the present experimental bounds on $m_{H^{++}}$ are in excess of
$410$~GeV or so and will be improved in the next LHC run.
These bounds are obtained under the assumption 
of same-sign di-lepton decays with branching ratio one
 \cite{ATLAS:2012hi, Chatrchyan:2012ya}, and can thus be
much weaker ($\sim 90$~GeV) if the $W^+ W^{+(*)}$ and/or $W^+ H^{+ (*)}$ decay channels of $H^{++}$ 
become important \cite{Kanemura:2013vxa, kang:2014jia, Kanemura:2014goa}. One should, however, keep in mind that this
weakening requires increasingly large values of $v_t$ that might 
become hardly consistent with the {\sl $H^0$-scenario}  whose
viability implies typically very small values
of this parameter, as we have shown in this section. For
$v_t \lesssim 10^{-4}$~GeV where the LEP constraints are totally
evaded, the present and future bounds from same-sign di-lepton
searches at the LHC fully apply, since  
$Br(H^{++} \to  l^+ l^+) \sim 1 $ in this case.
In the domain 
$4\times 10^{-4}$~GeV$\lesssim v_t \lesssim 1.3 \times 10^{-3}$~GeV for which the LEP exclusion domains are significantly reduced, except for a small region around the 
Z-boson mass, one finds 
$ 0.54 \lesssim Br(H^{++} \to  W^+  W^{+(*)}) \lesssim .99$
for $m_{H^{++} }\sim 400$GeV, signaling a reduction of the present
LHC bounds to roughly $m_{H^{++}} \gtrsim 160$~GeV. Note also
that the decay channel $H^{++} \to  W^+  H^{+ *}$ plays no role
in the {\sl $H^0$-scenario} due to the small mass splitting between the $H^+$ and $H^{++}$ states, 
Eqs.~(\ref{eq:mH+}, \ref{eq:mH++}).




\vspace{4cm} 

\section{Conclusion \label{sec:conclusion}}

There are mainly two 
dynamical regimes leading to electroweak scale states in the 
scalar sector of the {\sl type II seesaw} model. In this paper
we examined various 
phenomenological features of these regimes and highlighted 
in particular the viability of the {\sl $H^0$-scenario} where
 two electrically neutral CP-even 
and CP-odd scalar states are lighter than the  
discovered $125$~GeV Higgs-like state, and still 
compatible with LEP and present LHC constraints. The SM 
properties of the Higgs-like state are naturally accounted for due 
to the large hierarchy between the neutrino and the electroweak 
mass scales. Thus, future confirmation of the SM properties
of the $125$~GeV state with improved precision would 
not invalidate this scenario.
Even more, the diphoton and $Z\gamma$ decay channels can also
remain very close to their SM values due to a somewhat generic
screening of the effects of electroweak scale charged states.
Stringent lower bounds from future direct searches on the masses
of the latter states, combined with strict exclusion limits 
on invisible decays of the $125$~GeV state, will be eventually 
needed to disfavor this {\sl $H^0$-scenario}.    
  
\appendix
\section{\label{app:efp} effective fixed point in 
$\mathcal{H} \rightarrow \gamma\gamma, Z \gamma$. }

In section \ref{sec:HgagagaZ} we noted numerically a 
peculiar behavior of $\mathcal{H} \rightarrow \gamma\gamma, Z \gamma$
in the vicinity of the SM value. We give here a more detailed
quantitative study in the case of $\mathcal{H} \rightarrow \gamma\gamma$,
showing that this behavior is a direct 
consequence of the analytical structure of the (doubly-)charged
Higgs virtual contributions to the amplitude,
  
\begin{equation}
{\cal A} \equiv Q_+^2 \tilde{g}_{\mathcal{H} H^+\,H^-}
A_0^{{\mathcal{H}}}(\tau_{H^{+}})+
 Q_{++}^2 \tilde{g}_{\mathcal{H} H^{++}H^{--}}
A_0^{{\mathcal{H}}}(\tau_{H^{++}}) \label{eq:Ampcharged}
\end{equation}
entering Eq.(\ref{eq:Htogammagamma}), together with the form 
of their masses 
\begin{eqnarray}
m_{H^\pm}^2&=&\frac{(v_d^2+2 v_t^2)[2\sqrt{2}\mu- \lambda_4 v_t]}{4v_t} 
\label{eq:mHpm} \\
m_{H^{\pm\pm}}^2&=&\frac{\sqrt{2}\mu{v_d^2}- \lambda_4v_d^2v_t-2\lambda_3v_t^3}{2v_t}  \label{eq:mHpmpm}
\end{eqnarray} 
(see e.g.  \cite{Arhrib:2011vc} ).

\begin{figure}[!h]
\begin{tabular}{rr}
\hspace{-.8cm}\resizebox{86mm}{!}{\includegraphics{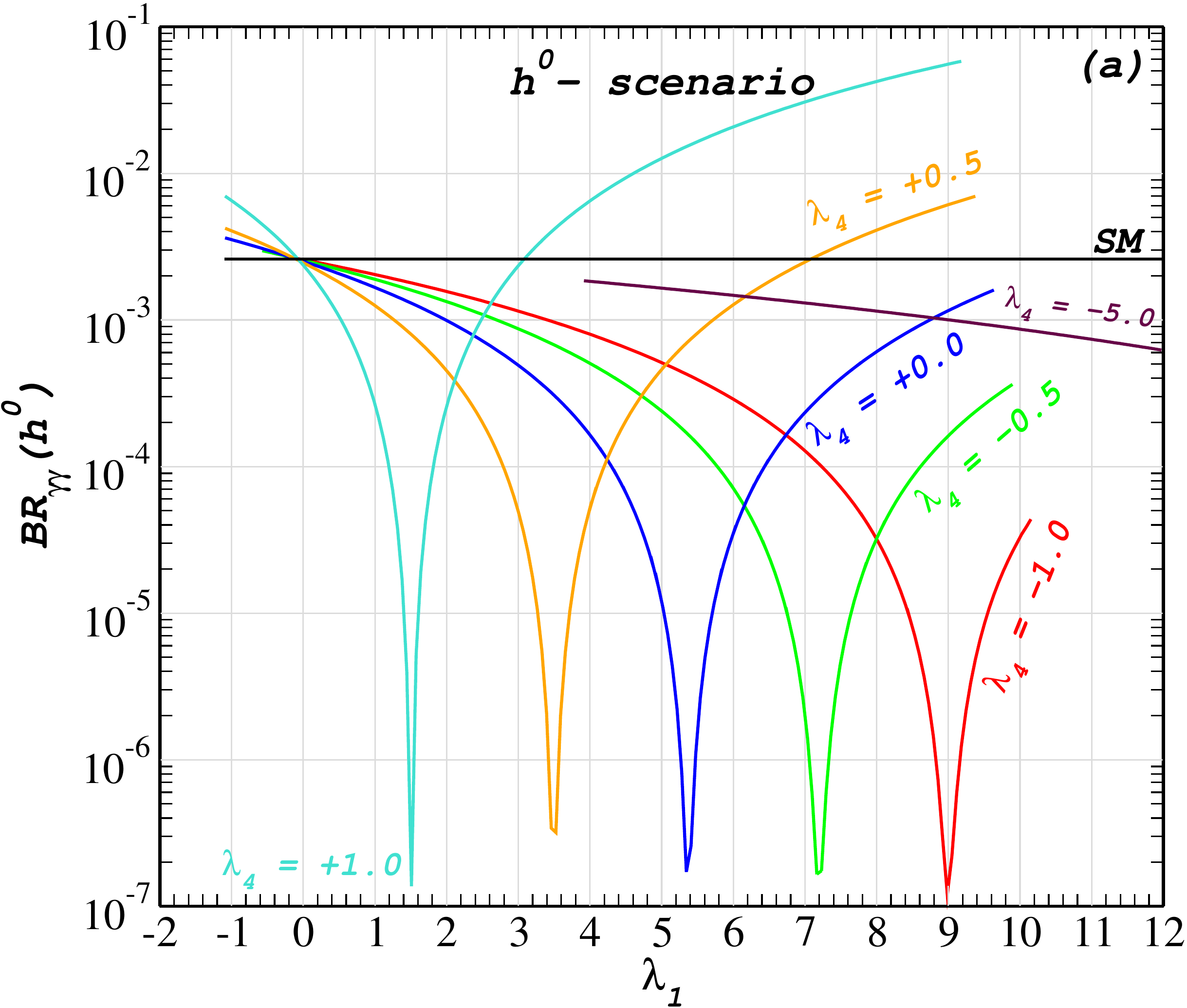}}&
\hspace{.1cm}\resizebox{86mm}{!}{\includegraphics{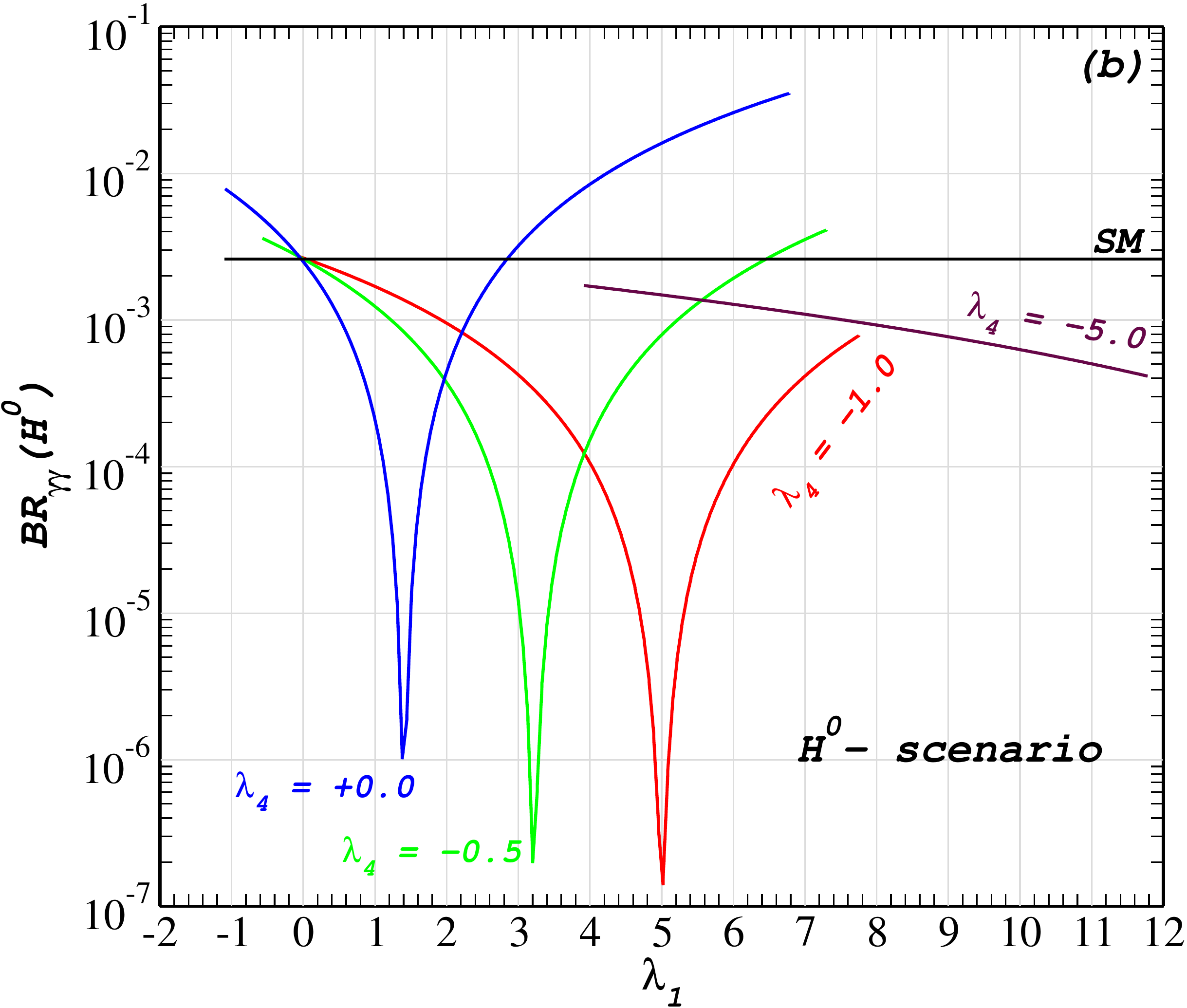}}\\
\hspace{-.8cm}\resizebox{86mm}{!}{\includegraphics{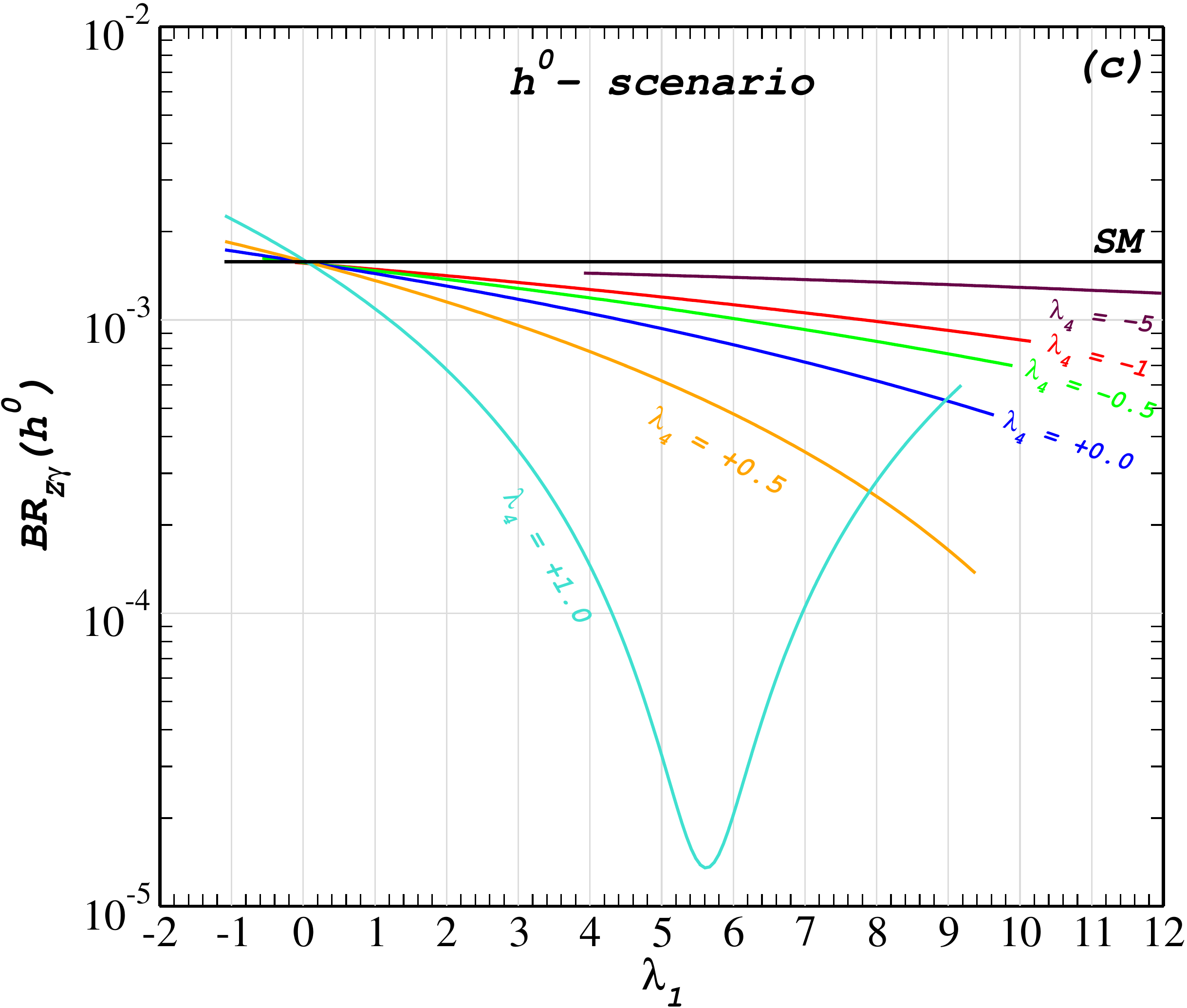}}&
\hspace{.1cm}\resizebox{86mm}{!}{\includegraphics{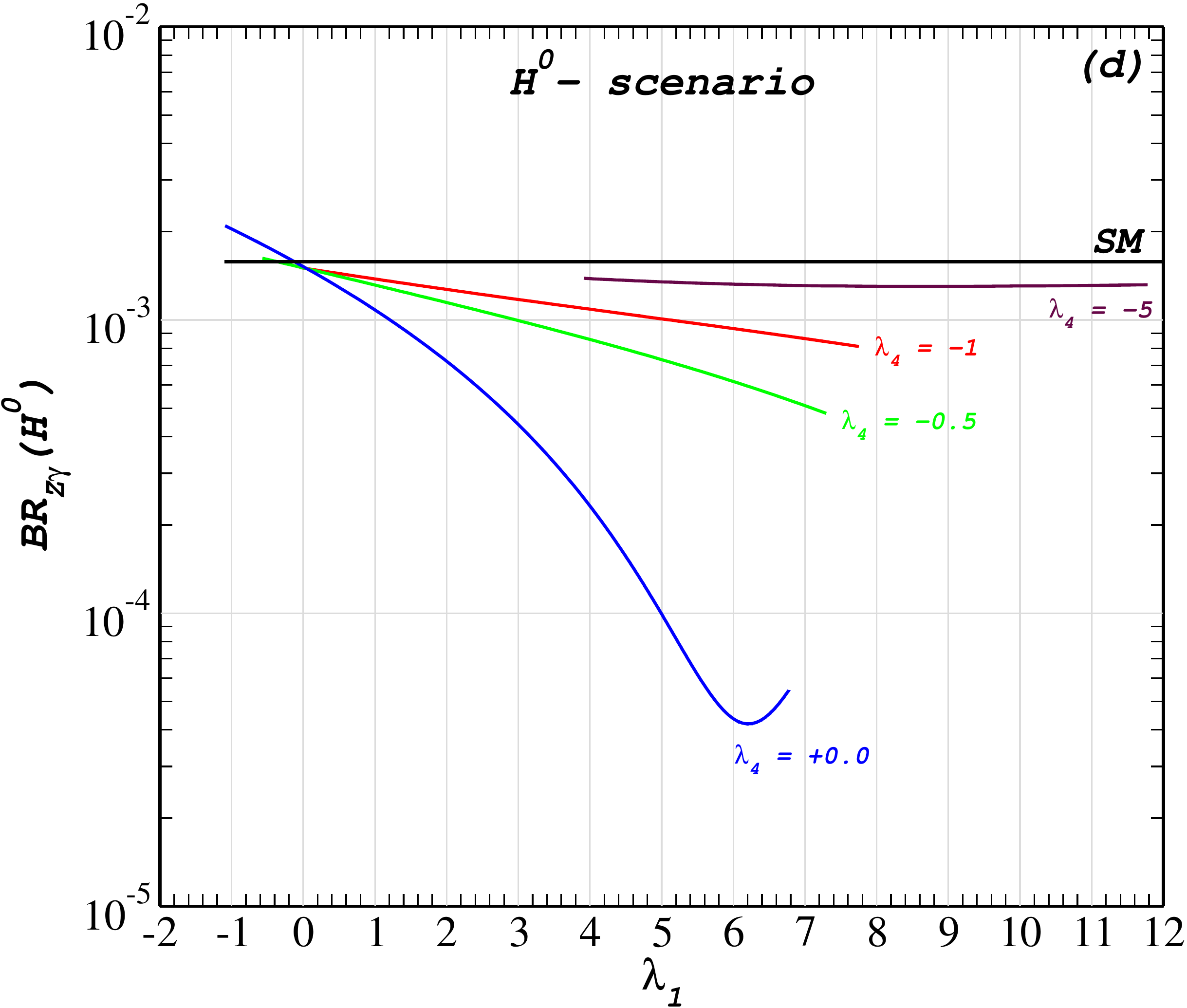}}
\end{tabular}
\caption{ 
The $\mathcal{H} \to \gamma\gamma,Z\gamma$ branching ratios as a function of $\lambda_1$ for various values of $\lambda_4$. 
We take $\lambda_3 = 2 \lambda_2$, $-5 \le \lambda_2 \le 5$, $-2 \le \lambda_1 \le 12$ and $v_t = 1$ GeV.
 figures (a),(c): `$h^0${\sl scenario}', $\mu=1$~GeV, 
 $\lambda = 0.521$,
 $m_{h^0}=125$--$125.6$~GeV, 
 $m_{H^0}\simeq m_{A^0} \approx 207$~GeV, 
$162~{\rm GeV} \lesssim m_{H^{+}} \lesssim 474$~GeV,
$97~{\rm GeV} \lesssim m_{H^{++}} \lesssim 637$~GeV
and $0.9 \leq \cos \alpha \leq 1$; 
figures (b),(d):
`$H^0${\sl scenario}', 
 $\mu=0.3$~GeV, $m_{H^0} = 125$--$126.5$~GeV, with $\lambda$
 as given by Eq.~(\ref{eq:lambda}), 
$m_{h^0} \simeq m_{A^0} \approx 113$~GeV, 
$100~{\rm GeV} \lesssim m_{H^{+}} \lesssim 440$~GeV,
$100~{\rm GeV} \lesssim m_{H^{++}} \lesssim 612$~GeV
and $0.9 \leq |\sin \alpha| \leq 1$.} 
\label{fig:BRgaga_BRgaZ}
\end{figure}

Treating ${\cal A}$ as a function of $\lambda_1, \lambda_4$, the 
observed effective fixed point in Figs.\ref{fig:BRgaga_BRgaZ}
(a), (b) can be understand as meaning that for 
$\lambda_1 = \lambda_1^0$ as defined in Eq.~(\ref{eq:SM-compatible}
), the gradient $\vec{ \nabla} {\cal A}$ in the $\lambda_1,
 \lambda_4$ space 
is essentially orthogonal to the displacement vector
$d\vec{\lambda} \equiv (d\lambda_1, d\lambda_4)$ in the directions
satisfying
$|d \lambda_1| \ll |d \lambda_4|$. This approximate
orthogonality occurs if 
$\partial {\cal A}/\partial \lambda_4 \ll \partial {\cal A}/\partial \lambda_1$, 
in which case one has
$d {\cal A} = \vec{ \nabla} {\cal A}.d\vec{\lambda} \simeq 0$
near the point $ {\cal A}_{\lambda_1 = \lambda_1^0} =0$,
thus leading to the observed very weak sensitivity to $\lambda_4$ 
when $\mathcal{H} \rightarrow \gamma\gamma$ coincides with the
SM value. Taking into account Eqs.~(\ref{eq:A0gg}, \ref{eq:ftau}, \ref{eq:mHpm}, 
\ref{eq:mHpmpm}) and the present
phenomenological bounds on $m_{H^+}, m_{H^{++}}$ that imply the occurrence of $\arcsin$ functions in Eqs.~(\ref{eq:Ampcharged},
\ref{eq:SM-compatible}), a somewhat lengthy but straightforward calculation leads
to the following expressions for 
$\frac{\partial {\cal A}/\partial \lambda_4}
{\partial {\cal A}/\partial \lambda_1}|_{\lambda_1 = \lambda_1^0}
$ in two relevant regimes.

\begin{itemize}
\item[ 1)] $\displaystyle \frac{v_t}{v_d} \ll 1$:
\end{itemize}
\begin{eqnarray}
\frac{\partial {\cal A}/\partial \lambda_4}
{\partial {\cal A}/\partial \lambda_1}|_{\lambda_1 = \lambda_1^0}
&=&
\frac{Q_{+}^2}{2 (Q_{+}^2+Q_{++}^2)}-
\frac{\lambda_4 Q_{+}^2 Q_{++}^2 }{2 \sqrt{2}  
(Q_{+}^2+Q_{++}^2)^2} \frac{v_t}{\mu}+ {\cal O}((\frac{v_t}{v_d})^
{\frac{3}{2}}) \nonumber \\
&\simeq& \frac{1}{10} - 5.6 \times 10^{-2} \lambda_4  \frac{v_t}{\mu}+ {\cal O}((\frac{v_t}{v_d})^
{\frac{3}{2}}) \label{eq:regime1}
\end{eqnarray}
with
\begin{eqnarray}
\lambda_1^0
&=&
-\frac{\lambda_4 Q_{+}^2}{2 (Q_{+}^2+Q_{++}^2)} +
\frac{\lambda_4^2 Q_{+}^2 Q_{++}^2 }{4 \sqrt{2}  
(Q_{+}^2+Q_{++}^2)^2} \frac{v_t}{\mu}+ {\cal O}((\frac{v_t}{v_d})^
{\frac{3}{2}}) \nonumber \\
&\simeq& -\frac{\lambda_4}{10} + 0.14 \times \lambda_4^2  \frac{v_t}{\mu}+ {\cal O}((\frac{v_t}{v_d})^
{\frac{3}{2}})
\end{eqnarray}
These expansions are valid for $\displaystyle \lambda_4  \frac{v_t}{\mu} \simeq
{\cal O}(1)$ or $\gg {\cal O}(1)$.


\begin{itemize}
\item[ 2)] $\displaystyle \frac{\mu}{v_t} \ll 1,\,\, 
\frac{v_t}{v_d} \ll 1$ and large $\lambda_4$:
\end{itemize}
\begin{eqnarray}
\frac{\partial {\cal A}/\partial \lambda_4}
{\partial {\cal A}/\partial \lambda_1}|_{\lambda_1 = \lambda_1^0}
&=&\frac{Q_{+}^2}{2 Q_{+}^2+Q_{++}^2}
+\frac{16 \sqrt{2}}{15} \frac{Q_{+}^2 Q_{++}^2 (Q_{+}^2+Q_{++}^2)}
{(2 Q_{+}^2+Q_{++}^2)^3 \lambda_4^2} 
\frac{m_{\cal H}^2}{v_d^2} \frac{\mu}{v_t} \nonumber \\
&~~& -
\frac{Q_{+}^2 Q_{++}^2 (362 Q_{+}^2+293 Q_{++}^2)}{1575 
 (2 Q_{+}^2+Q_{++}^2)^3 \lambda_4^2}  \frac{m_{\cal H}^4}{v_d^4}
 + {\cal O}(\frac{\mu^2}{v_t^2}, \frac{v_t^2}{v_d^2}, \lambda_4^{-\frac{5}{2}}) \nonumber \\
 &\simeq& \frac{1}{6} + (3.6 \times 10^{-2} \frac{\mu}{v_t}-
  1.2 \times 10^{-3}) \frac{1}{ \lambda_4^2} + {\cal O}(\frac{\mu^2}{v_t^2}, \frac{v_t^2}{v_d^2}, \lambda_4^{-\frac{5}{2}})
  \label{eq:regime2}
\end{eqnarray}
with
\begin{eqnarray}
\lambda_1^0
&=&-\frac{\lambda_4 Q_{+}^2}{2 Q_{+}^2+Q_{++}^2}
+ \frac{\sqrt{2} Q_+^2 Q_{++}^2}{(2 Q_{+}^2 + Q_{++}^2)^2
} ( \frac{16}{15} \frac{ (Q_{+}^2+Q_{++}^2)}
{(2 Q_{+}^2+Q_{++}^2) \lambda_4} 
\frac{m_{\cal H}^2}{v_d^2}- 1 )\frac{\mu}{v_t}\nonumber \\
&~~&
+\frac{4}{15} \frac{Q_{+}^2 Q_{++}^2}
{(2 Q_{+}^2+Q_{++}^2)^2 } 
\frac{m_{\cal H}^2}{v_d^2} -
\frac{Q_{+}^2 Q_{++}^2 (362 Q_{+}^2+293 Q_{++}^2)}{1575 
 (2 Q_{+}^2+Q_{++}^2)^3 \lambda_4}  \frac{m_{\cal H}^4}{v_d^4}
 + {\cal O}(\frac{\mu^2}{v_t^2}, \frac{v_t^2}{v_d^2}, \lambda_4^{-\frac{3}{2}}) \nonumber \\
 &\simeq& - \frac{\lambda_4}{6} + 
   (-0.16 + \frac{3.6 \times 10^{-2}}{\lambda_4}) \frac{\mu}{v_t}+ 
   7.6 \times 10^{-3} - \frac{1.2 \times 10^{-3}}{\lambda_4}  + 
   {\cal O}(\frac{\mu^2}{v_t^2}, \frac{v_t^2}{v_d^2}, \lambda_4^{-
   \frac{3}{2}})
\end{eqnarray}
Equations (\ref{eq:regime1}, \ref{eq:regime2}) illustrate the conditions under which the effective fixed point behavior is
reached. In particular in the regime
of large $\mu$, this behavior is
expected to be somewhat stronger than in the regime $
\mu \ll v_t$.
\section{\label{app:lambda1} $\lambda_1 <0$ versus $\lambda_1 > 0$ 
hyper-volumes}

We provide here the detailed evaluation of the result
stated in section \ref{sec:lam1lam4}.
We first recall the full set of U-BFB analytical constraints 
(see \cite{Arhrib:2011vc} for more details), recasting them 
here in separate sectors for the couplings:

{\sl $\lambda, \lambda_2, \lambda_3$ sector:}
\begin{eqnarray}
&& 0 \leq \lambda \leq \frac{2}{3} \kappa \pi  \label{eq:A1} \\
&& \lambda_2+\lambda_3 \geq 0  \;\;{\rm \&}  \;\;\lambda_2+\frac{\lambda_3}{2} \geq 0 \label{eq:A2}\\
&& \lambda_2 + 2 \lambda_3 \leq \frac{\kappa}{2} \pi \label{eq:A3}\\
&& 4 \lambda_2 + 3 \lambda_3 \leq  \frac{\kappa}{2} \pi 
\label{eq:A4}\\
&& 2 \lambda_2 - \lambda_3 \leq  \kappa \pi  \label{eq:A5}
\end{eqnarray}

{\sl  $\lambda_1, \lambda_4$ sector:}
\begin{eqnarray}
&&|\lambda_1 + \lambda_4| \leq \kappa \pi  \label{eq:A6} \\
&&|\lambda_1| \leq \kappa \pi \label{eq:A7} \\
&&|2 \lambda_1 + 3 \lambda_4| \leq 2 \kappa \pi \label{eq:A8} \\
&&|2 \lambda_1 - \lambda_4| \leq 2 \kappa \pi \label{eq:A9}
\end{eqnarray}

{\sl mixed sector:}
\begin{eqnarray}
&& \lambda_1+ \sqrt{\lambda(\lambda_2+\lambda_3)} \geq 0 \;\;{\rm \&}\;\;
\lambda_1+ \sqrt{\lambda(\lambda_2+\frac{\lambda_3}{2})} \geq 0  \label{eq:A10} \\
&&  \lambda_1+\lambda_4+\sqrt{\lambda(\lambda_2+\lambda_3)} \geq 0 \;\; {\rm \&} \;\; 
\lambda_1+\lambda_4+\sqrt{\lambda(\lambda_2+ \frac{\lambda_3}{2})} \geq 0 \label{eq:A11} \\
&&  |\lambda_4| \leq 
 \min\{M_+, M_-\} \label{eq:A12}\\
&& |2 \lambda_1 + \lambda_4| \leq \sqrt{2 (\lambda - \frac{2}{3} \kappa \pi) (4 \lambda_2 + 3 \lambda_3 
- \frac{\kappa}{2} \pi)} \label{eq:A13}
\end{eqnarray}
where 
\begin{eqnarray}
M_{\pm} \equiv \sqrt{(\lambda \pm 2 \kappa \pi) (\lambda_2 + 2 \lambda_3 \pm \frac{\kappa}{2} \pi)}
\end{eqnarray}

The correspondence with the equations of ref.\cite{Arhrib:2011uy}
is as follows:

\noindent
 (\ref{eq:A1}) -- (\ref{eq:A5}) $\leftrightarrow$ (6.14) -- (6.18); 
(\ref{eq:A6}) -- (\ref{eq:A8}) $\leftrightarrow$ (6.4) -- 
(6.6); 
(\ref{eq:A10}) $\leftrightarrow$ (6.2); 
(\ref{eq:A11}) $\leftrightarrow$ (6.3);
(\ref{eq:A12})$\leftrightarrow$ (6.19); 
(\ref{eq:A13})$\leftrightarrow$ (6.20) and (\ref{eq:A9}) $
\leftrightarrow$ (6.12). [Note that (6.12) was missing in the 
summary but included in the calculations in \cite{Arhrib:2011uy}.]

The domain delimited by the above equations is defined by 
intersections of hyperplanes and hyperbolas, thus in principle 
completely manageable analytically. The two-fold ambiguity in 
determining the minimum in Eq.(\ref{eq:A12}) can be easily lifted 
by noting that

\begin{eqnarray}
 &&\lambda + 4 (\lambda_2 + 2 \lambda_3) < 0 \Leftrightarrow 
 \; \mbox{$\min\{M_+, M_-\}=M_+$ } \nonumber\\
  &&\lambda + 4 (\lambda_2 + 2 \lambda_3) > 0 \Leftrightarrow 
 \; \mbox{$ \min\{M_+, M_-\}=M_-$ } \nonumber 
\end{eqnarray}   
Thus one just needs to add the simple hyperplane

\begin{equation}
\lambda + 4 (\lambda_2 + 2 \lambda_3) = 0 \label{eq:A14}
\end{equation}
to the set of boundary equations (\ref{eq:A1} - \ref{eq:A13}).

Moreover, to simplify the subsequent discussion without loss
of generality, 
we will take hereafter $\lambda = \displaystyle \frac{\pi}{6}$ 
which corresponds to $m_{\cal H} \simeq 126$GeV  
with $v_t/v_d \ll 1$.

The aim is to compare the relative sizes of the two hyper-volumes
$V_{\pm}$
\begin{eqnarray}
&&V_{\pm} \equiv \int_{{\cal D}_{\pm}} 
{\rm d}\lambda_2 \; {\rm d}\lambda_3 \; {\rm d}\lambda_4 
\; {\rm d}\lambda_1   
\end{eqnarray}
where ${\cal D}_{+}, {\cal D}_{-}$ denote the sub-domains
defined by Eqs.(\ref{eq:A1} - \ref{eq:A13}) and 
respectively 
$\lambda_1 >0$ and $\lambda_1<0$. Equations 
(\ref{eq:A2} - \ref{eq:A5}) 
can be worked out explicitly leading to piecewise integrals
over $\lambda_2, \lambda_3$. Taking into account
Eq.(\ref{eq:A14}), one can finally write $V_+$ in the form:
 
\begin{eqnarray}  
V_{+}&=& \int_0^{\frac{3 \kappa \pi}{10}} {\rm d}\lambda_3 \;
          \int_{-\frac{\lambda_3}{2}}^ {\frac{\kappa \pi}{8}  
          -\frac{3 \lambda_3}{4}} {\rm d}\lambda_2 \;
           \int_{0}^{\kappa \pi} {\rm d}\lambda_1 \;
            \int_{-M_{-}}^{M_{-}} {\rm d} \lambda_4 \; 
            {\cal B}[\lambda_1, \lambda_2, \lambda_3, \lambda_4] \nonumber \\
~&+&    \int_{\frac{3 \kappa \pi}{10}}^{\frac{\kappa \pi}{3}} {\rm d}\lambda_3 \;
           \int_{-\frac{\lambda_3}{2}}^{ \frac{\kappa \pi}{2}  -2 \lambda_3} {\rm d}\lambda_2 \;
            \int_{0}^{\kappa \pi} {\rm d}\lambda_1 \;
            \int_{-M_{-}}^{M_{-}} {\rm d} \lambda_4 \; 
            {\cal B}[\lambda_1, \lambda_2, \lambda_3, \lambda_4] \nonumber \\
~&+&        \int_{-\frac{\kappa \pi}{3}}^{-3 
\frac{\kappa \pi}{10}} {\rm d}\lambda_3 \;
            \int_{-\lambda_3}^{ \frac{\kappa \pi}{2} + 
            \frac{\lambda_3}{2}} {\rm d}\lambda_2 \;
            \int_{ 0}^{ \kappa \pi} {\rm d}\lambda_1 \;
           \int_{-M_{+}}^{M_{+}} {\rm d} \lambda_4 \; 
           {\cal B}[\lambda_1, \lambda_2, \lambda_3, \lambda_4]\nonumber \\
~&+& \int_{-3 \frac{\kappa \pi}{10}}^{-(1+ 3 \kappa) 
     \frac{\pi}{30}} {\rm d}\lambda_3 \;
      \int_{-\lambda_3}^{\frac{\kappa \pi}{8}  
      -\frac{3 \lambda_3}{4}} {\rm d}\lambda_2 \;
      \int_{ 0}^{ \kappa \pi} {\rm d}\lambda_1 \;
       \int_{-M_{+}}^{M_{+}} {\rm d} \lambda_4 \; 
       {\cal B}[\lambda_1, \lambda_2, \lambda_3, \lambda_4]\nonumber \\
~&+& \int_{-(1+ 3 \kappa) \frac{\pi}{30}}^{ -\frac{\pi}{24}} \!\!\!\!\!\!\!\!{\rm d}\lambda_3 \;
        \left \{ \int_{-\lambda_3}^{ - \frac{\pi}{24}  -2 \lambda_3} \!\!\!\!{\rm d}\lambda_2 \;
         \int_{0}^{\kappa \pi} {\rm d}\lambda_1 \;
           \int_{-M_{+}}^{M_{+}} {\rm d} \lambda_4 \; 
        {\cal B}[\lambda_1, \lambda_2, \lambda_3, \lambda_4]\right. \nonumber \\
 &&    ~~~~~~~~~~~~~~~~~~~~~~~~~~~~~~~~~~~~~~~~      +
\left. \int_{- \frac{\pi}{24}  -2 \lambda_3}^{\frac{\kappa \pi}{8}  -
\frac{3 \lambda_3}{4}} \!\!\!\!{\rm d} \lambda_2 \;
         \int_{ 0}^{ \kappa \pi} {\rm d}\lambda_1 \;
            \int_{-M_{-}}^{M_{-}} {\rm d} \lambda_4 \; 
        {\cal B}[\lambda_1, \lambda_2, \lambda_3, \lambda_4]  \right \} \nonumber \\
~&+& \int_{-\frac{\pi}{24}}^{0} {\rm d}\lambda_3 \;
 \int_{- \lambda_3}^{\frac{\kappa \pi}{8}  -3 
 \frac{\lambda_3}{4}} {\rm d}\lambda_2 \;
         \int_{ 0}^{\kappa \pi} {\rm d}\lambda_1 \;
            \int_{-M_{-}}^{M_{-}} {\rm d} \lambda_4 \; 
         {\cal B}[\lambda_1, \lambda_2, \lambda_3, \lambda_4]. 
         \nonumber
\end{eqnarray}
Here ${\cal B}[\lambda_1, \lambda_2, \lambda_3, \lambda_4]$ 
denotes the Boolean function for the remaining relevant constraints
Eqs.(\ref{eq:A6},\ref{eq:A8},\ref{eq:A9},\ref{eq:A11} \ref{eq:A13}). 

In the $\lambda_1 <0$ part one has to take also into account
that Eq.(\ref{eq:A10}) reduces to its first inequality for
$\lambda_3 <0$, and to its second inequality for $\lambda_3 >0$.
$V_-$ can then be written as,
\begin{eqnarray}  
V_{-}&=& \int_0^{\frac{3 \kappa \pi}{10}} {\rm d}\lambda_3 \;
          \int_{-\frac{\lambda_3}{2}}^ {\frac{\kappa \pi}{8}  
          -\frac{3 \lambda_3}{4}} {\rm d}\lambda_2 \;
\int_{-\sqrt{\lambda_2 + \frac{\lambda_3}{2}} \sqrt{\frac{\pi}{6}}}^{0} {\rm d} \lambda_1 \;
            \int_{-M_{-}}^{M_{-}} {\rm d} \lambda_4 \; 
            {\cal B}[\lambda_1, \lambda_2, \lambda_3, \lambda_4] \nonumber \\
~&+&    \int_{\frac{3 \kappa \pi}{10}}^{\frac{\kappa \pi}{3}} {\rm d}\lambda_3 \;
           \int_{-\frac{\lambda_3}{2}}^{ \frac{\kappa \pi}{2}  -2 \lambda_3} {\rm d}\lambda_2 \;
            \int_{-\sqrt{\lambda_2 + \frac{\lambda_3}{2}} \sqrt{\frac{\pi}{6}}}^{0} {\rm d} \lambda_1 \;
            \int_{-M_{-}}^{M_{-}} {\rm d} \lambda_4 \; 
            {\cal B}[\lambda_1, \lambda_2, \lambda_3, \lambda_4] \nonumber \\
~&+&        \int_{-\frac{\kappa \pi}{3}}^{-3 
\frac{\kappa \pi}{10}} {\rm d}\lambda_3 \;
            \int_{-\lambda_3}^{ \frac{\kappa \pi}{2} + 
            \frac{\lambda_3}{2}} {\rm d}\lambda_2 \;
            \int_{-\sqrt{\lambda_2 + {\lambda_3}} \sqrt{\frac{\pi}{6}}}^{0} {\rm d} \lambda_1 \;
           \int_{-M_{+}}^{M_{+}} {\rm d} \lambda_4 \; 
           {\cal B}[\lambda_1, \lambda_2, \lambda_3, \lambda_4]\nonumber \\
~&+& \int_{-3 \frac{\kappa \pi}{10}}^{-(1+ 3 \kappa) 
     \frac{\pi}{30}} {\rm d}\lambda_3 \;
      \int_{-\lambda_3}^{\frac{\kappa \pi}{8}  
      -\frac{3 \lambda_3}{4}} {\rm d}\lambda_2 \;
      \int_{-\sqrt{\lambda_2 + {\lambda_3}} \sqrt{\frac{\pi}{6}}}^{0} {\rm d} \lambda_1 \;
       \int_{-M_{+}}^{M_{+}} {\rm d} \lambda_4 \; 
       {\cal B}[\lambda_1, \lambda_2, \lambda_3, \lambda_4]\nonumber \\
~&+& \int_{-(1+ 3 \kappa) \frac{\pi}{30}}^{ -\frac{\pi}{24}} \!\!\!\!\!\!\!\!{\rm d}\lambda_3 \;
        \left \{ \int_{-\lambda_3}^{ - \frac{\pi}{24}  -2 \lambda_3} \!\!\!\!{\rm d}\lambda_2 \;
         \int_{-\sqrt{\lambda_2 + {\lambda_3}} \sqrt{\frac{\pi}{6}}}^{0} {\rm d} \lambda_1 \;
           \int_{-M_{+}}^{M_{+}} {\rm d} \lambda_4 \; 
        {\cal B}[\lambda_1, \lambda_2, \lambda_3, \lambda_4]\right. \nonumber \\
 &&    ~~~~~~~~~~~~~~~~~~~~~~~~~~~~~~~~~~~~~~~~      +
\left. \int_{- \frac{\pi}{24}  -2 \lambda_3}^{\frac{\kappa \pi}{8}  -
\frac{3 \lambda_3}{4}} \!\!\!\!{\rm d} \lambda_2 \;
         \int_{-\sqrt{\lambda_2 + \frac{\lambda_3}{2}} \sqrt{\frac{\pi}{6}}}^{0} {\rm d} \lambda_1 \;
            \int_{-M_{-}}^{M_{-}} {\rm d} \lambda_4 \; 
        {\cal B}[\lambda_1, \lambda_2, \lambda_3, \lambda_4]  \right \} \nonumber \\
~&+& \int_{-\frac{\pi}{24}}^{0} {\rm d}\lambda_3 \;
 \int_{- \lambda_3}^{\frac{\kappa \pi}{8}  -3 
 \frac{\lambda_3}{4}} {\rm d}\lambda_2 \;
         \int_{-\sqrt{\lambda_2 + {\lambda_3}} \sqrt{\frac{\pi}{6}}}^{0} {\rm d} \lambda_1 \;
            \int_{-M_{-}}^{M_{-}} {\rm d} \lambda_4 \; 
         {\cal B}[\lambda_1, \lambda_2, \lambda_3, \lambda_4]\nonumber
\end{eqnarray}
We stress that the same Boolean function ${\cal B}$ operates in 
both
$V_\pm$ domains. The reason is that the would-be extra constraint
$-\kappa \pi < \lambda_1$ of Eq.(\ref{eq:A7}) relevant for
$V_{-}$ can be shown to be always satisfied in the 
$(\lambda_2, \lambda_3)$
domain defined by Eqs.(\ref{eq:A2} - \ref{eq:A5}) when combined 
with Eqs.(\ref{eq:A10}). We note also that ${\cal B}$ can be
explicitly traded for further 
multiple piecewise integrations in the $(\lambda_1, \lambda_4)$, 
leading to highly involved but fully analytical integrations.
We refrain though from doing this here, since we are only 
interested in a numerical estimate of the hyper-volumes. The above
forms of $V_+$ and $V_-$ lend themselves easily to such an estimate
upon use of packages such as Mathematica.      

More general 
sub-volumes such as
$V_+^{\lambda_1 > \lambda_1^{min}>0}$ or $V_-^{\lambda_1 < \lambda_1^{max}<0}$, can be obtained respectively from $V_+$ 
through the substitution
$\int_{0}^{\kappa \pi} {\rm d}\lambda_1 \to \int_{\lambda_1^{min}}^{\kappa \pi} {\rm d}\lambda_1$, and from $V_-$ through
the substitutions
$\int_{-\sqrt{\lambda_2 + \frac{\lambda_3}{2}} \sqrt{\frac{\pi}{6}}}^{0} {\rm d} \lambda_1 \to \int_{-\sqrt{\lambda_2 + \frac{\lambda_3}{2}} \sqrt{\frac{\pi}{6}}}^{\lambda_1^{max}} {\rm d} \lambda_1$
and $\int_{-\sqrt{\lambda_2 + {\lambda_3}{}} \sqrt{\frac{\pi}{6}}}^{0} {\rm d} \lambda_1 \to
\int_{-\sqrt{\lambda_2 + {\lambda_3}{}} \sqrt{\frac{\pi}{6}}}^{\lambda_1^{max}} {\rm d} \lambda_1$.

The results of the numerical evaluation using Mathematica are given
in  table \ref{table:volumes}.

\begin{table}[!h]
\begin{center}
\renewcommand{\arraystretch}{1.5}
\begin{tabular}{|c|c|c|c|c|} \hline\hline
\ \ 
$\kappa$ \ \  & $8$ & $16$
\\ \hline\hline
$V_+$  &  $2514$  & $39796$  \ \\
\hline\hline
$V_-$  &  $275$  & $3027$  \ \\
\hline\hline
$V_+^{\lambda_1 > 10}$  &  $273$  & $18439$  \ \\
\hline\hline
$V_-^{\lambda_1 < - 0.5}$  &  $92$  & $1517$  \ \\
\hline\hline
$V_-^{\lambda_1 < - 1}$  &  $7$  & $433$  \ \\
\hline\hline
\end{tabular}
\end{center}
\caption{\label{table:volumes} Sizes of the various sub-volumes 
in the U-BFB four 
dimensional $\lambda_i$ parameter space region, 
for $\kappa = 8, 16$ and $\lambda = \frac{\pi}{6}$.}
\end{table}

\section{\label{app:QCD}QCD corrections to Higgs hadronic decays}
Hereafter $m$ denotes generically the $h^0, A^0$ masses.
We implemented the running quark masses $\overline{m}_b(\mu)$,
$\overline{m}_c(\mu)$ as well as $\alpha_s(\mu)$ up to 4-loop QCD order,  
relying partly on \cite{Chetyrkin:2000yt} and partly on our private
code, fixing the $\overline{\rm MS}$ b- and c-quark  masses to 
$\overline{m}_b(m_b)= m_b= 4.16$GeV,
$\overline{m}_c(m_c)= m_c= 1.28$GeV, and $\overline{\alpha}_s(M_Z)=
0.1184$ with $M_Z = 91.18$GeV.
The b- and c-quark pole masses are taken $M_b= 4.69$GeV, $M_c= 1.55$GeV,
and the top quark mass $M_t = 173$GeV. The other relevant parameters are fixed
as follows: $M_\tau= 1.777$GeV, $G_F= 1.16637 \times 10^{-5}$GeV${}^{-2}$,
$v_d = 246$GeV.

\vspace{.5 cm}
\noindent
{\sl  {-$b\bar{b}$ decay widths of $h^0, A^0$:}}
we use here the results of \cite{Chetyrkin:1995pd},

\begin{eqnarray}
\Gamma_{ S \to b\bar{b}} &=&  \frac{3 G_F}{4 \sqrt{2} \pi} C_{bb}^S \; m_S \; 
\overline{m}_b^2(m_S)  
\Big( 1 + \Delta\Gamma_{1,S} \frac{\overline{\alpha}_s(m_S)}{\pi} +   
(\Delta\Gamma_{2,S} + \frac{m_S^2}{M_{t}^2} \Delta\tilde{\Gamma}_{2,S}) 
\frac{\overline{\alpha}_s^2(m_S)}{\pi^2}      \nonumber \\
&&   + \; \frac{\overline{m}_b^2(m_S)}{m_S^2} \big(
   \Delta\Gamma_0^{(m_S)} + 
   \Delta\Gamma_1^{(m_S)} \frac{\overline{\alpha}_s(m_S)}{\pi} +  
   (\Delta\Gamma_2^{(m_S)} + 
   \frac{m_S^2}{M_t^2} \Delta\tilde{\Gamma}_2^{(m_S)}) 
   \frac{\overline{\alpha}_s^2(m_S)}{\pi^2} \big) \nonumber \\
&&   + \; {\cal O}(\frac{\overline{m}_b^4(m_S)}{m_S^4}) \Big)
\label{eq:Gammabbar}
\end{eqnarray} 
with $S=h^0, A^0$ and

\begin{eqnarray}
\Delta\Gamma_{1, h^0}&=& \frac{17}{3} \nonumber \\
\Delta\Gamma_{2, h^0} &=& 29.147 + k_{h^0} \big(1.57 - 
\frac{2}{3} \ln[\frac{m_{h^0}^2}{M_t^2}] + 
                      \frac{1}{9} \ln^2[\frac{\overline{m}_b^2(m_{h^0})}{m_{h^0}^2}] \big)
                      \nonumber \\ 
\Delta\tilde{\Gamma}_{2, h^0}&=& \frac{107}{675} - 
\frac{2}{45} \ln[\frac{m_{h^0}^2}{M_t^2}] + 
 k_{h^0} \big(-0.007 - \frac{41}{1620} \ln[\frac{m_{h^0}^2}{M_t^2}] + 
                   \frac{7}{1080} \ln^2[\frac{\overline{m}_b^2(m_{h^0})}{m_{h^0}^2}]\big) \nonumber \\
\Delta\Gamma_{0}^{(m_{h^0})} &=& -6 \nonumber \\
\Delta\Gamma_{1}^{(m_{h^0})} &=& -40 \nonumber \\ 
\Delta\Gamma_{2}^{(m_{h^0})} &=& -107.755 - 
0.98 \ln^2[\frac{\overline{m}_b^2(m_{h^0})}{m_{h^0}^2}] -
\frac{1}{12} \ln^4[\frac{\overline{m}_b^2(m_{h^0})}{m_{h^0}^2}] + 4 \nonumber \\
&& + \; k_{h^0} \big(-5.61 + 4 \ln[\frac{m_{h^0}^2}{M_t^2}] + 
\frac{16}{9} \ln[\frac{\overline{m}_b^2(m_{h^0})}{m_{h^0}^2}] -
\frac{4}{9} \ln^2[\frac{\overline{m}_b^2(m_{h^0})}{m_{h^0}^2}] \big) \nonumber \\
\Delta\tilde{\Gamma}_{2}^{(m_{h^0})} &=& -\frac{116}{75} + \frac{8}{45} \ln[\frac{m_{h^0}^2}{M_t^2}]   \nonumber \\
   &&   ~~~~~~~~~    + \; k_{h^0} \big(0.52 - \frac{7}{270} \ln[\frac{m_{h^0}^2}{M_t^2}] +
                     \frac{1}{135} \ln[\frac{\overline{m}_b^2(m_{h^0})}{m_{h^0}^2}] -
                     \frac{7}{270} \ln^2[\frac{\overline{m}_b^2(m_{h^0})}{m_{h^0}^2}]\big) \nonumber \\
\Delta\Gamma_{1, A^0} &=& \frac{17}{3} \nonumber \\
\Delta\Gamma_{2, A^0} &=& 29.147 + k_{A^0} \big(\frac{23}{6} -  
\ln[\frac{m_{A^0}^2}{M_t^2}] + 
                     \frac{1}{6} \ln^2[\frac{\overline{m}_b^2(m_{A^0})}{m_{A^0}^2}] \big) \nonumber
\end{eqnarray}
\begin{eqnarray}                      
\Delta\tilde{\Gamma}_{2,A^0}&=& \frac{107}{675} - \frac{2}{45} \ln[\frac{m_{A^0}^2}{M_t^2}] + 
 k_{A^0} \big(0.051 - \frac{7}{108} \ln[\frac{m_{A^0}^2}{M_t^2}] + 
                    \frac{1}{72} \ln[\frac{\overline{m}_b^2(m_{A^0})}{m_{A^0}^2}]^2 \big) \nonumber \\
\Delta\Gamma_{0}^{(m_{A^0})} &=& -2 \nonumber \\
\Delta\Gamma_{1}^{(m_{A^0})} &=& -\frac{8}{3} \nonumber \\                             
\Delta\Gamma_{2}^{(m_{A^0})} &=& 91.006 - 
26.32 \ln^2[\frac{\overline{m}_b^2(m_{A^0})}{m_{A^0}^2}] -
\frac{4}{3} \ln^4[\frac{\overline{m}_b^2(m_{A^0})}{m_{A^0}^2}] + 4 \nonumber \\
&& ~~~~~~~~~~~~~~~~~~~~~~~ + \; k_{A^0} \big(-5 + 2 \ln[\frac{m_{A^0}^2}{M_t^2}] - 
\frac{4}{3} \ln[\frac{\overline{m}_b^2(m_{A^0})}{m_{A_0}^2}] ) \nonumber \\
\Delta\tilde{\Gamma}_{2}^{(m_{A^0})}&=& -\frac{16}{25} + \frac{4}{15} \ln[\frac{m_{A^0}^2}{M_t^2}] + 
k_{A^0} \big(\frac{19}{108} - \frac{1}{18} \ln[\frac{m_{A^0}^2}{M_t^2}] -
                \frac{2}{9} \ln[\frac{\overline{m}_b^2(m_{A^0})}{m_{A^0}^2}] \big) \label{eq:DeltaGammas}
\end{eqnarray}
In the above, the number of quark flavors, $n_f=5$, has 
been assumed in the widths and in the running quantities,
since $m_b \ll m_S < m_t$.

\vspace{.5 cm}
\noindent
{\sl  {-$c\bar{c}$ decay widths of $h^0, A^0$:}}
these can be read from Eqs.(\ref{eq:Gammabbar}) to (\ref{eq:DeltaGammas}) by discarding
the extra contributions of finite c-quark mass, as well as
those originating from the heavy top limit 
\cite{Chetyrkin:1995pd},

\begin{eqnarray}
\Gamma_{S \to c\bar{c}} &=&  \frac{3 G_F}{4 \sqrt{2} \pi} 
C_{cc}^S \; m_S \; \overline{m}_c^2(m_S)   
\big( 1 +  5.67 \frac{\overline{\alpha}_s(m_S)}{\pi}  + 
(35.94 - 1.36 n_f + \delta_{c\bar{c}}^S) \frac{\overline{\alpha}_s^2(m_S)}{\pi^2} + {\cal O}(\frac{\overline{\alpha}_s^3(m_S)}{\pi^3}) \; \big)\nonumber
\end{eqnarray}
where
\begin{eqnarray}
\delta_{c\bar{c}}^{h^0}&=& k_{h^0} (1.57 - \frac{2}{3} \ln[\frac{m_{h^0}^2}{M_t^2}] + 
                      \frac{1}{9} \ln^2[\frac{\overline{m}_c^2(m_{h^0})}{m_{h^0}^2}])\nonumber \\                      
\delta_{c\bar{c}}^{A^0}&=& k_{A^0} (\frac{23}{6} -  \ln[\frac{m_{A^0}^2}{M_t^2}] + 
                     \frac{1}{6} \ln^2[\frac{\overline{m}_c^2(m_{A^0})}{m_{A^0}^2}] ) \nonumber
\end{eqnarray}
where again one should take $n_f=5$. Note that another known 
mass-independent ${\cal O}(\alpha_s^3)$ correction,
$\delta_{c\bar{c}}^{(3)} = (164.14 -25.77 n_f +0.26 n_f^2) 
\frac{\overline{\alpha}_s^3(m)}{\pi^3}$, has not been included
as it does not give significant contributions
 (see for instance 
\cite{Spira:1997dg} for a review of the QCD effects).

\vspace{.5 cm}
\noindent
{\sl  {-decay widths of $h^0, A^0$ in two gluons in the limit $m \ll M_t$}:}
we use the results of \cite{Chetyrkin:1997iv}, 

\begin{equation}
\Gamma_{S \to gg}= \Gamma_{S \to gg}^{\rm LO} K_{\rm factor}(n_f, m_S);
\end{equation}

\begin{equation}
\Gamma_{S \to gg}^{LO} = 
c_{S} C^{S} \frac{G_F m^3}{36 \sqrt{2} \pi} 
\frac{\overline{\alpha}_s^2(m_S)}{\pi^2} 
\end{equation}
with 
\begin{equation}
c_{h^0} =1, ~~~ c_{A^0} = 4
\end{equation}

\begin{eqnarray}
K_{\rm factor}(n_f, \mu) &=&
1 + \frac{\overline{\alpha}_s(m)}{\pi}  \big(\frac{95}{4} -\frac{7}{6} n_f \big) + 
\frac{\overline{\alpha}_s^2(m)}{\pi^2} \big(\frac{149533}{288} -\frac{363}{8} \zeta(2) 
-\frac{495}{8} \zeta(3) - 
 \frac{19}{8} \ln[\frac{M_t^2}{\mu^2}] \nonumber \\
&& + n_f (-\frac{4157}{72} +  \frac{11}{2} \zeta(2) 
+\frac{5}{4} \zeta(3) - \frac{2}{3} \ln[\frac{M_t^2}{\mu^2}]) +
n_f^2 (\frac{127}{108} -\frac{1}{6} \zeta(2)) \big) + {\cal O}(\frac{\overline{\alpha}_s^3(m)}{\pi^3})  \nonumber
\end{eqnarray}
with $\zeta(2) = \pi^2/6$ and $\zeta(3) \simeq 1.20206$.


The coefficients $C_{f f'}^S$, being defined as the product 
of 
the reduced couplings of $f$ and $f'$ to $S$, and $k_S$
defined as the ratios of these products, one has in the case
of $H^0${\sl-scenario} under consideration, see table \ref{table_couplings1}, 
 
\begin{eqnarray}
&& C_{bb}^{h_0} = C_{cc}^{h_0} = C^{h_0} = 
\frac{(\lambda_1+\lambda_4)^2}{\lambda^2} \times \frac{4 v_t^2}{v_d^2}, \nonumber \\
&&C_{bb}^{A_0} = C_{cc}^{A_0} = C^{A_0} =  \frac{4 v_t^2}{v_d^2}, \nonumber \\
&& k_{h^0} = k_{A^0} =1 .\nonumber
\end{eqnarray}
\section{\label{app:ZhA}relevant cross-section, width, decay length}

For completness we recall here the tree-level expressions of 
the Z-boson
decay width into a scalar and a pseudo-scalar states, as
well as the $e^+ e^- \to h^0 A^0$ cross-section for
a generic properly normalized coupling $c$, see also
section \ref{sec:lightS}, 

\begin{equation}
\Gamma_{Z \to h^0 A^0}= \frac{c^2 \sqrt{2} G_F m_Z^3}{48 \pi} \times \lambda[1, \frac{m_{h^0}^2}{m_Z^2}, \frac{m_{A^0}^2}{m_Z^2}]^{\frac{3}{2}}
\label{eq:ZtohA}
\end{equation}

\begin{equation}
\sigma(e^+ e^- \to h^0 A^0) =  \frac{c^2 G_F^2 m_Z^4 }{96 \pi s} (1 + (1 - 4 s_W^2)^2) 
\times (1 - \frac{m_Z^2}{s})^{-2} \lambda[1, \frac{m_{h^0}^2}{s}, \frac{m_{A^0}^2}{s}]^{\frac{3}{2}} 
\label{eq:eetohA}
\end{equation}
with the usual phase-space function defined as 
$$\lambda[x, y^2, z^2] \equiv (x - (y -z)^2)(x - (y+z)^2)$$

\noindent
The decay length $c\tau$ in the laboratory frame for a particle
of mass $m$, energy 
$\displaystyle E = \frac{\sqrt{s}}{2}$ and total
decay width $\Gamma$, is given, in the instantaneous decay approximation, by 
\begin{equation} 
c\tau = 9.86 \times 10^{-17} \times 
\frac{(s -4 \,m^2)^{\frac12}}{m \, \Gamma}
\end{equation}
where mass, energy and width are in GeV and $c\tau$ in meters.

\bibliography{references-triplet}

\ifx\mcitethebibliography\mciteundefinedmacro
\PackageError{unsrtM.bst}{mciteplus.sty has not been loaded}
{This bibstyle requires the use of the mciteplus package.}\fi
\begin{mcitethebibliography}{10}

\bibitem{Aad:2012tfa}
ATLAS Collaboration, G.~Aad {\em et~al.},
\newblock Phys.Lett. {\bf B716}, 1 (2012), 1207.7214\relax
\mciteBstWouldAddEndPuncttrue
\mciteSetBstMidEndSepPunct{\mcitedefaultmidpunct}
{\mcitedefaultendpunct}{\mcitedefaultseppunct}\relax
\EndOfBibitem
\bibitem{Chatrchyan:2012ufa}
CMS Collaboration, S.~Chatrchyan {\em et~al.},
\newblock Phys.Lett. {\bf B716}, 30 (2012), 1207.7235\relax
\mciteBstWouldAddEndPuncttrue
\mciteSetBstMidEndSepPunct{\mcitedefaultmidpunct}
{\mcitedefaultendpunct}{\mcitedefaultseppunct}\relax
\EndOfBibitem
\bibitem{ATLAS:2013sla}
ATLAS-CONF-2013-034,
\newblock (2013)\relax
\mciteBstWouldAddEndPuncttrue
\mciteSetBstMidEndSepPunct{\mcitedefaultmidpunct}
{\mcitedefaultendpunct}{\mcitedefaultseppunct}\relax
\EndOfBibitem
\bibitem{CMS:yva}
CMS-PAS-HIG-13-005,
\newblock (2013)\relax
\mciteBstWouldAddEndPuncttrue
\mciteSetBstMidEndSepPunct{\mcitedefaultmidpunct}
{\mcitedefaultendpunct}{\mcitedefaultseppunct}\relax
\EndOfBibitem
\bibitem{ATLAS:2013mla}
ATLAS-CONF-2013-040,
\newblock (2013)\relax
\mciteBstWouldAddEndPuncttrue
\mciteSetBstMidEndSepPunct{\mcitedefaultmidpunct}
{\mcitedefaultendpunct}{\mcitedefaultseppunct}\relax
\EndOfBibitem
\bibitem{CMS:xwa}
CMS-PAS-HIG-13-002,
\newblock (2013)\relax
\mciteBstWouldAddEndPuncttrue
\mciteSetBstMidEndSepPunct{\mcitedefaultmidpunct}
{\mcitedefaultendpunct}{\mcitedefaultseppunct}\relax
\EndOfBibitem
\bibitem{CMS:bxa}
CMS-PAS-HIG-13-003,
\newblock (2013)\relax
\mciteBstWouldAddEndPuncttrue
\mciteSetBstMidEndSepPunct{\mcitedefaultmidpunct}
{\mcitedefaultendpunct}{\mcitedefaultseppunct}\relax
\EndOfBibitem
\bibitem{Aad:2014aba}
ATLAS Collaboration, G.~Aad {\em et~al.},
\newblock (2014), 1406.3827\relax
\mciteBstWouldAddEndPuncttrue
\mciteSetBstMidEndSepPunct{\mcitedefaultmidpunct}
{\mcitedefaultendpunct}{\mcitedefaultseppunct}\relax
\EndOfBibitem
\bibitem{Khachatryan:2014ira}
CMS Collaboration, V.~Khachatryan {\em et~al.},
\newblock (2014), 1407.0558\relax
\mciteBstWouldAddEndPuncttrue
\mciteSetBstMidEndSepPunct{\mcitedefaultmidpunct}
{\mcitedefaultendpunct}{\mcitedefaultseppunct}\relax
\EndOfBibitem
\bibitem{Konetschny:1977bn}
W.~Konetschny and W.~Kummer,
\newblock Phys. Lett. {\bf B70}, 433 (1977)\relax
\mciteBstWouldAddEndPuncttrue
\mciteSetBstMidEndSepPunct{\mcitedefaultmidpunct}
{\mcitedefaultendpunct}{\mcitedefaultseppunct}\relax
\EndOfBibitem
\bibitem{Cheng:1980qt}
T.~P. Cheng and L.-F. Li,
\newblock Phys. Rev. {\bf D22}, 2860 (1980)\relax
\mciteBstWouldAddEndPuncttrue
\mciteSetBstMidEndSepPunct{\mcitedefaultmidpunct}
{\mcitedefaultendpunct}{\mcitedefaultseppunct}\relax
\EndOfBibitem
\bibitem{Lazarides:1980nt}
G.~Lazarides, Q.~Shafi, and C.~Wetterich,
\newblock Nucl. Phys. {\bf B181}, 287 (1981)\relax
\mciteBstWouldAddEndPuncttrue
\mciteSetBstMidEndSepPunct{\mcitedefaultmidpunct}
{\mcitedefaultendpunct}{\mcitedefaultseppunct}\relax
\EndOfBibitem
\bibitem{Schechter:1980gr}
J.~Schechter and J.~W.~F. Valle,
\newblock Phys. Rev. {\bf D22}, 2227 (1980)\relax
\mciteBstWouldAddEndPuncttrue
\mciteSetBstMidEndSepPunct{\mcitedefaultmidpunct}
{\mcitedefaultendpunct}{\mcitedefaultseppunct}\relax
\EndOfBibitem
\bibitem{Mohapatra:1980yp}
R.~N. Mohapatra and G.~Senjanovic,
\newblock Phys. Rev. {\bf D23}, 165 (1981)\relax
\mciteBstWouldAddEndPuncttrue
\mciteSetBstMidEndSepPunct{\mcitedefaultmidpunct}
{\mcitedefaultendpunct}{\mcitedefaultseppunct}\relax
\EndOfBibitem
\bibitem{Arhrib:2011uy}
A.~Arhrib {\em et~al.},
\newblock Phys.Rev. {\bf D84}, 095005 (2011), 1105.1925\relax
\mciteBstWouldAddEndPuncttrue
\mciteSetBstMidEndSepPunct{\mcitedefaultmidpunct}
{\mcitedefaultendpunct}{\mcitedefaultseppunct}\relax
\EndOfBibitem
\bibitem{Akeroyd:2012ms}
A.~Akeroyd and S.~Moretti,
\newblock Phys.Rev. {\bf D86}, 035015 (2012), 1206.0535\relax
\mciteBstWouldAddEndPuncttrue
\mciteSetBstMidEndSepPunct{\mcitedefaultmidpunct}
{\mcitedefaultendpunct}{\mcitedefaultseppunct}\relax
\EndOfBibitem
\bibitem{Wang:2012ts}
L.~Wang and X.-F. Han,
\newblock Phys.Rev. {\bf D87}, 015015 (2013), 1209.0376\relax
\mciteBstWouldAddEndPuncttrue
\mciteSetBstMidEndSepPunct{\mcitedefaultmidpunct}
{\mcitedefaultendpunct}{\mcitedefaultseppunct}\relax
\EndOfBibitem
\bibitem{Akeroyd:1900zz}
A.~Akeroyd, M.~Aoki, and H.~Sugiyama,
\newblock (2011)\relax
\mciteBstWouldAddEndPuncttrue
\mciteSetBstMidEndSepPunct{\mcitedefaultmidpunct}
{\mcitedefaultendpunct}{\mcitedefaultseppunct}\relax
\EndOfBibitem
\bibitem{Aoki:2011pz}
M.~Aoki, S.~Kanemura, and K.~Yagyu,
\newblock Phys.Rev. {\bf D85}, 055007 (2012), 1110.4625\relax
\mciteBstWouldAddEndPuncttrue
\mciteSetBstMidEndSepPunct{\mcitedefaultmidpunct}
{\mcitedefaultendpunct}{\mcitedefaultseppunct}\relax
\EndOfBibitem
\bibitem{Dev:2013ff}
P.~Bhupal~Dev, D.~K. Ghosh, N.~Okada, and I.~Saha,
\newblock JHEP {\bf 1303}, 150 (2013), 1301.3453\relax
\mciteBstWouldAddEndPuncttrue
\mciteSetBstMidEndSepPunct{\mcitedefaultmidpunct}
{\mcitedefaultendpunct}{\mcitedefaultseppunct}\relax
\EndOfBibitem
\bibitem{Akeroyd:2010je}
A.~G. Akeroyd and C.-W. Chiang,
\newblock Phys. Rev. {\bf D81}, 115007 (2010), 1003.3724\relax
\mciteBstWouldAddEndPuncttrue
\mciteSetBstMidEndSepPunct{\mcitedefaultmidpunct}
{\mcitedefaultendpunct}{\mcitedefaultseppunct}\relax
\EndOfBibitem
\bibitem{Perez:2008ha}
P.~Fileviez~Perez, T.~Han, G.-y. Huang, T.~Li, and K.~Wang,
\newblock Phys. Rev. {\bf D78}, 015018 (2008), 0805.3536\relax
\mciteBstWouldAddEndPuncttrue
\mciteSetBstMidEndSepPunct{\mcitedefaultmidpunct}
{\mcitedefaultendpunct}{\mcitedefaultseppunct}\relax
\EndOfBibitem
\bibitem{LHCHiggsCrossSectionWorkingGroup:2012nn}
LHC Higgs Cross Section Working Group, A.~David {\em et~al.},
\newblock (2012), 1209.0040\relax
\mciteBstWouldAddEndPuncttrue
\mciteSetBstMidEndSepPunct{\mcitedefaultmidpunct}
{\mcitedefaultendpunct}{\mcitedefaultseppunct}\relax
\EndOfBibitem
\bibitem{CMS:2013xfa}
CMS Collaboration,
\newblock (2013), 1307.7135\relax
\mciteBstWouldAddEndPuncttrue
\mciteSetBstMidEndSepPunct{\mcitedefaultmidpunct}
{\mcitedefaultendpunct}{\mcitedefaultseppunct}\relax
\EndOfBibitem
\bibitem{ATLAS-projections}
ATL-PHYS-PUB-2013-014\relax
\mciteBstWouldAddEndPuncttrue
\mciteSetBstMidEndSepPunct{\mcitedefaultmidpunct}
{\mcitedefaultendpunct}{\mcitedefaultseppunct}\relax
\EndOfBibitem
\bibitem{Nakamura:2010zzi}
Particle Data Group, K.~Nakamura {\em et~al.},
\newblock J.Phys.G {\bf G37}, 075021 (2010)\relax
\mciteBstWouldAddEndPuncttrue
\mciteSetBstMidEndSepPunct{\mcitedefaultmidpunct}
{\mcitedefaultendpunct}{\mcitedefaultseppunct}\relax
\EndOfBibitem
\bibitem{Kanemura:2012rs}
S.~Kanemura and K.~Yagyu,
\newblock Phys.Rev. {\bf D85}, 115009 (2012), 1201.6287\relax
\mciteBstWouldAddEndPuncttrue
\mciteSetBstMidEndSepPunct{\mcitedefaultmidpunct}
{\mcitedefaultendpunct}{\mcitedefaultseppunct}\relax
\EndOfBibitem
\bibitem{Kanemura:2012rj}
S.~Kanemura and H.~Sugiyama,
\newblock Phys.Rev. {\bf D86}, 073006 (2012), 1202.5231\relax
\mciteBstWouldAddEndPuncttrue
\mciteSetBstMidEndSepPunct{\mcitedefaultmidpunct}
{\mcitedefaultendpunct}{\mcitedefaultseppunct}\relax
\EndOfBibitem
\bibitem{Franceschini:2013aha}
R.~Franceschini and R.~Mohapatra,
\newblock (2013), 1306.6108\relax
\mciteBstWouldAddEndPuncttrue
\mciteSetBstMidEndSepPunct{\mcitedefaultmidpunct}
{\mcitedefaultendpunct}{\mcitedefaultseppunct}\relax
\EndOfBibitem
\bibitem{Banks:2010zn}
T.~Banks and N.~Seiberg,
\newblock Phys.Rev. {\bf D83}, 084019 (2011), 1011.5120\relax
\mciteBstWouldAddEndPuncttrue
\mciteSetBstMidEndSepPunct{\mcitedefaultmidpunct}
{\mcitedefaultendpunct}{\mcitedefaultseppunct}\relax
\EndOfBibitem
\bibitem{ATLAS:2013oma}
ATLAS-CONF-2013-012\relax
\mciteBstWouldAddEndPuncttrue
\mciteSetBstMidEndSepPunct{\mcitedefaultmidpunct}
{\mcitedefaultendpunct}{\mcitedefaultseppunct}\relax
\EndOfBibitem
\bibitem{CMS:ril}
CMS-PAS-HIG-13-001\relax
\mciteBstWouldAddEndPuncttrue
\mciteSetBstMidEndSepPunct{\mcitedefaultmidpunct}
{\mcitedefaultendpunct}{\mcitedefaultseppunct}\relax
\EndOfBibitem
\bibitem{Arhrib:2011vc}
A.~Arhrib, R.~Benbrik, M.~Chabab, G.~Moultaka, and L.~Rahili,
\newblock JHEP {\bf 1204}, 136 (2012), 1112.5453\relax
\mciteBstWouldAddEndPuncttrue
\mciteSetBstMidEndSepPunct{\mcitedefaultmidpunct}
{\mcitedefaultendpunct}{\mcitedefaultseppunct}\relax
\EndOfBibitem
\bibitem{Cahn:1978nz}
R.~Cahn, M.~S. Chanowitz, and N.~Fleishon,
\newblock Phys.Lett. {\bf B82}, 113 (1979)\relax
\mciteBstWouldAddEndPuncttrue
\mciteSetBstMidEndSepPunct{\mcitedefaultmidpunct}
{\mcitedefaultendpunct}{\mcitedefaultseppunct}\relax
\EndOfBibitem
\bibitem{Bergstrom:1985hp}
L.~Bergstrom and G.~Hulth,
\newblock Nucl.Phys. {\bf B259}, 137 (1985)\relax
\mciteBstWouldAddEndPuncttrue
\mciteSetBstMidEndSepPunct{\mcitedefaultmidpunct}
{\mcitedefaultendpunct}{\mcitedefaultseppunct}\relax
\EndOfBibitem
\bibitem{Bergstrom:1986err}
Erratum-ibid. {\bf B276}, 744 (1986)\relax
\mciteBstWouldAddEndPuncttrue
\mciteSetBstMidEndSepPunct{\mcitedefaultmidpunct}
{\mcitedefaultendpunct}{\mcitedefaultseppunct}\relax
\EndOfBibitem
\bibitem{Gunion:1989we}
J.~F. Gunion, H.~E. Haber, G.~L. Kane, and S.~Dawson,
\newblock Front.Phys. {\bf 80}, 1 (2000)\relax
\mciteBstWouldAddEndPuncttrue
\mciteSetBstMidEndSepPunct{\mcitedefaultmidpunct}
{\mcitedefaultendpunct}{\mcitedefaultseppunct}\relax
\EndOfBibitem
\bibitem{Spira:1997dg}
M.~Spira,
\newblock Fortsch. Phys. {\bf 46}, 203 (1998), hep-ph/9705337\relax
\mciteBstWouldAddEndPuncttrue
\mciteSetBstMidEndSepPunct{\mcitedefaultmidpunct}
{\mcitedefaultendpunct}{\mcitedefaultseppunct}\relax
\EndOfBibitem
\bibitem{Djouadi:2005gj}
A.~Djouadi,
\newblock Phys. Rept. {\bf 459}, 1 (2008), hep-ph/0503173\relax
\mciteBstWouldAddEndPuncttrue
\mciteSetBstMidEndSepPunct{\mcitedefaultmidpunct}
{\mcitedefaultendpunct}{\mcitedefaultseppunct}\relax
\EndOfBibitem
\bibitem{Chen:2013dh}
C.-S. Chen, C.-Q. Geng, D.~Huang, and L.-H. Tsai,
\newblock (2013), 1302.0502\relax
\mciteBstWouldAddEndPuncttrue
\mciteSetBstMidEndSepPunct{\mcitedefaultmidpunct}
{\mcitedefaultendpunct}{\mcitedefaultseppunct}\relax
\EndOfBibitem
\bibitem{Chen:2013vi}
C.-S. Chen, C.-Q. Geng, D.~Huang, and L.-H. Tsai,
\newblock (2013), 1301.4694\relax
\mciteBstWouldAddEndPuncttrue
\mciteSetBstMidEndSepPunct{\mcitedefaultmidpunct}
{\mcitedefaultendpunct}{\mcitedefaultseppunct}\relax
\EndOfBibitem
\bibitem{Hahn:2000kx}
T.~Hahn,
\newblock Comput. Phys. Commun. {\bf 140}, 418 (2001), hep-ph/0012260\relax
\mciteBstWouldAddEndPuncttrue
\mciteSetBstMidEndSepPunct{\mcitedefaultmidpunct}
{\mcitedefaultendpunct}{\mcitedefaultseppunct}\relax
\EndOfBibitem
\bibitem{Hahn:1998yk}
T.~Hahn and M.~Perez-Victoria,
\newblock Comput. Phys. Commun. {\bf 118}, 153 (1999), hep-ph/9807565\relax
\mciteBstWouldAddEndPuncttrue
\mciteSetBstMidEndSepPunct{\mcitedefaultmidpunct}
{\mcitedefaultendpunct}{\mcitedefaultseppunct}\relax
\EndOfBibitem
\bibitem{ATLAS:2012hi}
ATLAS Collaboration, G.~Aad {\em et~al.},
\newblock Eur.Phys.J. {\bf C72}, 2244 (2012), 1210.5070\relax
\mciteBstWouldAddEndPuncttrue
\mciteSetBstMidEndSepPunct{\mcitedefaultmidpunct}
{\mcitedefaultendpunct}{\mcitedefaultseppunct}\relax
\EndOfBibitem
\bibitem{Chatrchyan:2012ya}
CMS Collaboration, S.~Chatrchyan {\em et~al.},
\newblock Eur.Phys.J. {\bf C72}, 2189 (2012), 1207.2666\relax
\mciteBstWouldAddEndPuncttrue
\mciteSetBstMidEndSepPunct{\mcitedefaultmidpunct}
{\mcitedefaultendpunct}{\mcitedefaultseppunct}\relax
\EndOfBibitem
\bibitem{ATLAS:2013rma}
ATLAS-CONF-2013-009,
\newblock (2013)\relax
\mciteBstWouldAddEndPuncttrue
\mciteSetBstMidEndSepPunct{\mcitedefaultmidpunct}
{\mcitedefaultendpunct}{\mcitedefaultseppunct}\relax
\EndOfBibitem
\bibitem{Aad:2014fia}
ATLAS Collaboration, G.~Aad {\em et~al.},
\newblock Phys.Lett. {\bf B732}, 8 (2014), 1402.3051\relax
\mciteBstWouldAddEndPuncttrue
\mciteSetBstMidEndSepPunct{\mcitedefaultmidpunct}
{\mcitedefaultendpunct}{\mcitedefaultseppunct}\relax
\EndOfBibitem
\bibitem{Chatrchyan:2013vaa}
CMS Collaboration, S.~Chatrchyan {\em et~al.},
\newblock (2013), 1307.5515\relax
\mciteBstWouldAddEndPuncttrue
\mciteSetBstMidEndSepPunct{\mcitedefaultmidpunct}
{\mcitedefaultendpunct}{\mcitedefaultseppunct}\relax
\EndOfBibitem
\bibitem{Melfo:2011nx}
A.~Melfo, M.~Nemevsek, F.~Nesti, G.~Senjanovic, and Y.~Zhang,
\newblock (2011), 1108.4416\relax
\mciteBstWouldAddEndPuncttrue
\mciteSetBstMidEndSepPunct{\mcitedefaultmidpunct}
{\mcitedefaultendpunct}{\mcitedefaultseppunct}\relax
\EndOfBibitem
\bibitem{Chatrchyan:2013lba}
CMS Collaboration, S.~Chatrchyan {\em et~al.},
\newblock JHEP {\bf 06}, 081 (2013), 1303.4571\relax
\mciteBstWouldAddEndPuncttrue
\mciteSetBstMidEndSepPunct{\mcitedefaultmidpunct}
{\mcitedefaultendpunct}{\mcitedefaultseppunct}\relax
\EndOfBibitem
\bibitem{ATLAS:2012soa}
ATLAS-CONF-2012-079,
\newblock (2012)\relax
\mciteBstWouldAddEndPuncttrue
\mciteSetBstMidEndSepPunct{\mcitedefaultmidpunct}
{\mcitedefaultendpunct}{\mcitedefaultseppunct}\relax
\EndOfBibitem
\bibitem{Aad:2014iia}
ATLAS Collaboration, G.~Aad {\em et~al.},
\newblock Phys.Rev.Lett. {\bf 112}, 201802 (2014), 1402.3244\relax
\mciteBstWouldAddEndPuncttrue
\mciteSetBstMidEndSepPunct{\mcitedefaultmidpunct}
{\mcitedefaultendpunct}{\mcitedefaultseppunct}\relax
\EndOfBibitem
\bibitem{CMS:2013yda}
CMS Collaboration, CMS-PAS-HIG-13-018,
\newblock (2013)\relax
\mciteBstWouldAddEndPuncttrue
\mciteSetBstMidEndSepPunct{\mcitedefaultmidpunct}
{\mcitedefaultendpunct}{\mcitedefaultseppunct}\relax
\EndOfBibitem
\bibitem{CMS:HIG-13-013}
CMS Collaboration, CMS-PAS-HIG-13-013,
\newblock (2013)\relax
\mciteBstWouldAddEndPuncttrue
\mciteSetBstMidEndSepPunct{\mcitedefaultmidpunct}
{\mcitedefaultendpunct}{\mcitedefaultseppunct}\relax
\EndOfBibitem
\bibitem{Chatrchyan:2014tja}
CMS Collaboration, S.~Chatrchyan {\em et~al.},
\newblock Eur.Phys.J. {\bf C74}, 2980 (2014), 1404.1344\relax
\mciteBstWouldAddEndPuncttrue
\mciteSetBstMidEndSepPunct{\mcitedefaultmidpunct}
{\mcitedefaultendpunct}{\mcitedefaultseppunct}\relax
\EndOfBibitem
\bibitem{Englert:2011aa}
C.~Englert, T.~Plehn, M.~Rauch, D.~Zerwas, and P.~M. Zerwas,
\newblock Phys.Lett. {\bf B707}, 512 (2012), 1112.3007\relax
\mciteBstWouldAddEndPuncttrue
\mciteSetBstMidEndSepPunct{\mcitedefaultmidpunct}
{\mcitedefaultendpunct}{\mcitedefaultseppunct}\relax
\EndOfBibitem
\bibitem{Cheung:2013kla}
K.~Cheung, J.~S. Lee, and P.-Y. Tseng,
\newblock JHEP {\bf 1305}, 134 (2013), 1302.3794\relax
\mciteBstWouldAddEndPuncttrue
\mciteSetBstMidEndSepPunct{\mcitedefaultmidpunct}
{\mcitedefaultendpunct}{\mcitedefaultseppunct}\relax
\EndOfBibitem
\bibitem{Belanger:2013kya}
G.~Belanger, B.~Dumont, U.~Ellwanger, J.~Gunion, and S.~Kraml,
\newblock Phys.Lett. {\bf B723}, 340 (2013), 1302.5694\relax
\mciteBstWouldAddEndPuncttrue
\mciteSetBstMidEndSepPunct{\mcitedefaultmidpunct}
{\mcitedefaultendpunct}{\mcitedefaultseppunct}\relax
\EndOfBibitem
\bibitem{Belanger:2013xza}
G.~Belanger, B.~Dumont, U.~Ellwanger, J.~Gunion, and S.~Kraml,
\newblock Phys.Rev. {\bf D88}, 075008 (2013), 1306.2941\relax
\mciteBstWouldAddEndPuncttrue
\mciteSetBstMidEndSepPunct{\mcitedefaultmidpunct}
{\mcitedefaultendpunct}{\mcitedefaultseppunct}\relax
\EndOfBibitem
\bibitem{Espinosa:2012vu}
J.~R. Espinosa, M.~Muhlleitner, C.~Grojean, and M.~Trott,
\newblock JHEP {\bf 1209}, 126 (2012), 1205.6790\relax
\mciteBstWouldAddEndPuncttrue
\mciteSetBstMidEndSepPunct{\mcitedefaultmidpunct}
{\mcitedefaultendpunct}{\mcitedefaultseppunct}\relax
\EndOfBibitem
\bibitem{Englert:2012wf}
C.~Englert, M.~Spannowsky, and C.~Wymant,
\newblock Phys.Lett. {\bf B718}, 538 (2012), 1209.0494\relax
\mciteBstWouldAddEndPuncttrue
\mciteSetBstMidEndSepPunct{\mcitedefaultmidpunct}
{\mcitedefaultendpunct}{\mcitedefaultseppunct}\relax
\EndOfBibitem
\bibitem{Ablikim:2011es}
BESIII Collaboration, M.~Ablikim {\em et~al.},
\newblock Phys.Rev. {\bf D85}, 092012 (2012), 1111.2112\relax
\mciteBstWouldAddEndPuncttrue
\mciteSetBstMidEndSepPunct{\mcitedefaultmidpunct}
{\mcitedefaultendpunct}{\mcitedefaultseppunct}\relax
\EndOfBibitem
\bibitem{Lees:2012iw}
BaBar collaboration, J.~Lees {\em et~al.},
\newblock Phys.Rev. {\bf D87}, 031102 (2013), 1210.0287\relax
\mciteBstWouldAddEndPuncttrue
\mciteSetBstMidEndSepPunct{\mcitedefaultmidpunct}
{\mcitedefaultendpunct}{\mcitedefaultseppunct}\relax
\EndOfBibitem
\bibitem{Lees:2013vuj}
BaBar Collaboration, J.~Lees {\em et~al.},
\newblock Phys.Rev. {\bf D88}, 031701 (2013), 1307.5306\relax
\mciteBstWouldAddEndPuncttrue
\mciteSetBstMidEndSepPunct{\mcitedefaultmidpunct}
{\mcitedefaultendpunct}{\mcitedefaultseppunct}\relax
\EndOfBibitem
\bibitem{CMS:2013aga}
CMS-PAS-HIG-13-007,
\newblock (2013)\relax
\mciteBstWouldAddEndPuncttrue
\mciteSetBstMidEndSepPunct{\mcitedefaultmidpunct}
{\mcitedefaultendpunct}{\mcitedefaultseppunct}\relax
\EndOfBibitem
\bibitem{ATLAS:2013qma}
ATLAS-CONF-2013-010, ATLAS-COM-CONF-2013-003,
\newblock (2013)\relax
\mciteBstWouldAddEndPuncttrue
\mciteSetBstMidEndSepPunct{\mcitedefaultmidpunct}
{\mcitedefaultendpunct}{\mcitedefaultseppunct}\relax
\EndOfBibitem
\bibitem{Abbiendi:2000ug}
OPAL Collaboration, G.~Abbiendi {\em et~al.},
\newblock Eur.Phys.J. {\bf C18}, 425 (2001), hep-ex/0007040\relax
\mciteBstWouldAddEndPuncttrue
\mciteSetBstMidEndSepPunct{\mcitedefaultmidpunct}
{\mcitedefaultendpunct}{\mcitedefaultseppunct}\relax
\EndOfBibitem
\bibitem{Abdallah:2004wy}
DELPHI, J.~Abdallah {\em et~al.},
\newblock Eur. Phys. J. {\bf C38}, 1 (2004), hep-ex/0410017\relax
\mciteBstWouldAddEndPuncttrue
\mciteSetBstMidEndSepPunct{\mcitedefaultmidpunct}
{\mcitedefaultendpunct}{\mcitedefaultseppunct}\relax
\EndOfBibitem
\bibitem{Abbiendi:2004gn}
OPAL Collaboration, G.~Abbiendi {\em et~al.},
\newblock Eur.Phys.J. {\bf C40}, 317 (2005), hep-ex/0408097\relax
\mciteBstWouldAddEndPuncttrue
\mciteSetBstMidEndSepPunct{\mcitedefaultmidpunct}
{\mcitedefaultendpunct}{\mcitedefaultseppunct}\relax
\EndOfBibitem
\bibitem{LEPHiggsWorking:2001ab}
LEP Higgs Working Group, ALEPH collaboration, DELPHI collaboration, L3
  collaboration, OPAL Collaboration,
\newblock (2001), hep-ex/0107030\relax
\mciteBstWouldAddEndPuncttrue
\mciteSetBstMidEndSepPunct{\mcitedefaultmidpunct}
{\mcitedefaultendpunct}{\mcitedefaultseppunct}\relax
\EndOfBibitem
\bibitem{Beringer:1900zz}
Particle Data Group, J.~Beringer {\em et~al.},
\newblock Phys.Rev. {\bf D86}, 010001 (2012)\relax
\mciteBstWouldAddEndPuncttrue
\mciteSetBstMidEndSepPunct{\mcitedefaultmidpunct}
{\mcitedefaultendpunct}{\mcitedefaultseppunct}\relax
\EndOfBibitem
\bibitem{Ackerstaff:1998ms}
OPAL Collaboration, K.~Ackerstaff {\em et~al.},
\newblock Eur.Phys.J. {\bf C5}, 19 (1998), hep-ex/9803019\relax
\mciteBstWouldAddEndPuncttrue
\mciteSetBstMidEndSepPunct{\mcitedefaultmidpunct}
{\mcitedefaultendpunct}{\mcitedefaultseppunct}\relax
\EndOfBibitem
\bibitem{Ackerstaff:1997cza}
OPAL Collaboration, K.~Ackerstaff {\em et~al.},
\newblock Eur.Phys.J. {\bf C1}, 425 (1998), hep-ex/9709003\relax
\mciteBstWouldAddEndPuncttrue
\mciteSetBstMidEndSepPunct{\mcitedefaultmidpunct}
{\mcitedefaultendpunct}{\mcitedefaultseppunct}\relax
\EndOfBibitem
\bibitem{Kanemura:2013vxa}
S.~Kanemura, K.~Yagyu, and H.~Yokoya,
\newblock Phys.Lett. {\bf B726}, 316 (2013), 1305.2383\relax
\mciteBstWouldAddEndPuncttrue
\mciteSetBstMidEndSepPunct{\mcitedefaultmidpunct}
{\mcitedefaultendpunct}{\mcitedefaultseppunct}\relax
\EndOfBibitem
\bibitem{kang:2014jia}
Z.~Kang, J.~Li, T.~Li, Y.~Liu, and G.-Z. Ning,
\newblock (2014), 1404.5207\relax
\mciteBstWouldAddEndPuncttrue
\mciteSetBstMidEndSepPunct{\mcitedefaultmidpunct}
{\mcitedefaultendpunct}{\mcitedefaultseppunct}\relax
\EndOfBibitem
\bibitem{Kanemura:2014goa}
S.~Kanemura, M.~Kikuchi, K.~Yagyu, and H.~Yokoya,
\newblock (2014), 1407.6547\relax
\mciteBstWouldAddEndPuncttrue
\mciteSetBstMidEndSepPunct{\mcitedefaultmidpunct}
{\mcitedefaultendpunct}{\mcitedefaultseppunct}\relax
\EndOfBibitem
\bibitem{Chetyrkin:2000yt}
K.~Chetyrkin, J.~H. Kuhn, and M.~Steinhauser,
\newblock Comput.Phys.Commun. {\bf 133}, 43 (2000), hep-ph/0004189\relax
\mciteBstWouldAddEndPuncttrue
\mciteSetBstMidEndSepPunct{\mcitedefaultmidpunct}
{\mcitedefaultendpunct}{\mcitedefaultseppunct}\relax
\EndOfBibitem
\bibitem{Chetyrkin:1995pd}
K.~Chetyrkin and A.~Kwiatkowski,
\newblock Nucl.Phys. {\bf B461}, 3 (1996), hep-ph/9505358\relax
\mciteBstWouldAddEndPuncttrue
\mciteSetBstMidEndSepPunct{\mcitedefaultmidpunct}
{\mcitedefaultendpunct}{\mcitedefaultseppunct}\relax
\EndOfBibitem
\bibitem{Chetyrkin:1997iv}
K.~Chetyrkin, B.~A. Kniehl, and M.~Steinhauser,
\newblock Phys.Rev.Lett. {\bf 79}, 353 (1997), hep-ph/9705240\relax
\mciteBstWouldAddEndPuncttrue
\mciteSetBstMidEndSepPunct{\mcitedefaultmidpunct}
{\mcitedefaultendpunct}{\mcitedefaultseppunct}\relax
\EndOfBibitem
\end{mcitethebibliography}
\bibliographystyle{h-physrev4}
\newpage

\begin{figure}[!h]
\begin{tabular}{rr}
\hspace{-.8cm}\resizebox{86mm}{!}{\includegraphics{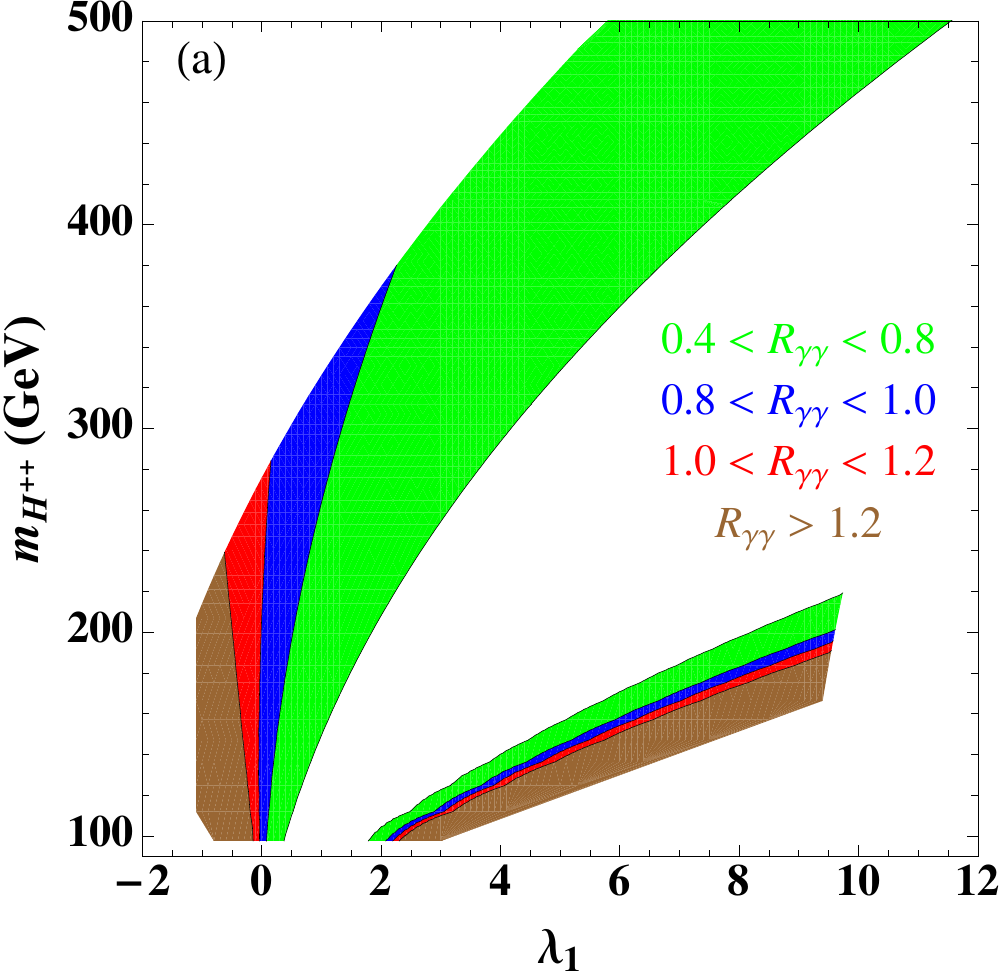}}&
\hspace{.1cm}\resizebox{86mm}{!}{\includegraphics{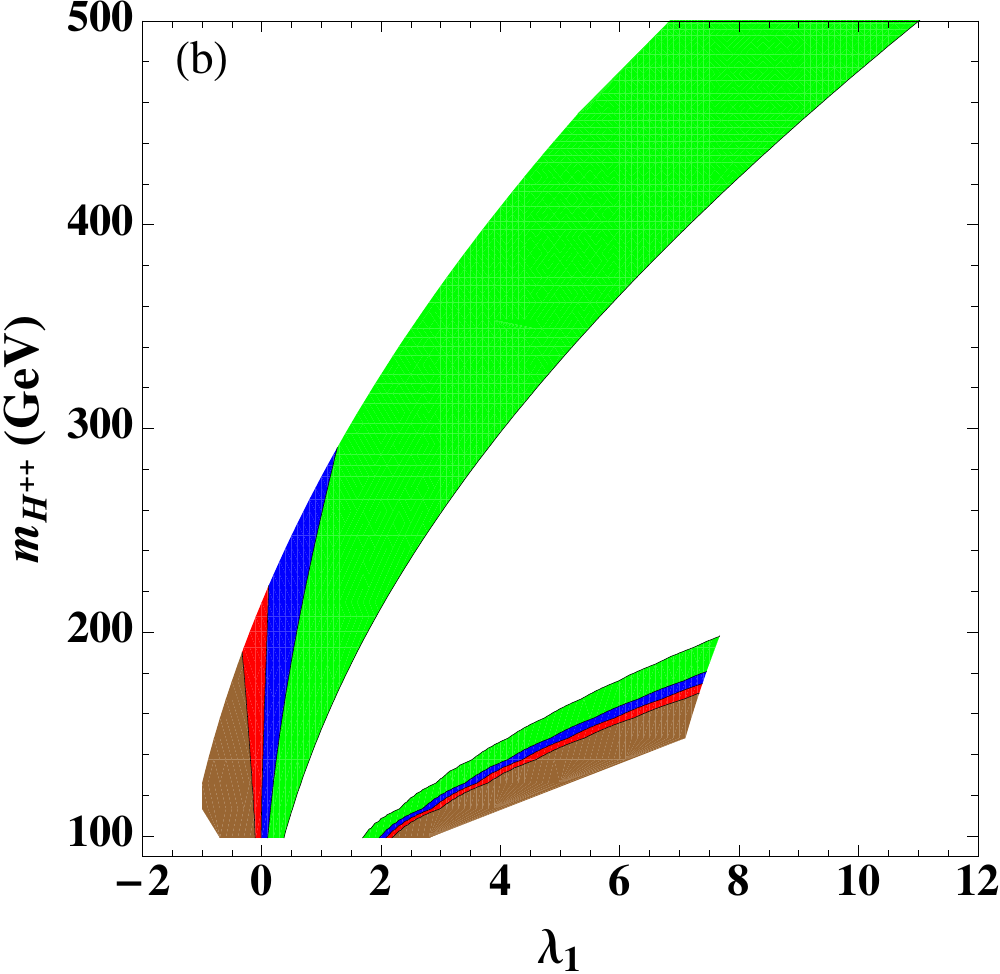}}\\
\hspace{-.8cm}\resizebox{86mm}{!}{\includegraphics{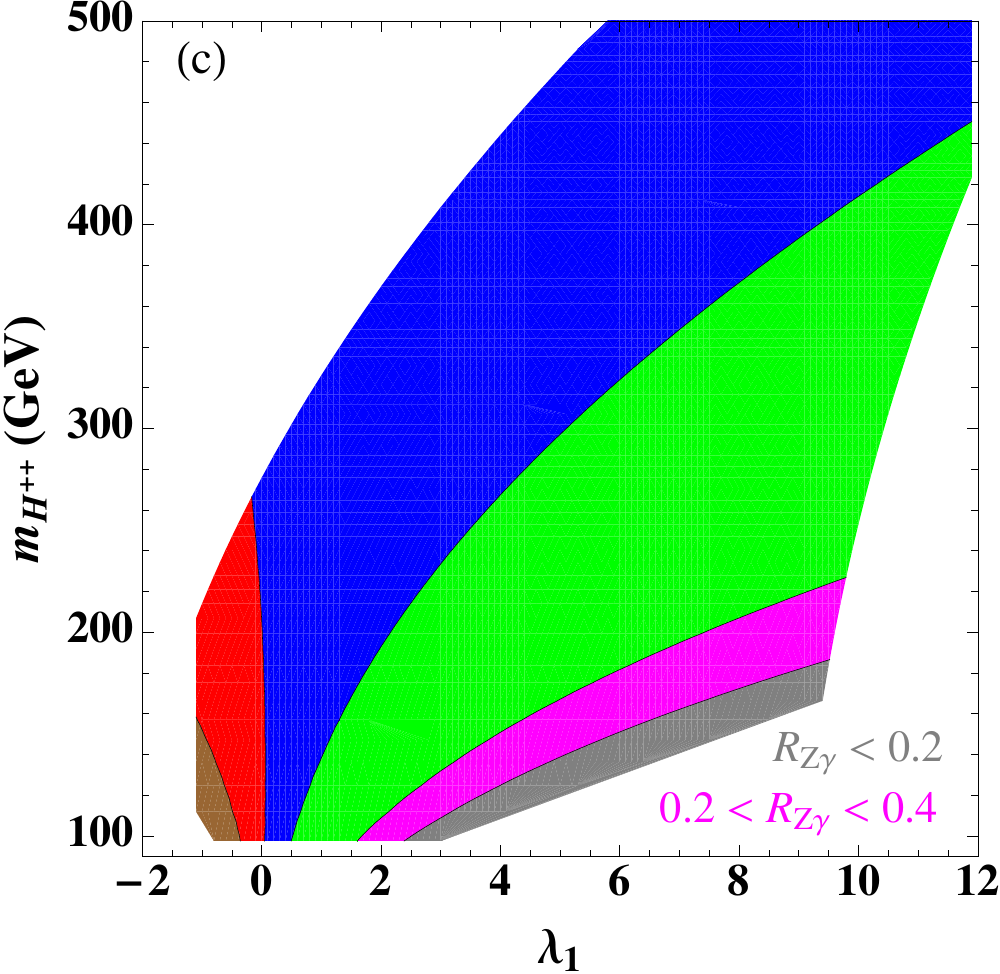}}&
\hspace{.1cm}\resizebox{86mm}{!}{\includegraphics{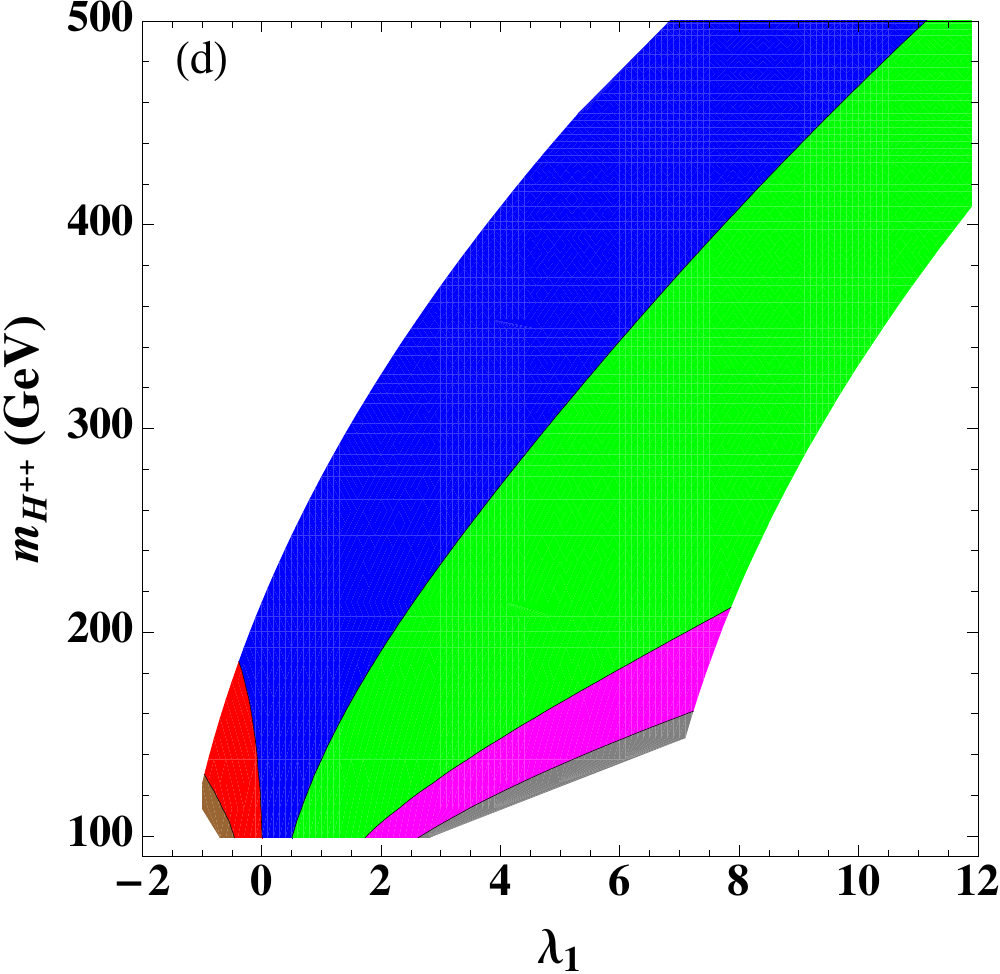}}
\end{tabular}
\caption{
Scatter plots in the $[\lambda_1,m_{H^{++}}]$ showing the ratios $R_{\gamma\gamma}$ (upper) and $R_{Z\gamma}$ (lower).
(a), (c) correspond to the `$h^0${\sl scenario}'
and (b), (d) to the `$H^0${\sl scenario}'. The color code is the
same for the four figures. The scan is in the range
 $-12 \le \lambda_4 \le 2$ and all other parameter and mass
 values are as in  Fig.\ref{fig:BRgaga_BRgaZ}}
\label{fig:Rgaga_RgaZ_scatter}
\end{figure}

\begin{figure}[!h]
\begin{tabular}{rr}
\hspace{-.8cm}\resizebox{86mm}{!}{\includegraphics{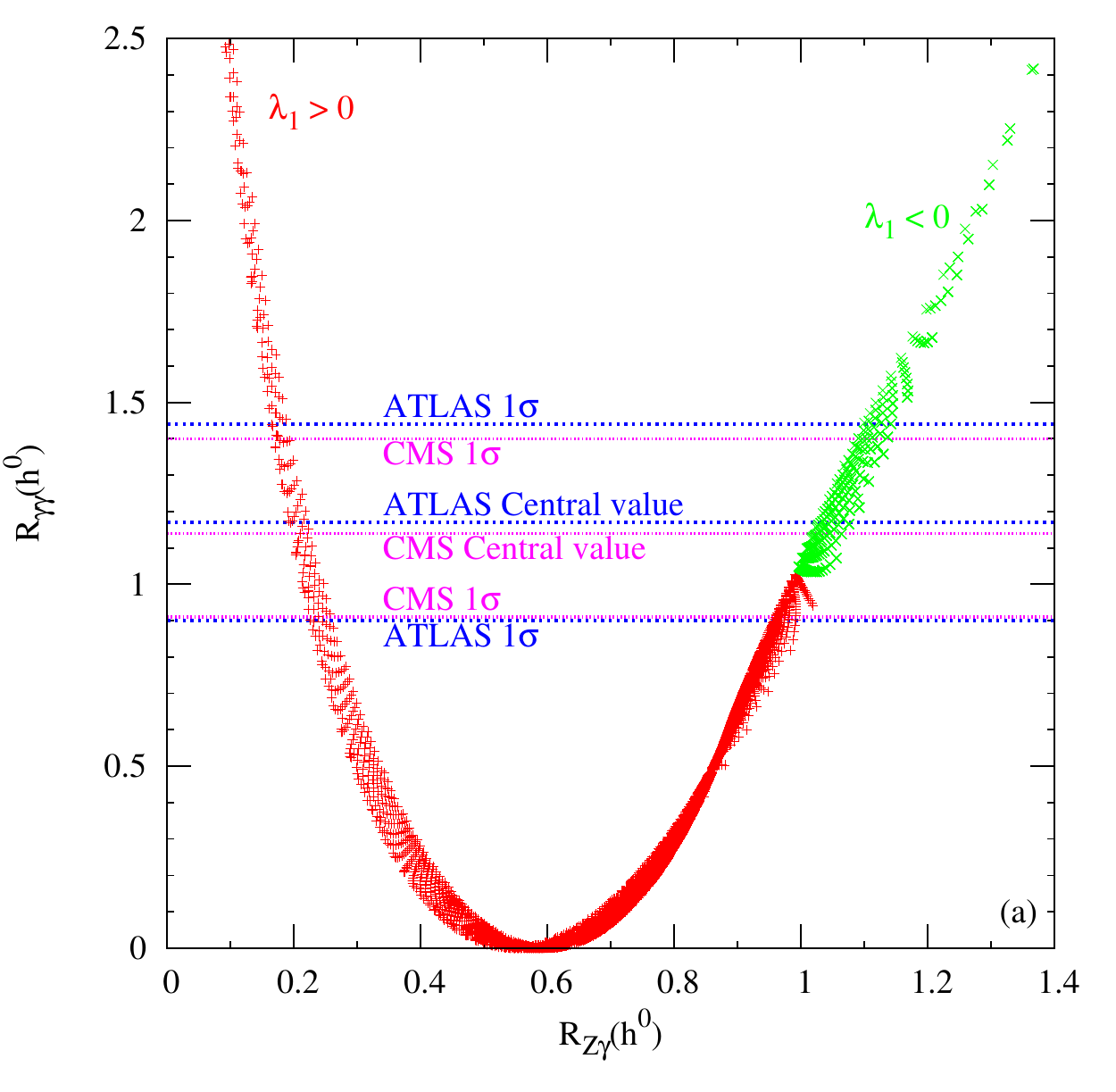}}&
\hspace{.1cm}\resizebox{86mm}{!}{\includegraphics{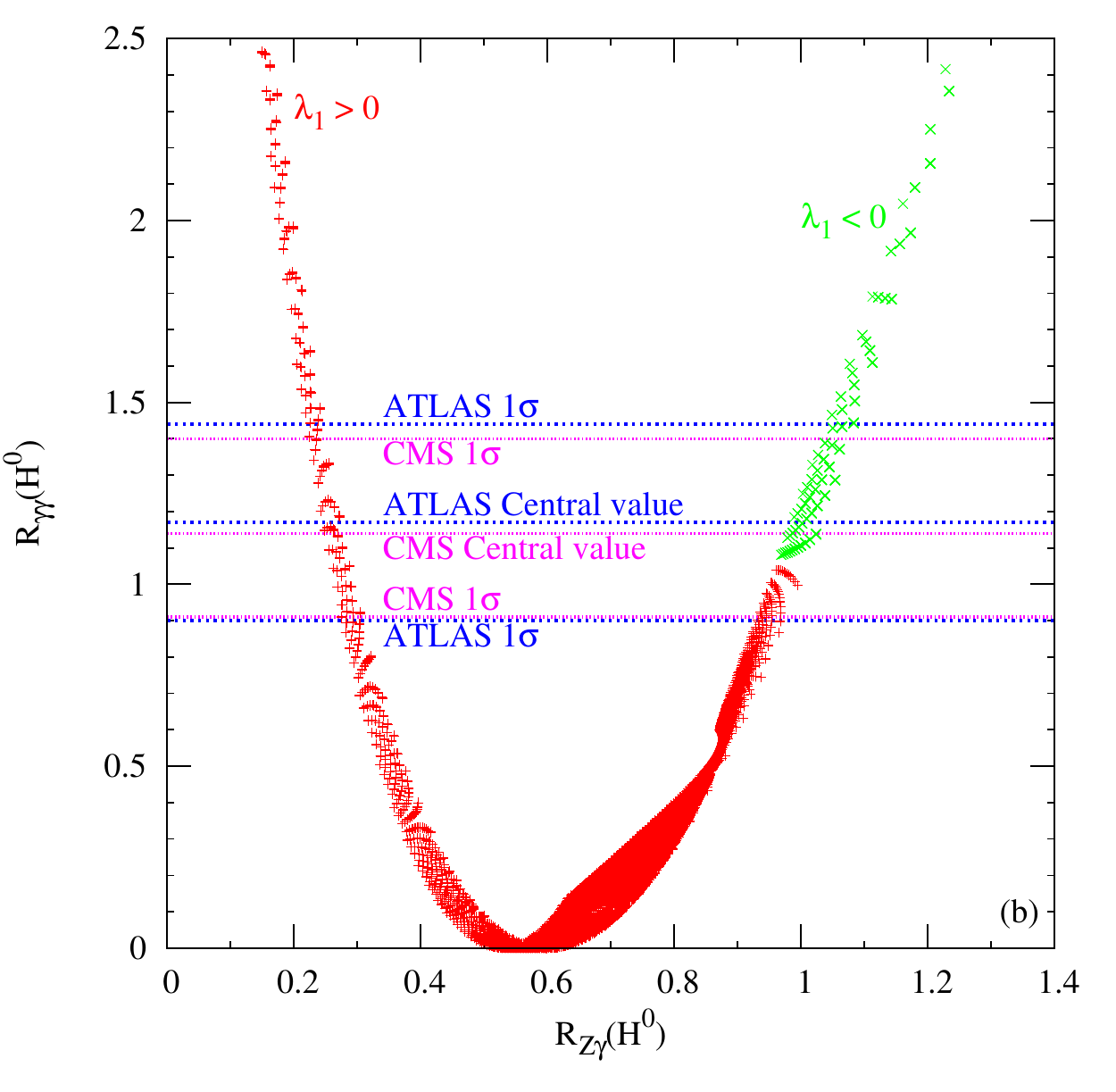}}
\end{tabular}
\caption{Correlation between $R_{Z\gamma}$ and $R_{\gamma\gamma}$ observables, (a) `$h^0${\sl scenario}', (b) `$H^0${\sl scenario}';
the parameter scan and mass ranges are as in 
Figs.\ref{fig:BRgaga_BRgaZ}, \ref{fig:Rgaga_RgaZ_scatter};
the scatter points  
 correspond to $\lambda_1<0$ (green) and $\lambda_1 >0$ (red).
 We also show the central values and $1\sigma$ bands of
 the recent ATLAS~\cite{Aad:2014aba} and CMS~\cite{Khachatryan:2014ira} results. See text for further discussion.}

\label{fig:gg_gZ_correlation}
\end{figure}


\begin{figure}[!h]
\begin{tabular}{rr}
\hspace{-.8cm}\resizebox{86mm}{!}{\includegraphics{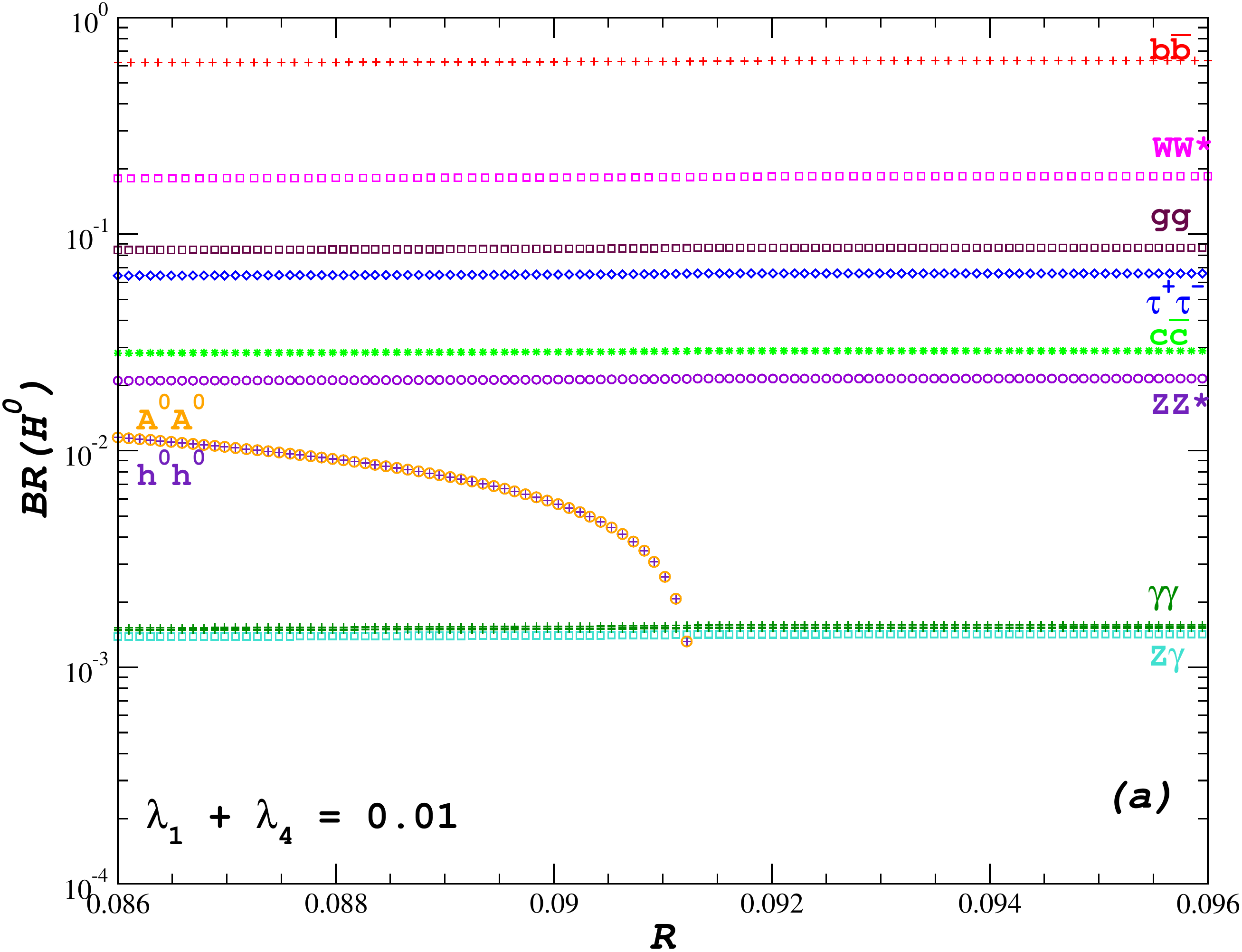}}&
\hspace{.1cm}\resizebox{86mm}{!}{\includegraphics{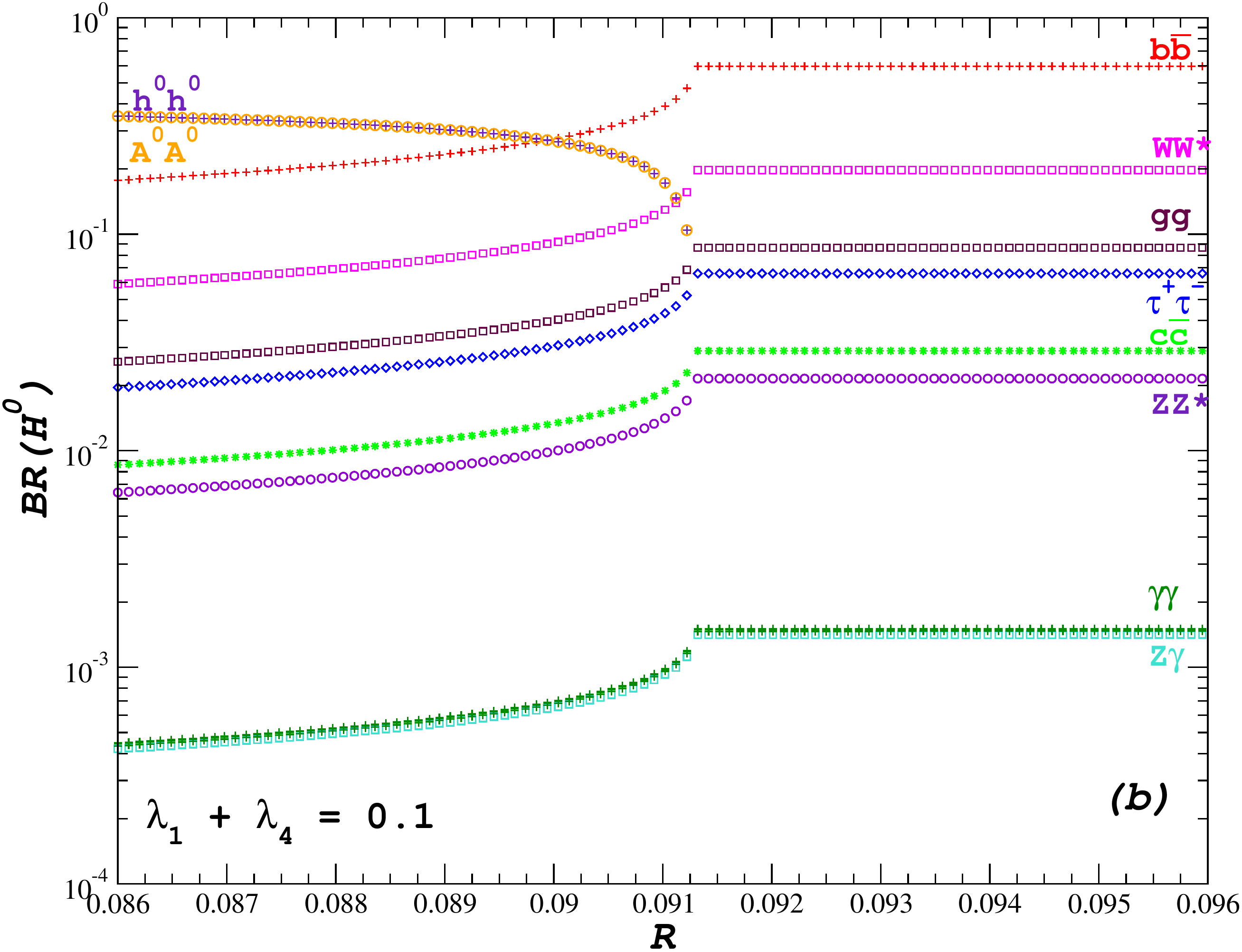}}\\\\
\hspace{-.8cm}\resizebox{86mm}{!}{\includegraphics{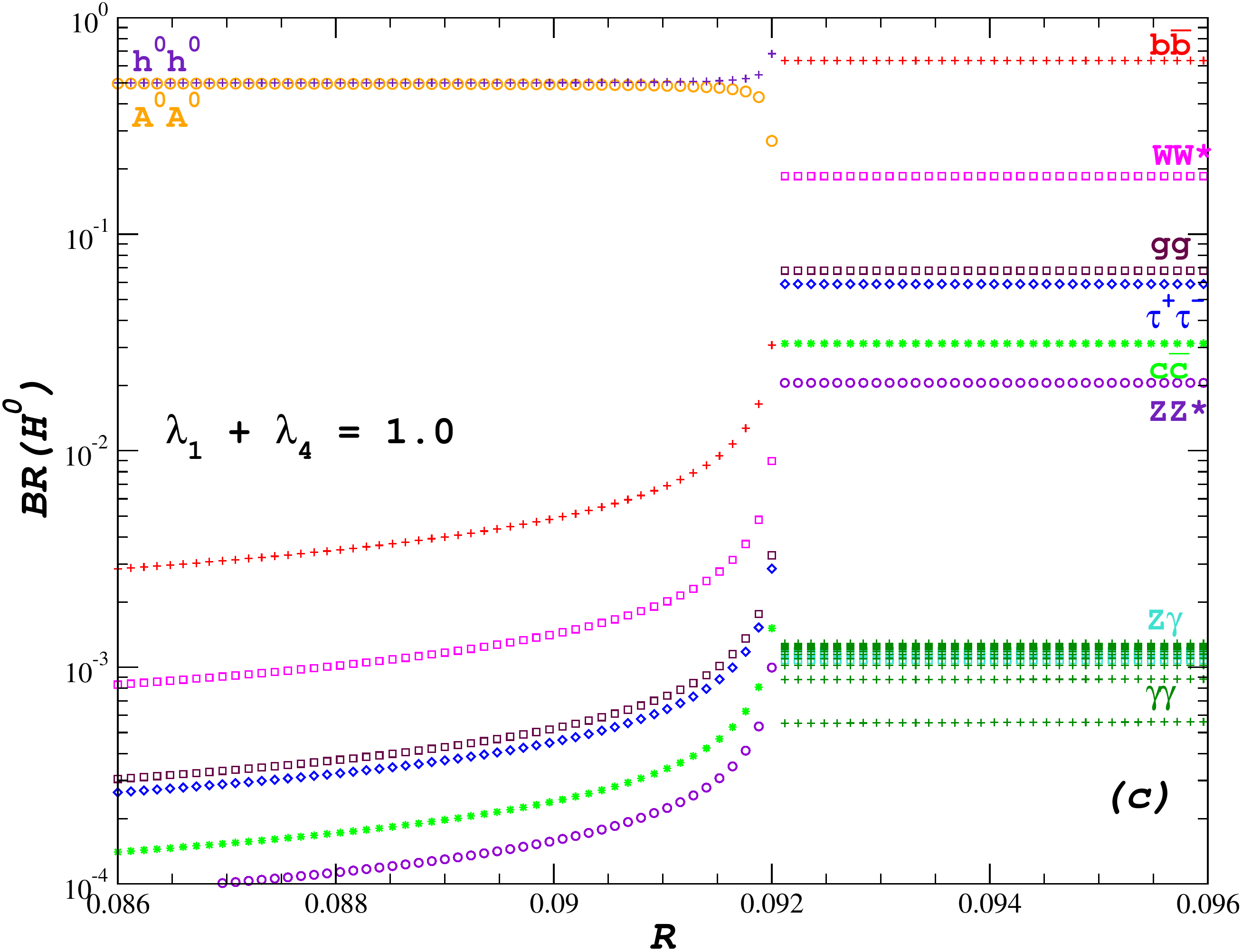}}&
\hspace{.1cm}\resizebox{86mm}{!}{\includegraphics{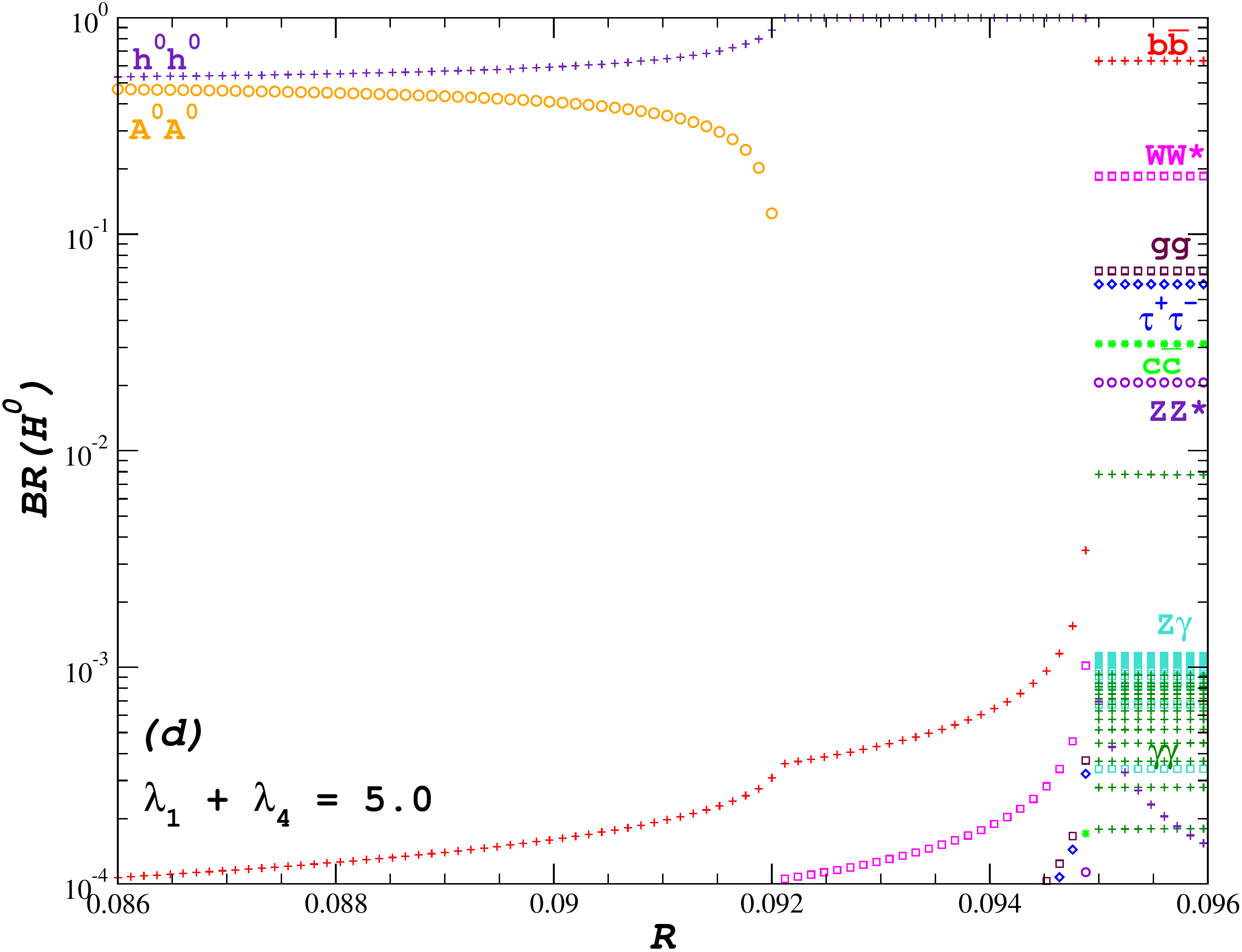}}
\end{tabular}
\caption{The branching ratios $BR(H^0)$ as a function of 
the ratio $R\equiv \frac{\mu}{v_t}$, for $m_{H^0} = 125.5$ GeV,  $\lambda_2 = 0.1$, $\lambda_3=2\lambda_2$, $-10 \le\lambda_4\le2$ and various values of
 $\lambda_1+\lambda_4$.}
\label{fig:BRH}
\end{figure}

\begin{figure}[!h]
\hspace{-.8cm}\resizebox{86mm}{!}{\includegraphics{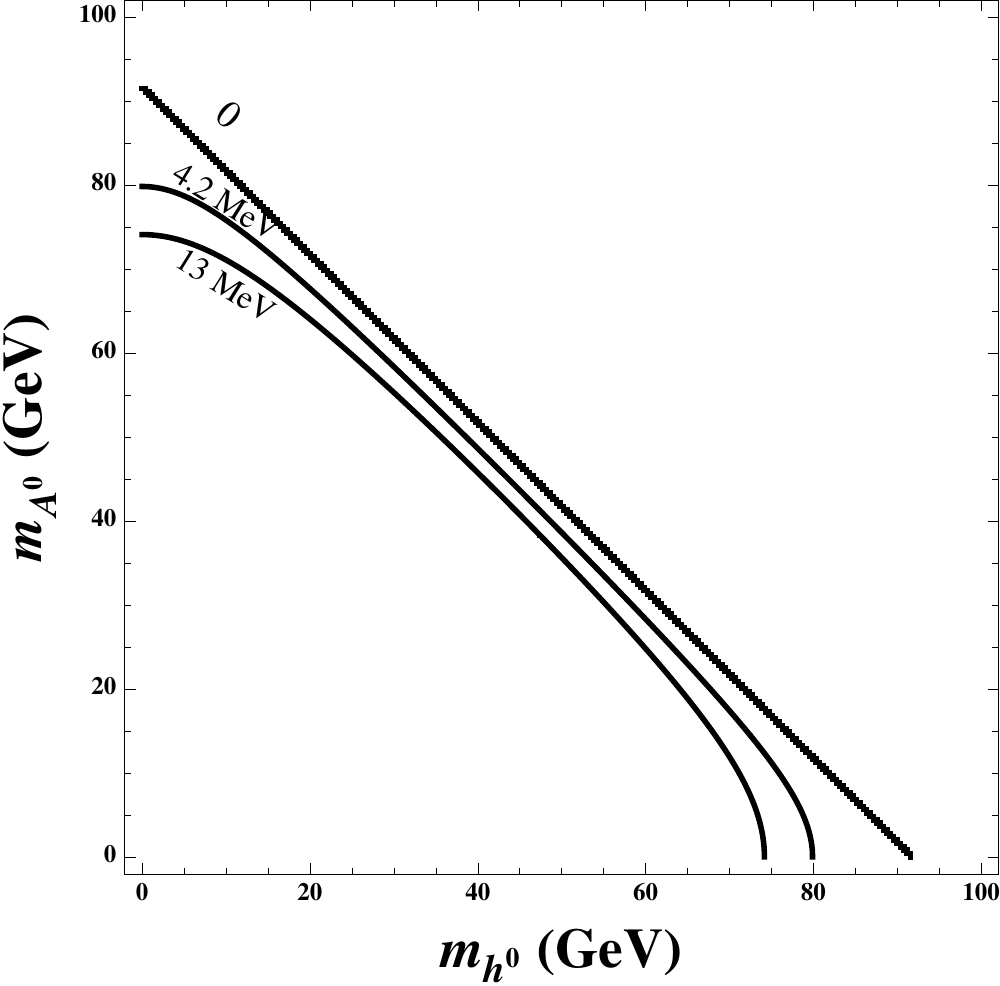}}
\caption{constraint from the $Z\to h^0 A^0$ contribution to the  $Z$-boson total width.}
\label{fig:ZAhwidth}
\end{figure}

\begin{figure}[!h]
\hspace{-.8cm}\resizebox{86mm}{!}{\includegraphics{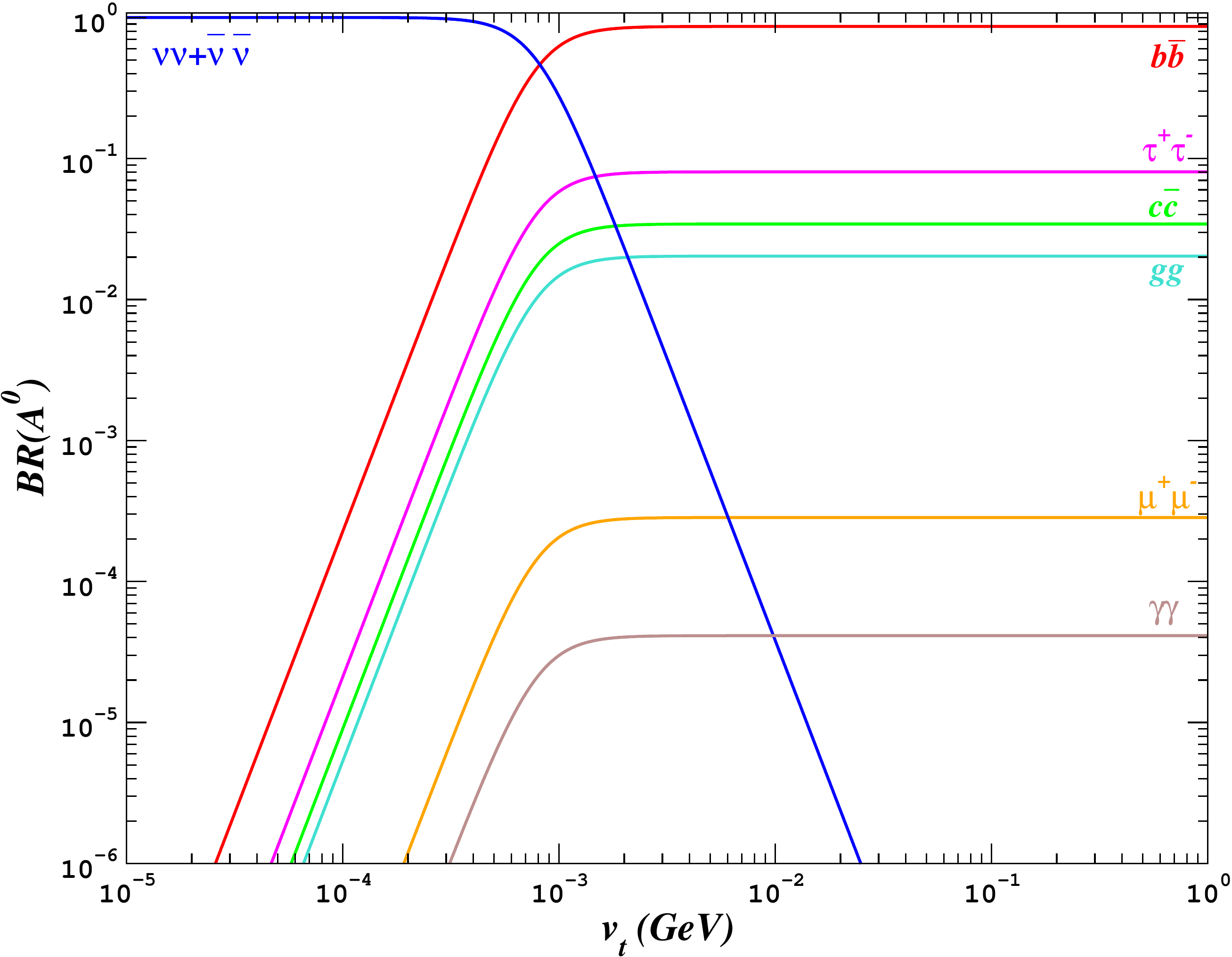}}
\caption{Branching ratios of $A^0$ as a function of $v_t$ for
$m_{A^0} \approx 55$GeV. }
\label{fig:BrA0}
\end{figure}
\begin{figure}[!h]
\begin{tabular}{rr}
\hspace{-.8cm}\resizebox{84mm}{!}{\includegraphics{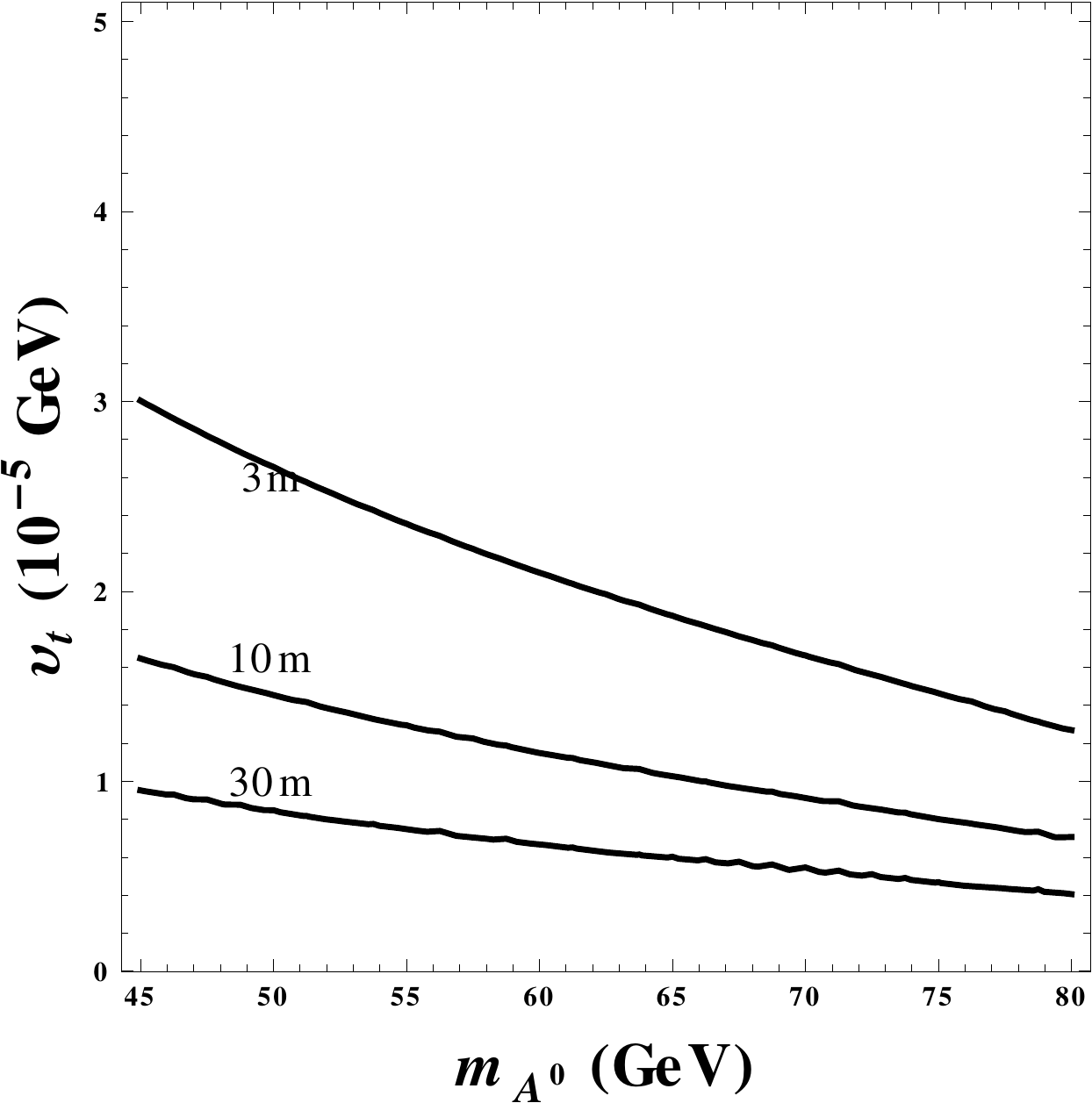}}

\resizebox{84mm}{!}{\includegraphics{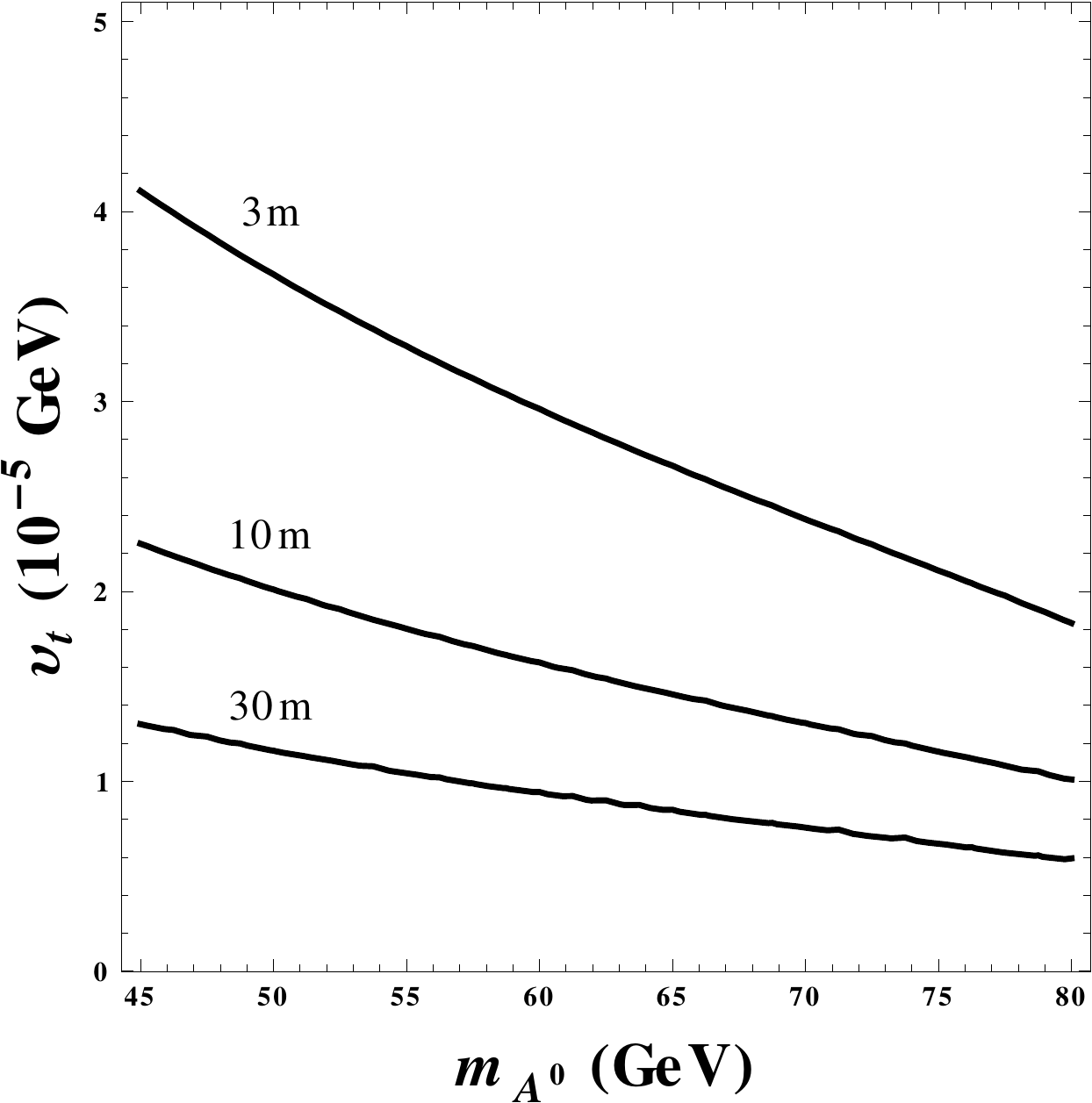}}
\end{tabular}
\caption{$A^0$ decay length $c\tau$-contours in meters in the $(m_{A^0}, v_t)$ plane: 
 tree-level (left), including QCD corrections (right), assuming  $A^0$ 
 produced through $e^+ e^- \to h^0 A^0$ at the LEP2 C.M. energy $\sqrt{s}=183$GeV 
 and a visible decay mainly into $b \bar{b}$.}
\label{fig:distance}
\end{figure}

\begin{figure}[!h]
\hspace{-.8cm}\resizebox{86mm}{!}{\includegraphics{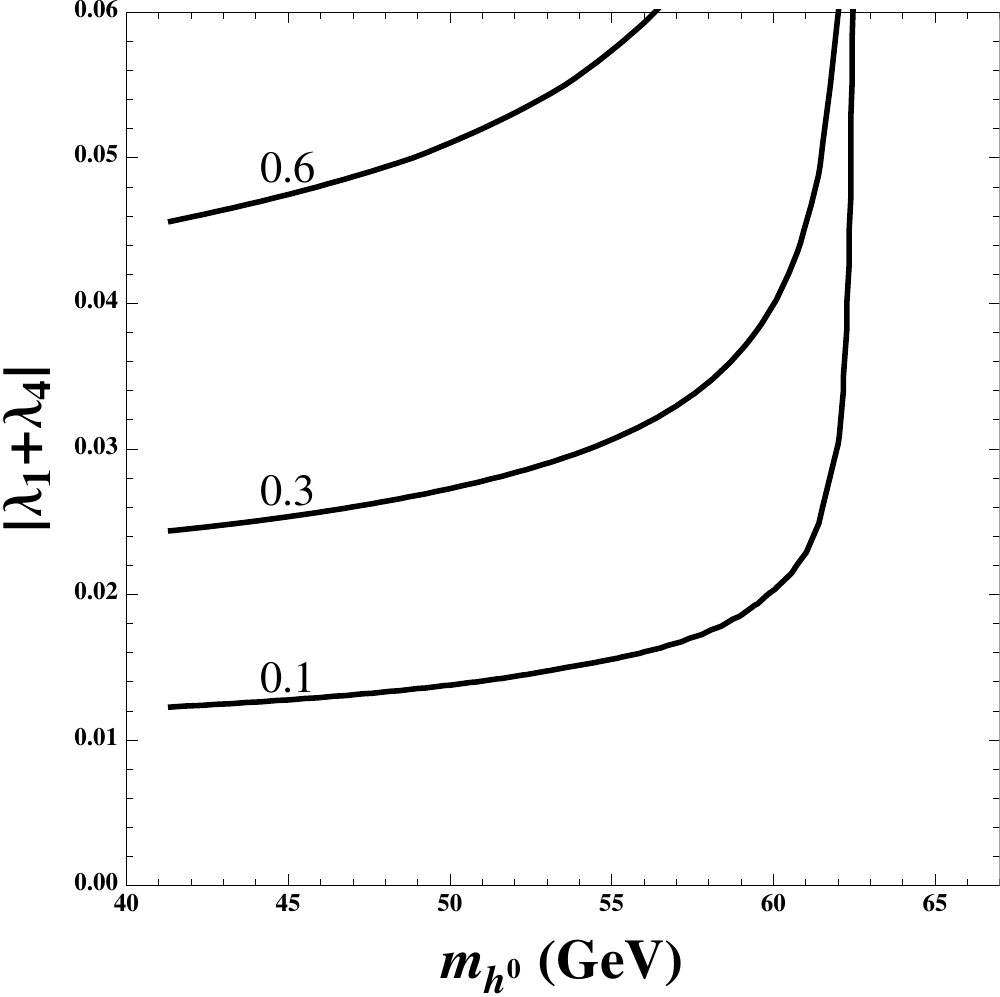}}
\caption{Branching ratio contours for the invisible/undetected
 ${\cal H} \to h^0 h^0 + A^0 A^0$ decays in the $m_{h^0}$ versus
  $|\lambda_1+\lambda_4|$ plane, with 
$g_{{\cal H} h h}=g_{{\cal H} AA}=(\lambda_1+\lambda_4) v_d$, $v_d=246$ GeV, in the limit
$\sin \alpha=1$,   $ \sin \beta=0$, taking
$\Gamma_{{\cal H}}^{visible}=4$ MeV and 
$m_{H^0}=125$GeV.}
\label{fig:invisible/undetected}
\end{figure}
\begin{figure}[h]
\hspace{-1.cm}\resizebox{90mm}{!}{\includegraphics{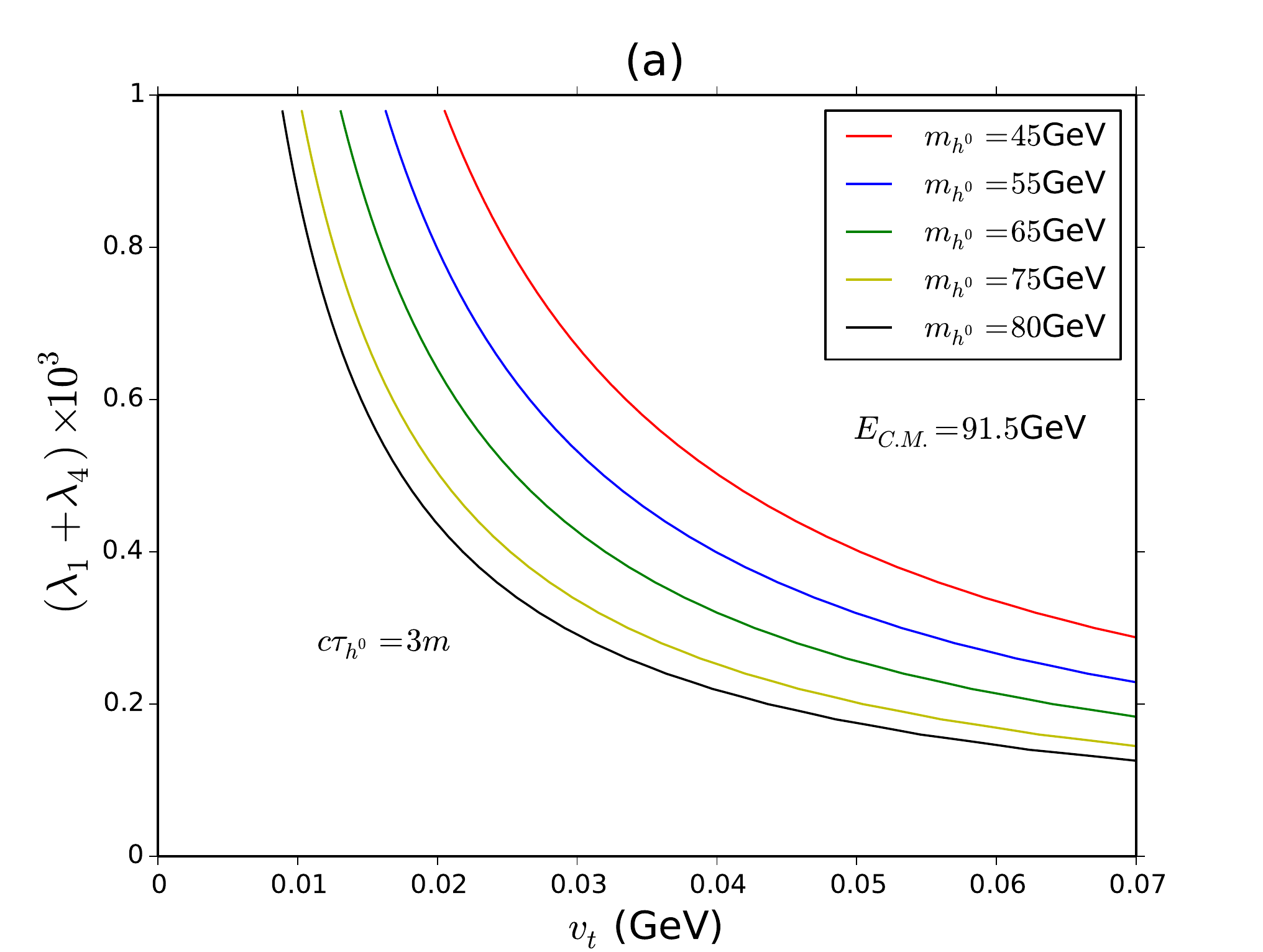}}

\hspace{-1.cm}\resizebox{90mm}{!}{\includegraphics{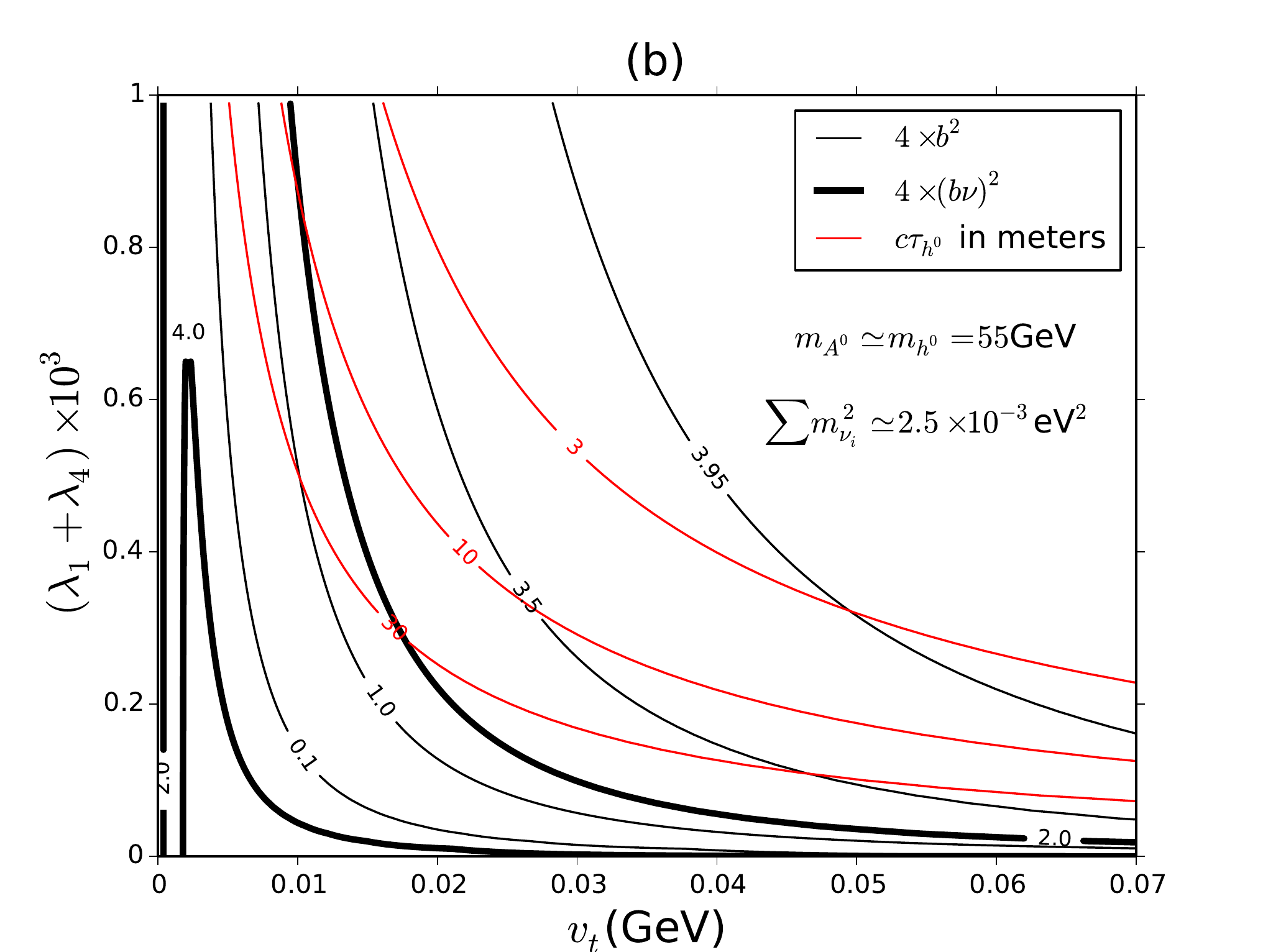}} 
\hspace{-.8cm}\resizebox{90mm}{!}{\includegraphics{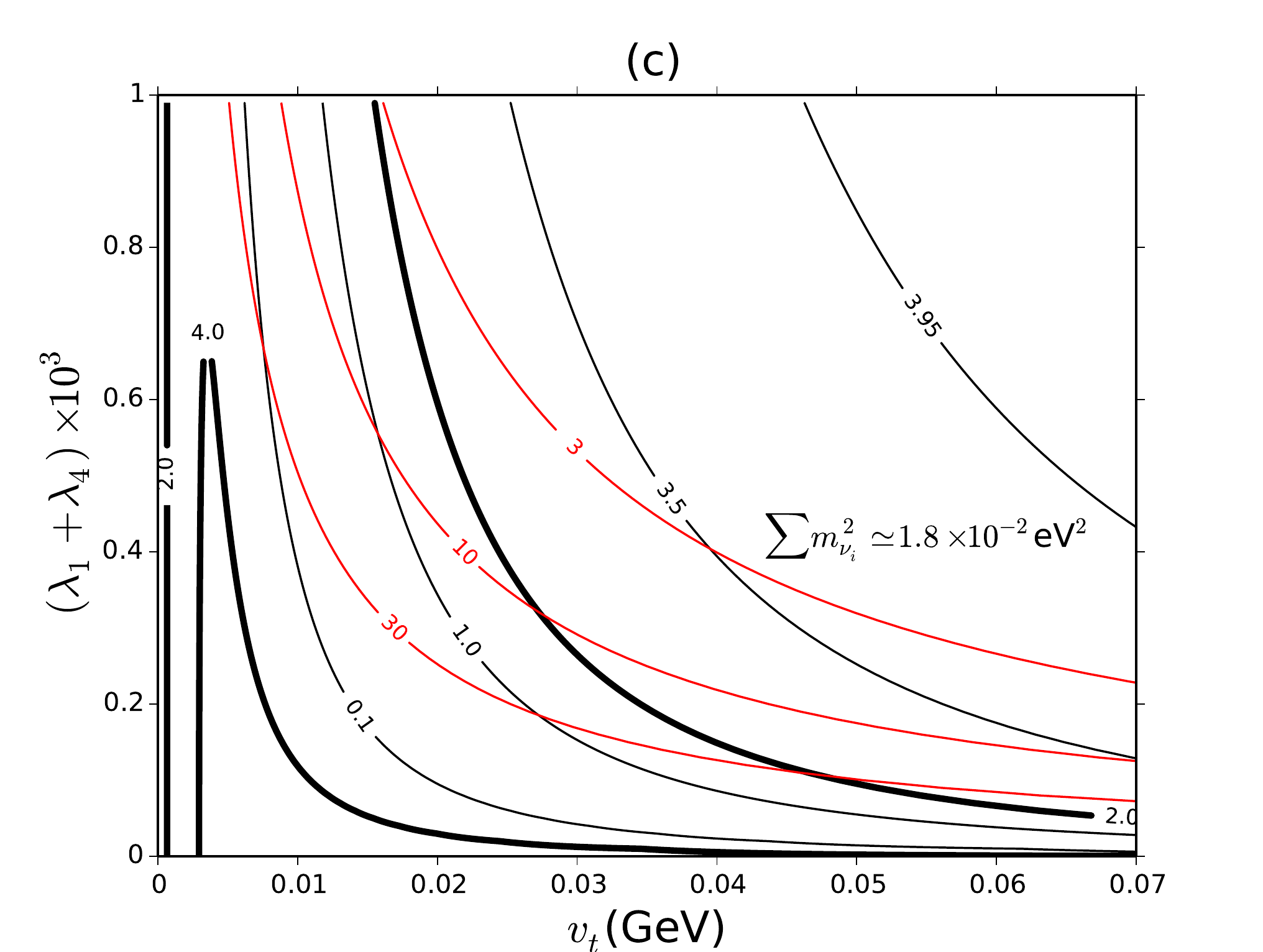}}
\caption{(a) decay length contours in the $(v_t, \lambda_{1 4}^+ \times 10^3)$ plane, with a fiducial value c$\tau=3$~meters for $h^0$ decaying visibly with
energy $E_{C.M.} =91.5$ GeV
, shown for different $h^0$ masses; (b) contour plots for
$4\times Br(A^0 \to b \bar{b}) \times Br(h^0  \to b \bar{b}) = 0.1, 1, 3.5, 3.95$  (thin black lines), $4 \times  
Br(A^0 \to b \bar{b}) \times Br(h^0  \to \nu \nu +
\bar{\nu} \bar{\nu}
)= 2, 4$, (thick black lines), and $h^0$ decay length in meters, $c\tau=3,10, 30$m (red 
lines),    in the $(v_t, \lambda_{1 4}^+ \times 10^3)$ plane, 
for $m_{A^0} \simeq m_{h^0} = 55$GeV and $ \sum m_{\nu_i}^2|_{\rm min} \simeq 2.5 \times 10^{-3}$eV${}^2$, in the limit of SM-like 
$H^0$. (c) same as (b) but with $ \sum m_{\nu_i}^2|_{\rm max} \simeq 1.8 \times 10^{-2}$eV${}^2$. }
\label{fig:branchings-LEP2}
\end{figure}



\begin{figure}[!h]
\hspace{-.8cm}\resizebox{86mm}{!}{\includegraphics{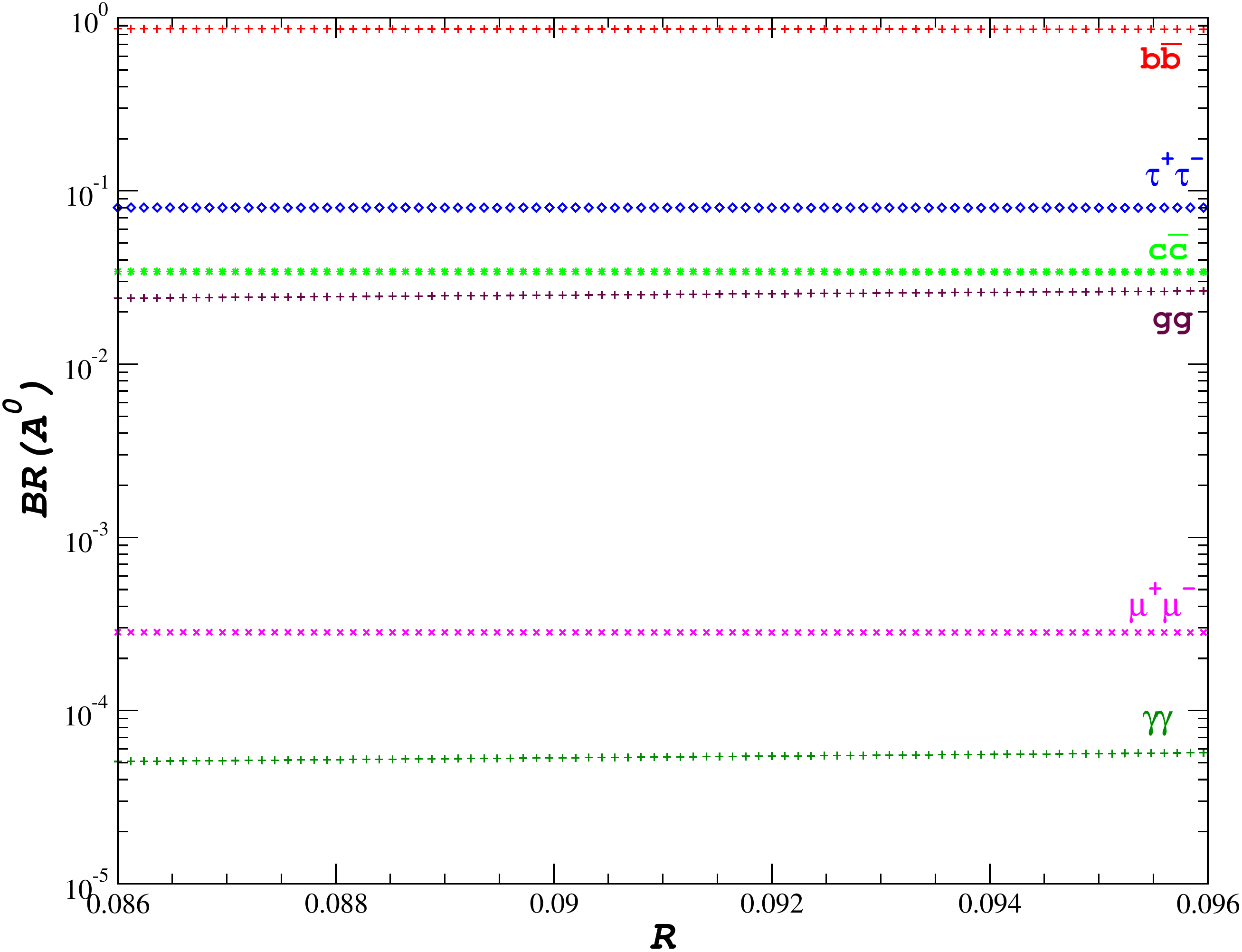}}
\caption{The branching ratios $BR(A^0)$ as a function of 
the ratio $R\equiv \frac{\mu}{v_t}$ for fixed 
$m_{H^0} = 125.5$ GeV and various $\lambda_1+\lambda_4$, with
$\lambda_2 = 0.1$, $\lambda_3=2\lambda_2$, $-10 \le\lambda_4\le2
$.}
\label{fig:BRA0}
\end{figure}


\begin{figure}[!h]
\begin{tabular}{rr}
\hspace{-.8cm}\resizebox{86mm}{!}{\includegraphics{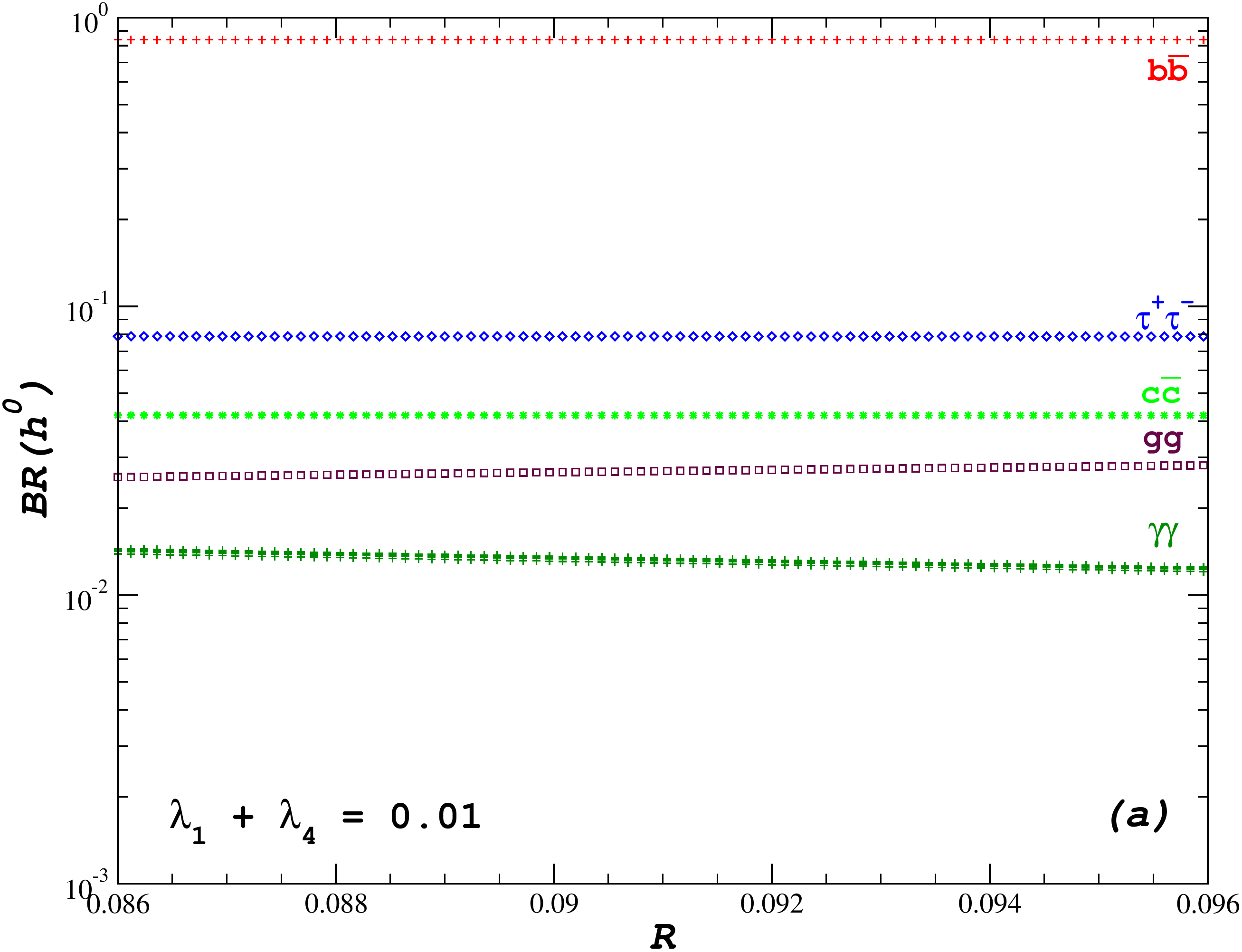}}&
\hspace{.1cm}\resizebox{86mm}{!}{\includegraphics{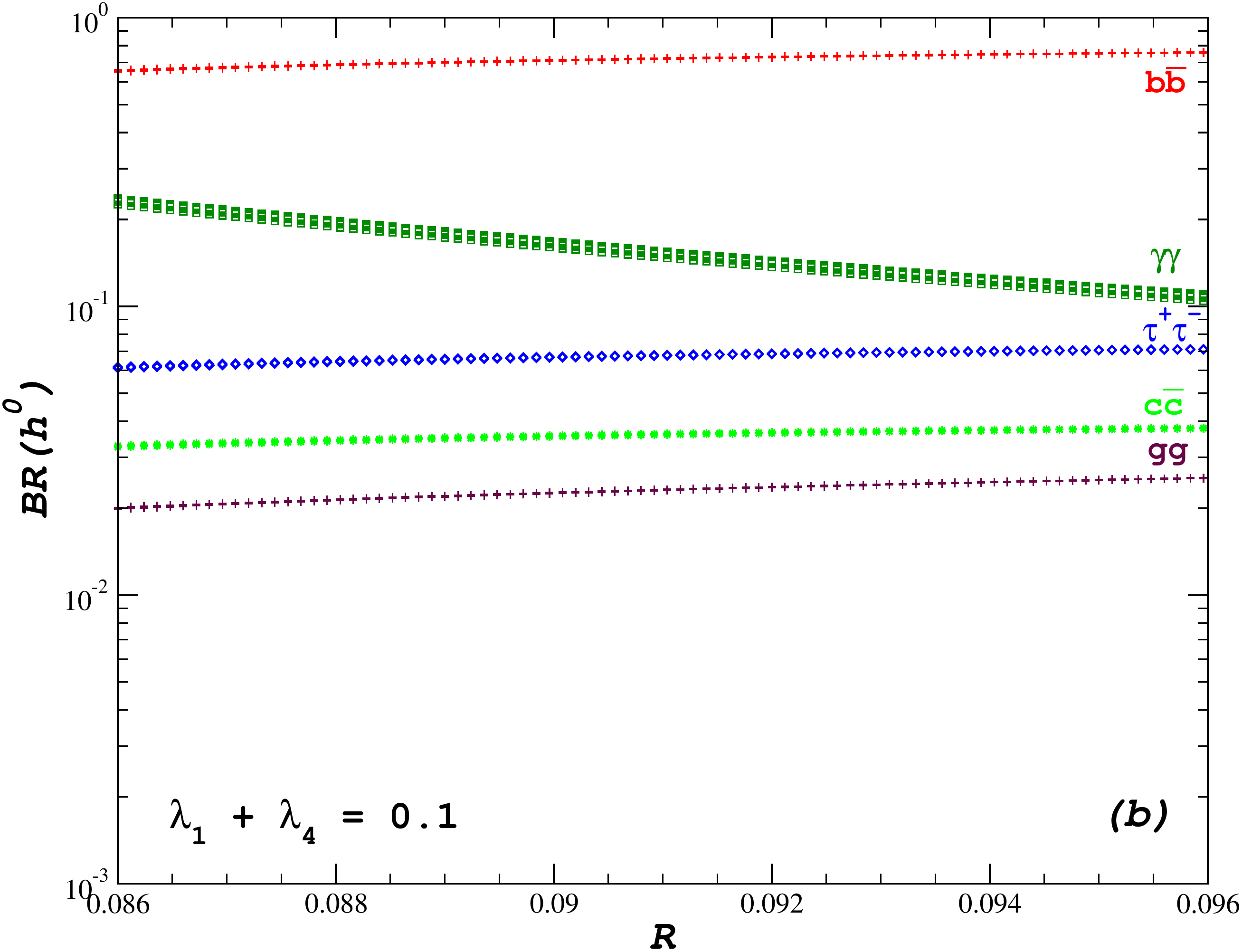}}\\\\
\hspace{-.8cm}\resizebox{86mm}{!}{\includegraphics{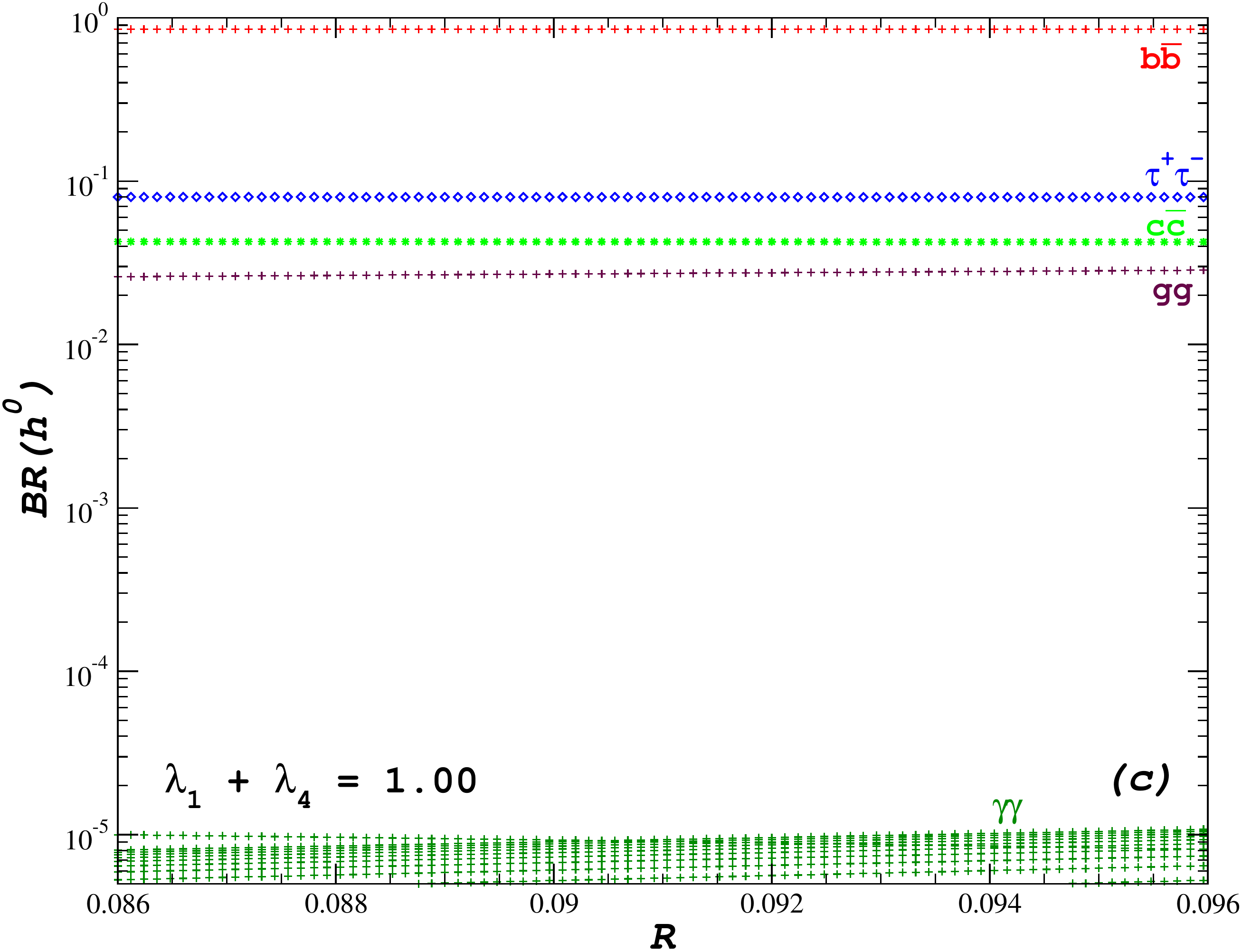}}&
\hspace{.1cm}\resizebox{86mm}{!}{\includegraphics{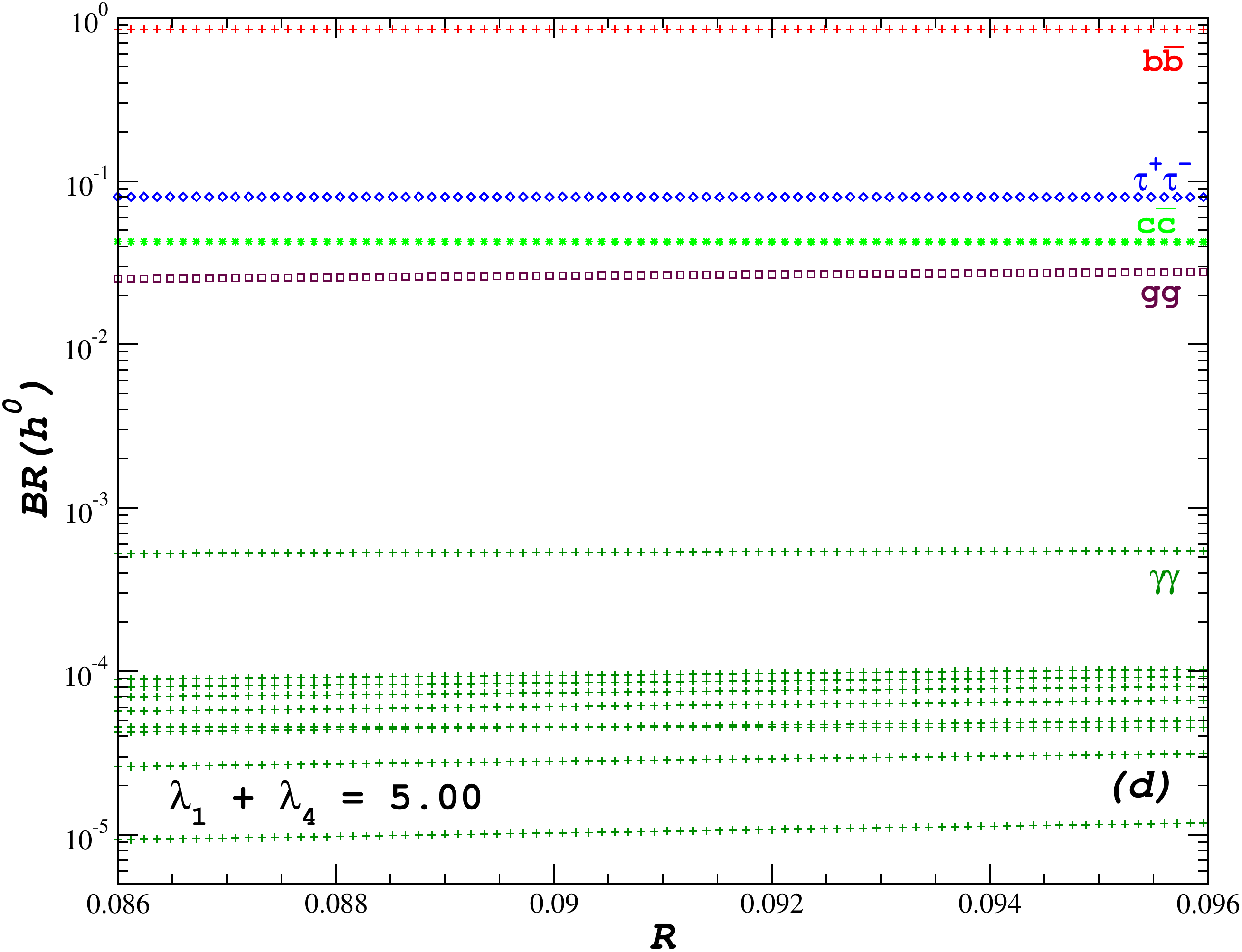}}
\end{tabular}
\caption{The branching ratios $BR(h^0)$ as a function of
the ratio $R\equiv \frac{\mu}{v_t}$
 for fixed $H^0$ mass $m_{H^0} = 125.5$ GeV and various $\lambda_1+\lambda_4$, $\lambda_2 = 0.1$, $\lambda_3=2\lambda_2$, $-10 \le\lambda_4\le2$.}
\label{fig:BRh0}
\end{figure}


\bibliographystyle{h-physrev2}
\end{document}